\documentclass[12pt,twoside,a4paper]{report}

\usepackage{amsmath}       
\usepackage{amsfonts}      
\usepackage{amssymb}       
\usepackage{amsthm}        
\usepackage{mathrsfs}      
\usepackage{yhmath}        
\usepackage{empheq}        
\usepackage{bm}            
\usepackage{mathtools}     
\usepackage{nccmath}
\usepackage{epigraph}  
\usepackage{setspace}      
\usepackage{geometry}      
\usepackage{hyperref}      

\hypersetup{
    colorlinks,
    linkcolor={blue},
    citecolor={magenta},
    urlcolor={blue},
    pdfpagelayout=TwoPageRight         
}

\usepackage{microtype}
\usepackage{xcolor}
\usepackage{amsmath}
\usepackage{graphicx}
\usepackage{subcaption}
\usepackage[utf8]{inputenc}

    \usepackage[backend=biber,style=numeric,sorting=none]{biblatex}
\usepackage[nottoc]{tocbibind}     
\usepackage{appendix}      
\usepackage{caption}       
\usepackage{booktabs}      
\usepackage{graphicx}      
\graphicspath{{./Images/}} 
\usepackage{multicol}      
\usepackage{longtable}     

\usepackage{fancyhdr}      

\usepackage[UKenglish,cleanlook]{isodate}

\usepackage{xcolor}        

\usepackage{tikz}          

 \usepackage{fontawesome}       
 \usepackage{siunitx}     
\usepackage[utf8]{inputenc}
\usepackage{titlesec}

\titleformat{\chapter}[display]
  {\normalfont\huge\bfseries}
  {\chaptername\ \thechapter}
  {20pt}
  {\Huge}
  [\vspace{1ex}\titlerule]

\theoremstyle{plain}

\theoremstyle{definition}

\theoremstyle{remark}

\newcommand{\RegNumber}{[Registration Number]}
\newcommand{\Degree}{}
\newcommand{\Department}{[Department Name]}
\newcommand{\ProjectSupervisor}{[Project Supervisor]}
\newcommand{\ThesisCommitteeMemberA}{[TC Member 1]}

\newcommand{\ThesisCommitteeMemberB}{[TC Member 2]}
\newcommand{\TCMBFacultyPosition}{[Faculty Position]}
\newcommand{\ThesisCommitteeMemberC}{[TC Member 3]}

\newcommand{\ThesisAuthor}{John Doe}
\newcommand{\ThesisTitle}{My Thesis Title}
\newcommand{\ThesisSubject}{\Degree\ Thesis, IISER Thiruvananthapuram}
\newcommand{\ThesisKeywords}{IISER Thiruvananthapuram}

\newcommand{\SetThesisTitle}[1]{\renewcommand{\ThesisTitle}{#1}}
\newcommand{\SetThesisAuthor}[1]{\renewcommand{\ThesisAuthor}{#1}}

\newcommand{\SetRegNumber}[1]{\renewcommand{\RegNumber}{#1}}
\def\msc{MSc}
\def\phd{PhD}
\def\minor{Minor}
\newcommand{\SetDegree}[1]{%
    \let\Degree=#1%
    \def\DegreeDisplay{}%
    \edef\tempDegree{#1}%
    \ifx\tempDegree\msc
        \def\DegreeDisplay{Master of Science}%
    \else
        \ifx\tempDegree\phd
            \def\DegreeDisplay{Doctor of Philosophy}%
        \else
            \ifx\tempDegree\minor
                \def\DegreeDisplay{Minor Degree}%
            \else
                \def\DegreeDisplay{#1}%
            \fi
        \fi
    \fi
}
\newcommand{\SetDepartment}[1]{\renewcommand{\Department}{#1}}
\newcommand{\SetProjectSupervisor}[1]{\renewcommand{\ProjectSupervisor}{#1}}
\newcommand{\SetThesisCommitteeMemberA}[1]{\renewcommand{\ThesisCommitteeMemberA}{#1}}

\newcommand{\SetThesisCommitteeMemberB}[1]{\renewcommand{\ThesisCommitteeMemberB}{#1}}
\newcommand{\SetTCMBFacultyPosition}[1]{\renewcommand{\TCMBFacultyPosition}{#1}}
\newcommand{\SetThesisCommitteeMemberC}[1]{\renewcommand{\ThesisCommitteeMemberC}{#1}}

\newcommand{\SetThesisKeywords}[1]{\renewcommand{\ThesisKeywords}{#1}}

\definecolor{mygrey}{RGB}{204,0,204}   

\usepackage{pgffor} 

\foreach \X in {A,...,Z} {%
    \expandafter\newcommand\csname hat\X\endcsname{\widehat{\X}}%
}

\foreach \x in {a,...,z} {%
    \expandafter\newcommand\csname hat\x\endcsname{\widehat{\x}}%
}   

\SetThesisTitle{Gravitating Tubes Beyond the World-Line Paradigm In General Relativity}
\SetThesisAuthor{Abhi Savaliya}
\SetRegNumber{I21PH040}
\SetDegree{\msc}                         
\SetDepartment{Department of Physics}
\SetProjectSupervisor{Dr. Vikash Kumar Ojha}
\SetThesisCommitteeMemberA{[TC Member 1]}
\SetThesisCommitteeMemberB{[TC Member 2]}
\SetTCMBFacultyPosition{[Faculty Position]}
\SetThesisCommitteeMemberC{[TC Member 3]}
\SetThesisKeywords{\small ABC,...}                     

\hypersetup{
    pdfauthor={\ThesisAuthor},
    pdftitle={\ThesisTitle},
    pdfsubject={\ThesisSubject},
    pdfkeywords={SVNIT, Surat, \ThesisKeywords}
}

\usepackage{etoolbox}
\makeatletter
\pretocmd{\tableofcontents}{%
  \setlength{\parskip}{0pt}%
}{}{}

\begin{document}


\newgeometry{
  inner=45pt,
  outer=48pt,
  top=1.5in
}

\pagenumbering{roman}
\setcounter{page}{3}

\begin{titlepage}
\enlargethispage{4cm}

\begin{center}

\vspace*{-1cm}

{\bf \Large \MakeUppercase{\ThesisTitle}}\\[10pt]

\vspace{10cm}
{\em  by} \\ \vspace{1mm}
{\large \bf \ThesisAuthor} \\

\vspace{4mm}

{\large  Under the supervision of }\\
\vspace{2mm}
{\large \textbf{Dr. Vikash Kumar Ojha (SVNIT, India)}}

\vspace{2mm}

{\large \textit{Submitted in partial fulfillments of the requirements for the degree of}}\\
\vspace{4mm}
{\large \textbf{Master of Science in Physics}}\\
\vspace{2mm}
{\large \textit{at}}

{\large \MakeUppercase{\Department}} \\
{ \large SARDAR VALLABHBHAI NATIONAL INSTITUTE OF TECHNOLOGY}\\
{\large SURAT 395007, GUJARAT, INDIA}\\
{\large 2025-2026}

\end{center}

\end{titlepage}
\thispagestyle{empty} 
\cleardoublepage 

\begin{flushleft} 
    \setlength{\parskip}{0pt}
    {\centering{{\Large{\bf{ABSTRACT}}}} \par} 
\end{flushleft}
\noindent
The simplest point-particle description of classical matter is incompatible with Einstein's \textit{General Relativity}, because the stress-energy tensor of a point particle is distributional and concentrated on a one-dimensional worldline. For such higher-codimension sources, smooth spacetime solutions generally do not exist, and the standard notion of energy conservation becomes problematic. This obstruction was first clarified by Geroch and Traschen for sources of codimension \(\geq 2\) in general relativity.

Respecting the Geroch--Traschen obstruction, this thesis proposes codimension-zero tubes as a more fundamental notion of gravitating matter than the standard worldline. We construct timelike tubes geometrically inside the tubular neighbourhood of an auxiliary timelike curve. To obtain a well-behaved action functional for the tube, we draw on ideas from foliation theory and foliate the finite interior region of the tubular neighbourhood by timelike codimension-one leaves, whose dynamics are governed by a brane-like action. We show that the stress-energy tensor corresponding to the collective action of these leaves is smooth and well behaved, in contrast to the distributional stress-energy tensor of a point particle.

For a broad class of tension and potential profiles, the strong energy condition is violated in the tube interior, while the null and weak energy conditions remain satisfied. In the ultraviolet limit, in which the transverse size of the tube shrinks to zero, and after an appropriate rescaling of the Lagrangian density, the tube action reduces to the worldline action together with a canonical self-force-like contribution. In the same limit, the rest mass of the point particle emerges as an effective quantity rather than being introduced as a fundamental localized parameter.

We also study the perturbative stability of the tube at two levels. First, standard field-perturbation analysis yields an infinite squared sound speed, indicating that the foliation-generating scalar field is non-dynamical in the usual wave-propagation sense and instead behaves as a geometry-constraining, \textit{cuscuton}-like field. Second, we analyse small deformations of the intrinsic geometry of the leaves, which leads to the Jacobi equation for the congruence of timelike hypersurfaces and further constrains the admissible class of tension and potential profiles.

Thus, this thesis establishes the notion of gravitating tubes not only at the geometric level, but also at the dynamical level, while consistently respecting the Geroch--Traschen obstruction.


\cleardoublepage

\tableofcontents
\clearpage
\listoffigures
\listoftables
\chapter*{List of Preprints}
\addcontentsline{toc}{chapter}{List of Pre-prints}

\begin{itemize}

    \item \textbf{Pre-prints during thesis period}
    \begin{enumerate}

        \item \textbf{A.~Savaliya}
        ``Dynamical Dark Energy from a Massive Vector Field in Generalized Proca Theory''.
        Inspire:\href{https://inspirehep.net/literature/3076204}{3076204},
        Arxiv: \href{https://arxiv.org/abs/2511.01700}{2511.01700}
        \end{enumerate}
\end{itemize}

\newpage
\restoregeometry
\pagenumbering{arabic}
\setcounter{page}{1}

\chapter{Introduction}

Einstein's theory of General Relativity \cite{Einstein1916English}, as a generalisation of Newtonian gravity, opened a fundamentally new perspective on gravitation, in which gravity is no longer interpreted as a force but rather as a manifestation of spacetime geometry. In this geometric formulation, the acceleration produced by gravitating sources is encoded in the metric structure of spacetime itself, in accordance with the equivalence principle. By virtue of this principle, a freely falling test particle in the presence of a gravitational field follows a timelike curve \(\gamma(\tau)\) on the spacetime manifold, determined by the geodesic equation
\begin{align}\label{1}
    &\nabla_{\dot\gamma}\dot\gamma =0
     \quad \text{or}\quad \frac{d^2\gamma^\lambda}{d\tau^2}+\Gamma^\lambda_{\mu\nu}\frac{d\gamma^\mu}{d\tau}\frac{d\gamma^\nu}{d\tau} =0.
\end{align}
For a given set of initial conditions, such as an initial Cauchy hypersurface together with a specified set of connection coefficients \(\Gamma^\lambda_{\mu\nu}\) associated with a pseudo-Riemannian metric \(g_{\mu\nu}\), the curve \(\gamma(\tau)\) is uniquely determined \cite{Wald:1984rg}. More generally, the curve may be regarded as a map \(\gamma:\mathbb{R}\rightarrow \mathcal{M}_{4}\), where \(\mathcal{M}_{4}\) is a four-dimensional spacetime manifold.\footnote{Strictly speaking, the map \(\gamma\) does not directly map into \(\mathbb{R}^{d}\). Rather, if \(\phi:\mathcal{M}\to\mathbb{R}^{d}\) is a coordinate chart and \(\gamma:\mathbb{R}\to\mathcal{M}\), then the coordinate representation of the curve is given by \(\phi\circ\gamma:\mathbb{R}\to\mathbb{R}^{d}\). The notation \(\gamma^\mu\) is simply shorthand for the coordinate functions associated with \(\phi\circ\gamma\).} 

If, in addition to gravity, non-gravitational interactions are present, then the motion of the particle is no longer exactly geodesic. Instead, its trajectory still lies on the spacetime manifold, but satisfies
\begin{align}\label{2}
 &\nabla_{\dot\gamma}\dot\gamma =a
     \quad \text{or}\quad \frac{d^2\gamma^\lambda}{d\tau^2}+\Gamma^\lambda_{\mu\nu}\frac{d\gamma^\mu}{d\tau}\frac{d\gamma^\nu}{d\tau} = Q^\lambda.
\end{align}
Here \(Q^\lambda\) denotes the acceleration induced by non-gravitational matter fields, including radiation, scalar fields, vector fields, tensor fields, spinor fields, and other possible interactions. In other words, if a classical particle carries a set of charges \(\{q^i\}\), including gauge charges and a gravitational coupling parameter such as its mass, and couples to a corresponding set of classical fields \(\{\Phi_i\}\), with \(i\in\mathbb{Z}^{+}\),\footnote{These fields may be scalars, vectors, spinors, higher-rank tensors, or suitable combinations thereof.} then the resulting spacetime trajectory is governed by Eq.~\eqref{2}. This represents the generic situation for a classical particle in spacetime.\footnote{For an explicit example in which \(Q^\lambda\) arises from the electromagnetic field, see Ref.~\cite{Weinberg:1972kfs}.} For freely falling observers, one has \(Q^\lambda=0\). By contrast, massless particles follow null geodesics, which are again solutions of Eq.~\eqref{1}, but parametrised by an affine parameter \(\lambda\) rather than proper time \(\tau\). In the present work, our primary focus will be on timelike trajectories.

A particle, viewed as a zero-dimensional object, traces out a one-dimensional timelike curve in spacetime. Restricting for the moment to the case \(Q^\lambda=0\), Eq.~\eqref{1} follows from extremising the point-particle action
\begin{equation}\label{Pointparticleaction}
    S_{\rm PP}[x] = -m\int d\tau =-m\int d\lambda \sqrt{-g_{\mu\nu}\frac{dx^\mu}{d\lambda}\frac{dx^\nu}{d\lambda}}.
\end{equation}
At this stage the formalism appears mathematically well defined. The corresponding stress-energy tensor derived from the point-particle action is, however, distributional:
\begin{equation}\label{pointparticlestressenergy}
    T^{\mu\nu}(x) =\frac{2}{\sqrt{-g}}\frac{\delta S_{\rm PP}}{\delta g_{\mu\nu}} = m\int d\tau\, u^\mu u^\nu\frac{\delta^{D}(x-\gamma(\tau))}{\sqrt{-g}}.
\end{equation}
This expression simply states that the matter source is concentrated on the one-dimensional worldline traced by the particle in the spacetime manifold \(\mathcal{M}\). The difficulty arises once the point-particle action \eqref{Pointparticleaction} is coupled to the Einstein--Hilbert action:
\begin{align}
    S_{\rm total} =S_{\rm EH} +S_{\rm PP} =\int d^4x \sqrt{-g}\, R  -m\int d \tau,
\end{align}
since the resulting Einstein equations now contain a delta-function source, or more generally a distributional source,\footnote{See Appendix \eqref{A.1DISTRIBUTION} for the precise definition of a distribution.}
\begin{equation}\label{distributionaleq}
    R_{\mu\nu} -\frac{1}{2}g_{\mu\nu} R =T_{\mu\nu} = m\int d\tau\, u^\mu u^\nu\frac{\delta^{(D)}(x-\gamma(\tau))}{\sqrt{-g}}.
\end{equation}

Here the source is supported on a one-dimensional curve in spacetime. As shown in the work of Geroch and Traschen \cite{GEROCHandtraschen}, the Einstein equations do not admit well-defined solutions in any sufficiently smooth class of metrics when the matter source is concentrated on sets of codimension greater than one. In particular, distributional sources supported on worldlines are incompatible with the nonlinear structure of Einstein's equations. More generally, this obstruction is not restricted to one-dimensional sources in four-dimensional spacetime. Even if one attempts to regularise a point particle by replacing it with a classical string, whose history sweeps out a two-dimensional worldsheet, the resulting source is still too singular to admit a smooth metric solution of the Einstein equations. In a \(D\)-dimensional spacetime, any source supported on a set of dimension smaller than \(D-1\) fails to admit a solution within the relevant smooth metric class. This is the well-known \textit{Geroch--Traschen obstruction}.

The original work \cite{GEROCHandtraschen} also identifies a class of metrics, known as \textit{regular metrics}, for which Einstein's equations with distributional sources can still be meaningfully discussed, at least in the sense that the curvature tensors exist as distributions. These are precisely the metrics that are locally bounded, locally square-integrable, globally invertible, and possess weak first derivatives. Even in this class, however, the resulting curvature does not in general satisfy the Bianchi identities, and consequently the conservation of the stress-energy tensor cannot be guaranteed.

A second, closely related issue associated with concentrated sources, especially worldline-supported sources, is the so-called \textit{self-force problem} in General Relativity. Because Einstein's theory is nonlinear, gravity itself gravitates. The test-particle approximation ignores this fact by assuming that the particle moves on the geodesics of a fixed background metric. Real particles, however, perturb the spacetime through their own gravitational field, and this backreaction modifies the trajectory \(\gamma(\tau)\) itself \cite{Pound_2015}. The point-particle approximation is therefore inadequate whenever the particle's mass is large enough for its gravitational backreaction to be non-negligible \cite{Barack_2018}. A sufficiently massive distribution would collapse into a black hole and can no longer be approximated as a point particle. This naturally raises the question of how to incorporate self-force effects into the dynamics of matter. There exist both perturbative and non-perturbative approaches to this problem; for pedagogical discussions, see \cite{Pound_2015,Barack_2018,Wald:2009ue}. Although we will not pursue the self-force problem further here, it is important to emphasise that the point-particle idealisation introduces deep conceptual and mathematical difficulties.

A more successful framework, consistent with the Geroch--Traschen obstruction, is Israel's thin-shell formalism \cite{Israel:1966rt,Mansouri:1996ps}. In this approach, matter is localised on a codimension-one hypersurface \(\Sigma\), across which the metric remains continuous while its first derivatives are allowed to jump, thereby generating delta-function contributions in the second derivatives. The metric in the interior region, \(\mathbf g^{-}\), is matched to the exterior metric, \(\mathbf g^{+}\), at the shell \(\Sigma\), so that continuity of the induced metric is maintained:
\begin{equation}
    [\mathbf g]_{\Sigma}=0.
\end{equation}
This is the first junction condition. The second junction condition determines the surface stress-energy tensor in terms of the jump in the extrinsic curvature,
\begin{equation}
    [K_{ab}]_\Sigma=K_{ab}^+|_\Sigma-K_{ab}^{-}|_\Sigma.
\end{equation}
A shell is absent when \([K_{ab}]_\Sigma=0\).

One may attempt to regularise a point-particle worldline using the thin-shell formalism, with the expectation that the worldline should be recovered in the limit of vanishing shell radius. Suppose, for example, that the particle worldline is replaced by a spherical shell of small radius \(R\), with Minkowski spacetime inside and Schwarzschild spacetime outside. If the particle of rest mass \(m\) is smeared over the shell and treated as pressureless matter, then the surface stress-energy tensor can be written as
\begin{equation}
    S^{ab}= \sigma U^a U^b.
\end{equation}
Evaluating the junction conditions, one finds that the surface density is inversely proportional to the shell area,
\begin{equation}
    \sigma =\frac{m}{4\pi R^2},
\end{equation}
where \(m\) is the shell's rest mass, assumed constant during the evolution. The corresponding gravitational mass of the shell is
\begin{equation}
    M_G= m\sqrt{1+\dot R^2}-\frac{m^2}{2R}.
\end{equation}
In the regime \(m>M_G\), the collapse becomes irreversible. In particular, for a weakly expanding shell satisfying \(M_G<m\), collapse is unavoidable. Thus, the thin-shell regularisation of a point particle with sufficiently large rest mass is itself gravitationally unstable.

Other attempts to avoid the Geroch--Traschen obstruction model matter as genuinely extended rather than concentrated. A notable class of approaches \cite{Dixon,papapetrou} treats the particle as a physically extended body whose stress-energy tensor is smooth within a finite worldtube rather than localised on a worldline. In such formalisms, the lowest-order dynamics yields the effective geodesic motion of the centre of mass, while higher-order terms encode effects such as spin, internal deformation, and other multipole corrections. Dixon's formalism is, in this sense, closer to physical reality, since actual matter is extended and possesses internal structure that the point-particle approximation cannot capture.

However, because Dixon's framework permits a completely general stress-energy tensor inside the worldtube, it does not by itself provide detailed information about whether the strong energy condition is satisfied there, although the weak energy condition is typically expected to hold when the matter density remains positive. Moreover, while matter is described by a smooth stress-energy tensor supported on a finite worldtube, the background spacetime metric is usually treated as fixed. The emphasis is therefore on the effective motion of extended bodies, rather than on constructing self-consistent solutions of Einstein's equations with finite tubular sources.

At this point it is useful to step briefly away from General Relativity and consider classical electrodynamics, especially Dirac's work on the classical theory of radiating electrons \cite{Dirac}. There Dirac introduced a thin auxiliary tube of sufficiently small radius around the electron's worldline in order to derive its equation of motion. In that context, the tube functions as a regulator for the divergent Coulomb self-energy associated with a worldline-supported charge. Importantly, this tube is not interpreted as a physically extended electron, but merely as a mathematical device. Dirac's analysis also considers the interaction of an extended electron with an electromagnetic pulse and finds that signal transfer through the interior of the electron may, in principle, occur faster than the speed of light, where the ``interior'' corresponds to a region in which certain elementary spacetime properties fail. Thus, even in electrodynamics, similar foundational difficulties emerge in the simplest source models.

Motivated by these considerations, the present thesis introduces \textit{tubes} as a more fundamental notion for describing matter sources than traditional worldlines, while remaining consistent with the Geroch--Traschen obstruction. The tube considered here is neither a physically extended particle in the sense of Dixon, nor merely a mathematical regulator of the kind introduced by Dirac. Rather, it is defined geometrically as a finite tubular neighbourhood surrounding an auxiliary timelike curve \(\gamma\), obtained by foliating a region in the transverse vicinity of that curve. One of the central goals of this thesis is to show that the worldline emerges as an ultraviolet limit of the tube, not merely at the geometric level, but dynamically, through the behaviour of the corresponding action functionals. This is one of the main reasons why a foliation-based construction of spacetime is required.

The foliation of a \(D\)-dimensional spacetime manifold \(\mathcal{M}\) generates a family of space-filling leaves, each of which is a \(d\)-dimensional submanifold embedded in the ambient manifold \(\mathcal{M}\). Each leaf may then be assigned dynamics through the simplest canonical action, namely the area functional. In this way, the dynamics of a leaf is governed by an area-minimising principle. In the standard literature \cite{Duff:1987cs,POLYAKOV1981207,Zwiebach_2009,Vergara:2000D/}, such dynamics is described by the Nambu--Goto action for \(d\)-dimensional surfaces or branes. We make use of this fact to construct a full action for all leaves in the foliated region by integrating the individual leaf action along the flow generated by the gradient vector field of the foliation-defining scalar field \(\Phi\). Since the leaves are taken to be of dimension \(D-1\), there is only one transverse direction along which this integration must be carried out. This construction avoids an ad hoc prescription for the action and instead provides a natural mathematical framework for associating a canonical action functional to the tube.

The resulting tube action reproduces the worldline action in the ultraviolet limit, together with an additional self-force-like correction, thereby establishing the mathematical consistency of the framework. In the same limit, the point-particle mass appears as an emergent quantity. In contrast with Dixon's background-dependent treatment of extended bodies, the interior spacetime of our tube remains dynamical with respect to the spacetime metric and satisfies Einstein's equations sourced by the stress-energy tensor derived from the tube action. This stress-energy tensor allows us to investigate the interior geometry of the tube through the usual energy conditions, and, rather strikingly, we find that the strong energy condition is violated for a broad class of tube action functionals.

Our purpose in this thesis is not to solve the full Einstein equations inside the tube, as that would require a separate and substantial analysis of its own. Rather, the immediate objective is to construct the tube geometrically, couple it consistently to gravity, and do so in a way that respects the Geroch--Traschen obstruction. In this sense, the thesis aims to replace the fundamental notion of worldline-supported matter with that of a gravitating tube, a construction which, to the best of our knowledge, has not previously been developed in this form.

\section{Structure of the Thesis}

Since the tube must first be constructed geometrically, the thesis begins with a substantial set of mathematical preliminaries. These include an introduction to hypersurfaces and their geometry, the theory of foliations and the geometry of leaves, geodesic congruences, congruences of \(d\)-dimensional surfaces, tubular neighbourhoods and their construction, and the concept of pseudo-Riemannian submersions. All of this material is presented in Chapter \ref{chapter2}.

Chapter \ref{chapter3} uses the tools developed in Chapter \ref{chapter2} to construct the tube from timelike foliations and to study the geometry of its leaves using the associated tensorial data. In this chapter, we show that a regular foliation-generating scalar field is, by itself, insufficient to incorporate the auxiliary curve that forms the core of the tube. This issue is resolved by supplementing the construction with an interpretation of the same scalar field as a Morse--Bott function. This viewpoint not only incorporates the core curve in a mathematically natural way, but also fixes the mass dimension of the scalar field.

In Chapter \ref{chapter4}, the tube is coupled to the gravitational sector by adding the newly constructed tube action to the Einstein--Hilbert action. We study the corresponding stress-energy tensor both in the interior of the tube and at its core, and we analyse the resulting energy conditions in the tube interior. In Subsection \eqref{4.1.1UVIR}, we explicitly demonstrate the emergence of the worldline action together with a canonical self-force-like term, as well as the emergence of the point-particle mass in the ultraviolet limit of the tube. Later in the chapter, in Subsection \eqref{4.2.2perturbativestability}, we investigate the perturbative stability of the tube from two distinct perspectives and use this to impose stability constraints on the admissible class of action functionals. Finally, in Subsection \eqref{Hamiltoniandynamics}, we study the Hamiltonian dynamics of the system together with the associated Dirac constraint algebra.

\chapter{Mathematical Preliminaries}\label{chapter2}

This chapter develops the mathematical framework required for the construction presented in this thesis. The notation and conventions are chosen to follow the standard literature as closely as possible, drawing primarily from Refs.~\cite{BlauGRNotes,Poisson:2009pwt,Capovilla,Capovilla_1995,mukherjee2015differential,submersion,Whitney1}. Wherever necessary, the relevant definitions, identities, and derivations will be presented explicitly and then used in the subsequent chapters.

Section \eqref{2.1geometryofhypersurface} develops the construction of an isolated hypersurface and studies both its intrinsic and extrinsic geometry in detail. Section \eqref{2.2Foliations} formally introduces the theory of foliations on pseudo-Riemannian manifolds and addresses the question of how one foliates spacetime, with particular reference to standard ADM foliations. Section \eqref{2.3 congruence} discusses congruences, including geodesic congruences and congruences of general \(d\)-dimensional submanifolds, best described through generalised Raychaudhuri equations. Section \eqref{2.4tubularneighbourhood} is comparatively more abstract and deals with the notion of tubular neighbourhoods in pseudo-Riemannian ambient manifolds. Section \eqref{2.5 submersions} introduces general submersions and pseudo-Riemannian submersions. Finally, we summarise the main ideas of these sections and explain their relevance for the constructions developed in the later chapters.

\section{Geometry of the Hypersurfaces}\label{2.1geometryofhypersurface}

Mathematically, a hypersurface \(\Sigma\) is a \((D-1)\)-dimensional submanifold of a \(D\)-dimensional ambient manifold \(\mathcal{M}\). In other words, it is a codimension-one object.\footnote{Codimension \(=\dim(\mathcal{M})-\dim(\Sigma)\).} In this section, we discuss the differential geometry of such surfaces as one of the central mathematical preliminaries for the constructions to follow. In particular, the machinery developed here will be used in Chapter 3 to construct a one-parameter family of tubular hypersurfaces around an auxiliary timelike curve \(\gamma(\tau)\). The ambient manifold \((\mathcal{M},g)\) will be taken to be a \(D\)-dimensional spacetime manifold equipped with a Lorentzian metric of mostly plus signature,
\((- ,+,+,+,\dots,+)\). The hypersurfaces to be studied are therefore \((D-1)\)-dimensional spacetime hypersurfaces. Throughout this section, we will frequently refer to Refs.~\cite{Poisson:2009pwt,BlauGRNotes} for standard constructions and results concerning both intrinsic and extrinsic geometry.

\subsection{Method of Construction}

The most convenient way to construct a hypersurface is through the level-set method. Let \((\mathcal{M},g)\) be a Lorentzian \(D\)-dimensional spacetime manifold. A hypersurface \(\Sigma\) of codimension one may be defined as the zero level set of a smooth scalar function \(\Phi:\mathcal{M}\to\mathbb{R}\), provided its gradient does not vanish on that level set.\footnote{Strictly speaking, the gradient of a function is a covector. However, in the presence of a metric tensor, one can freely raise the index and speak of the corresponding gradient vector without loss of structure.} More precisely,
\begin{equation}\label{hypersurfacedef}
    \Sigma=\bigg\{x\in\mathcal{M}\,\bigg|\,\Phi(x)=0,\quad \nabla\Phi\big|_{\Phi=0}\neq0\bigg\},
\end{equation}
so that \(\dim(\Sigma)=D-1\). The non-vanishing of the gradient in Eq.~\eqref{hypersurfacedef} is directly related to the existence of a well-defined normal vector. Any tangent vector \(V\in T(\Sigma)\) at a point \(p\in\Sigma\) must be orthogonal to the gradient of the hypersurface at that point. Explicitly,
\begin{equation}
    g(V,\nabla\Phi)=0 \qquad \text{or equivalently} \qquad V^\mu\nabla_\mu\Phi=0,
\end{equation}
since \(\Phi\) is constant along tangential directions.\footnote{By construction, \(\Phi\) does not vary along directions tangent to the hypersurface; only displacements along the gradient change its value.} Thus, the unnormalised normal to the hypersurface is given by\footnote{For the normal to be expressible in this way, it must satisfy the Frobenius integrability condition \(\nabla_\nu n_\mu-\nabla_\mu n_\nu=0\), so that \(n_\mu=\nabla_\mu\Phi\).}
\begin{equation}
    n^\mu=\nabla^\mu\Phi.
\end{equation}
The associated unit normal is therefore
\begin{equation}\label{normalvector}
    N^\mu=\frac{\nabla^\mu\Phi}{\sqrt{|\nabla_\nu\Phi\nabla^\nu\Phi|}}.
\end{equation}

Since the ambient manifold is Lorentzian, the normal vector can be timelike, spacelike, or null, according to
\begin{align}
   N^\mu \text{ is }
\begin{cases}
\text{timelike}, & \text{if } N_\mu N^\mu < 0, \\[6pt]
\text{spacelike}, & \text{if } N_\mu N^\mu > 0, \\[6pt]
\text{null}, & \text{if } N_\mu N^\mu = 0.
\end{cases}
\end{align}
The null case requires a separate treatment, which we will not consider here. The construction is invariant under smooth redefinitions \(\Phi\to\tilde\Phi=f(\Phi)\), provided \(f(0)=0\) on \(\Sigma\) and \(f'(\Phi)\big|_{\Phi=0}\neq0\). In that case,
\begin{equation}
 \tilde n^\mu = \frac{\nabla^\mu\tilde\Phi}{ \sqrt{|\nabla_\nu\tilde \Phi \, \nabla^\nu \tilde\Phi|} }
 = \frac{f'(0)\nabla^{\mu}\Phi}{\sqrt{f'(0)^2|\nabla_\nu \Phi \, \nabla^\nu \Phi|}}
 = \pm n^\mu.
\end{equation}
Thus, any such smooth redefinition changes the normal at most by a sign and therefore preserves the intrinsic and extrinsic geometry up to orientation. If one further imposes
\begin{equation}
    \nabla_\nu \Phi \,\nabla^\nu \Phi = \epsilon = \text{constant},
\end{equation}
then this becomes the eikonal equation in curved spacetime, or equivalently a Hamilton--Jacobi-type equation \cite{Misner:1973prb},
\begin{equation}
    P^\mu P_\mu-\epsilon=0,
\end{equation}
with the identification \(P^\mu=n^\mu\). This equation describes geodesics normal to the hypersurface \(\Sigma\). Indeed, taking the covariant derivative of
\(\nabla_\nu \Phi \,\nabla^\nu \Phi-\epsilon=0\) gives
\begin{align*}
    \nabla_\nu \big(\nabla_\mu \Phi \,\nabla^\mu \Phi\big)=0
    \quad \Longrightarrow \quad
    \nabla^\mu\Phi\,\nabla_\nu\nabla_\mu\Phi
    =\nabla^\mu\Phi\,\nabla_\mu\nabla_\nu\Phi=0
    \quad \Longrightarrow \quad
    n^\mu\nabla_\mu n_\nu=0.
\end{align*}
For non-constant \(\epsilon\), the normal curves are accelerated, and one instead has
\begin{equation}
    n^\mu \nabla_\mu n_\nu = \nabla_\nu \epsilon.
\end{equation}

An equivalent construction of a hypersurface is obtained by parametrising the ambient coordinates using intrinsic coordinates on \(\Sigma\). The ambient coordinates \(x^\mu\), with \(\mu\in[0,D-1]\), may be expressed as parametrised maps \(x^\mu=X^\mu(\xi^a)\), where \(\xi^a\), with \(a\in[0,D-2]\), are intrinsic coordinates on \(\Sigma\). Before proceeding, let us consider a few examples to build geometric intuition.

\begin{itemize}
    \item[] \textbf{Example 1:} Let the ambient manifold be \(\mathbb{R}^{3}\) with Euclidean metric \(g_{\mu\nu}=\mathrm{diag}(1,1,1)\) and coordinates \(x^\mu=[x^1,x^2,x^3]\). Let the embedded hypersurface be the two-sphere \(S^2\). Then
    \begin{equation}\label{Eg1}
        \Sigma=S^2\subset\mathbb{R}^{3}
        =\bigg\{x\in\mathbb{R}^{3}\,\bigg|\, \Phi(x^\mu)=(x^1)^2+(x^2)^2+(x^3)^2-a^2=0\bigg\}.
    \end{equation}
    In parametric form, with intrinsic coordinates \(\xi^a=[\theta^1,\theta^2]\),
    \begin{align}
        &x^1(\theta^1,\theta^2)= a\sin \theta^1 \cos\theta^2,\\
        &x^2(\theta^1,\theta^2)= a\sin \theta^1 \sin\theta^2,\\
        &x^3(\theta^1,\theta^2)= a\cos \theta^1.
    \end{align}
    The normal vector is then
    \begin{equation}\label{Normal1`}
        n^\mu =2[x^1,x^2,x^3]
        =2a\Big[\sin\theta^1\cos\theta^2,\quad\sin\theta^1\sin\theta^2,\quad\cos\theta^1 \Big].
    \end{equation}
    Both descriptions are equivalent and useful. One immediately verifies that the gradient of \(\Phi\) in Eq.~\eqref{Eg1} does not vanish on \(\Sigma\), and hence the unit normal \(N^\mu\) is well defined.

    \item[] \textbf{Example 2:} Consider now a hypersurface \(\Sigma\) embedded in the three-dimensional Minkowski spacetime \(M_{1,2}=\mathbb{R}_{1,2}\) with metric \(g_{\mu\nu}=\mathrm{diag}(-1,1,1)\) and coordinates \(x^\mu=[t,x^1,x^2]\). Let the hypersurface be a constant proper-time surface defined by
    \begin{equation}
        \Sigma=\bigg\{x\in M_{1,2}\,\bigg|\, \Phi(x^\mu)=a^2-t^2+(x^1)^2+(x^2)^2=0\bigg\}.
    \end{equation}
    In parametric form,
    \begin{align}
        &t(\rho,\theta)=a\sinh\rho,\\
        &x^1(\rho,\theta)=a\cosh\rho \sin\theta,\\
        &x^2(\rho,\theta)=a\cosh\rho \cos\theta,
    \end{align}
    where \(\xi^a=[\rho,\theta]\) are intrinsic coordinates on \(\Sigma\). The unit normal is then
    \begin{equation}
        N^\mu = \frac{(-t ,x^1,x^2)}{\sqrt{|-t^2+(x^1)^2+(x^2)^2|}}
        = \Big[{-\sinh\rho},\quad {\cosh(\rho )\sin\theta}, \quad{\cosh(\rho )\cos\theta}\Big].
    \end{equation}

    Figure \ref{fig:1} shows a schematic representation of this hypersurface.
    \begin{figure}[t]
    \centering
    \includegraphics[width=0.6\linewidth]{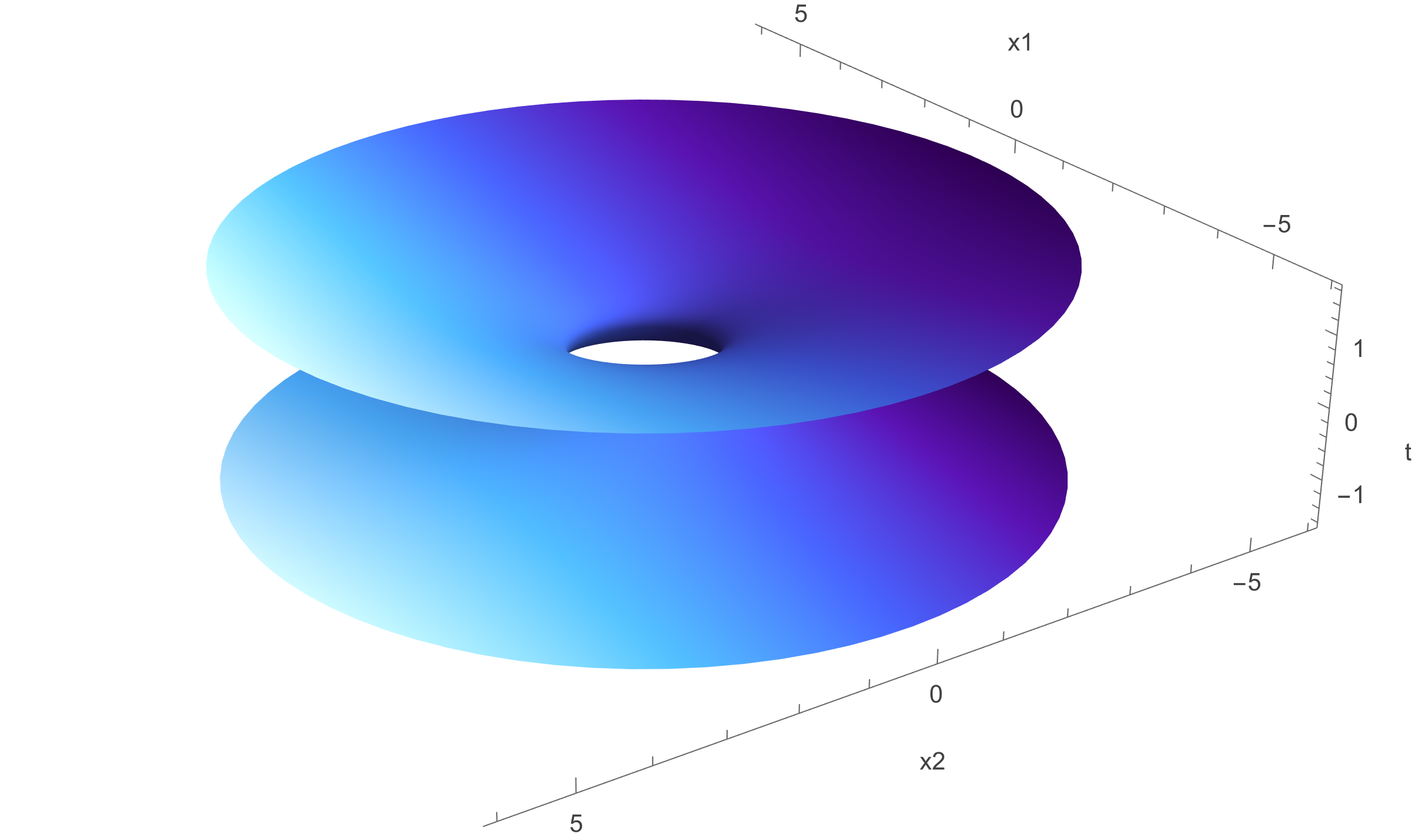}
    \caption{Hypersurface \(\Sigma\) of constant proper time \(\tau=a\).}
    \label{fig:1}
\end{figure}
\end{itemize}

Both descriptions can be unified by introducing adapted coordinates \cite{BlauGRNotes} on \(\Sigma\), namely
\begin{equation}
    X^\mu=[\Phi,\xi^a],
\end{equation}
so that the hypersurface is described by \([X^0=0, X^a(\xi)=\xi^a]\). Up to this point, the hypersurface has been constructed using only minimal data on the manifold,\footnote{The construction uses only the metric structure, the ambient and intrinsic coordinates, and the existence of smooth level sets.} but the next important step is to distinguish between tensors defined on the ambient manifold \(\mathcal{M}\) and tensors intrinsic to the hypersurface \(\Sigma\). As discussed in Blau's notes \cite{BlauGRNotes}, \(\Sigma\)-tensors are invariant under ambient-coordinate transformations, while \(\mathcal{M}\)-tensors are invariant under transformations of the intrinsic coordinates on \(\Sigma\), with each class respecting its own tensorial transformation law.\footnote{This distinction becomes clearer in the next subsection when discussing the pullback of the ambient metric tensor \(g\).} We shall refer to \(\Sigma\)-tensors as \textit{intrinsic tensors} and to \(\mathcal{M}\)-tensors as \textit{ambient tensors}.

\subsection{Intrinsic Geometry}\label{secintrinsic}

The intrinsic geometry of a hypersurface is encoded in tensors and scalars defined entirely on \(\Sigma\). Since \(\Sigma\) is an embedded submanifold of \(\mathcal{M}\), the ambient metric \(g\), together with the embedding functions \(X^\mu(\xi^a)\), naturally induces a metric tensor \(h\) on \(\Sigma\). This induced metric measures lengths, areas, and volumes intrinsic to the hypersurface. The construction is straightforward: one restricts both the ambient line element and the differentials to \(\Sigma\):
\begin{equation}
    ds^2\Big|_\Sigma
    =g_{\mu\nu}(x)dx^\mu dx^\nu\Big|_\Sigma
    =g_{\mu\nu}(X)dX^\mu dX^\nu
    =g_{\mu\nu}(X)\frac{\partial X^\mu}{\partial \xi^a}\frac{\partial X^\nu}{\partial \xi^b}d\xi^ad\xi^b
    = h_{ab}d\xi^ad\xi^b.
\end{equation}
Thus the induced metric on \(\Sigma\) is
\begin{equation}\label{Inducedmetric}
    h_{ab}(\xi)= g_{\mu\nu}(X(\xi))\frac{\partial X^\mu(\xi)}{\partial \xi^a}\frac{\partial X^\nu(\xi)}{\partial \xi^b}
    \quad \text{or equivalently} \quad
    h_{ab}=g_{\mu\nu}e^\mu_ae^\nu_b.
\end{equation}
Here \(e^\mu_a=\partial_a X^\mu\) is the Jacobian of the embedding \(\Psi:\Sigma\to\mathcal{M}\). This is not a coordinate transformation in the usual sense, even though it looks similar. Rather, it is an instance of the pullback mechanism, which restricts ambient tensors to the embedded hypersurface.\footnote{For an actual coordinate transformation, \(\Psi\) would be a local diffeomorphism \(\Psi:\mathcal{M}\to\mathcal{M}\).} We now formalise this construction via push-forward and pullback.

\textbf{A. Push-forward}

\begin{itemize}
    \item[] \(\rm Def:\) Let \(\Psi:\mathcal{N}\to\mathcal{M}\) be a smooth map between manifolds. The push-forward \(\Psi_*\) at a point \(p\in\mathcal{N}\) is the linear map
    \begin{equation}
        \Psi_* :T_p(\mathcal{N}) \rightarrow T_{\Psi(p)}(\mathcal{M}),
    \end{equation}
    which sends a vector at \(p\in\mathcal{N}\) to a vector at \(\Psi(p)\in\mathcal{M}\). Thus, for an injective map \(\Psi\),
    \begin{equation}
       \forall \ \mathbf{V}\in T_p(\mathcal{N}), \qquad \exists \ \Psi_*\mathbf{V} \in T_{\Psi(p)}(\mathcal{M}).
    \end{equation}
\end{itemize}

The push-forward acts on vectors. It therefore answers the question: given a vector \(\mathbf{V}\in T_p(\mathcal{N})\), what is its image in \(T_{\Psi(p)}(\mathcal{M})\) under the map \(\Psi\)? For our purposes, \(\mathcal{N}=\Sigma\), and the push-forward of the tangent basis on \(\Sigma\) into the ambient spacetime \(\mathcal{M}\) is
\begin{equation}\label{pushforward}
    e_a=\Psi_* \partial_{\xi^a}
    = \frac{\partial\Psi}{\partial\xi^a}\partial_{\Psi}
    = \frac{\partial X^\mu}{\partial\xi^a}\partial_{\mu}
    = e^\mu_a\partial_\mu.
\end{equation}
Here the smooth injective map \(\Psi\) is the embedding map \(X:\Sigma\to\mathcal{M}\). Thus, starting from the basis \(\partial_a\) of \(T_p(\Sigma)\), we obtain its image \(e_a\in T_{X(p)}(\mathcal{M})\). Accordingly,
\begin{equation}
    \mathbf{V}=V^a\partial_a \qquad \text{and} \qquad \mathbf{U}=V^a e_a.
\end{equation}
The objects \(e^\mu_a\) are linearly independent tangent vectors to \(\Sigma\), and since they are tangent to the hypersurface, they are orthogonal to the normal:
\begin{equation}
    g(e_a,n)=g_{\mu\nu}e^\mu_aN^\nu=0.
\end{equation}
Without a metric structure, the normal would be treated as a covector and the orthogonality condition would instead be \(N_\mu e^\mu_a=0\).

\begin{itemize}
    \item[] \textbf{Example 1a:} For the configuration in Example 1, the tangent vectors \(e^\mu_a\) on \(\Sigma=S^2\) are
    \begin{equation}\label{tangents}
        e^\mu{}_a=
        \begin{pmatrix}
        a\cos\theta^1\cos\theta^2 & -a\sin\theta^1\sin\theta^2\\
        a\cos\theta^1\sin\theta^2 & a\sin\theta^1\cos\theta^2\\
        -a\sin\theta^1 & 0
    \end{pmatrix}.
    \end{equation}
    Contracting these with the normal vector in Eq.~\eqref{Normal1`}, one directly checks that the orthogonality condition \(N_\mu e^\mu_a=0\) holds. From the push-forward formula \eqref{pushforward}, and using orthogonality, one also finds \(e_a\Phi=0\).
\end{itemize}

The push-forward, however, does not act on covectors. For that, one requires the pullback mechanism, which also allows us to derive the induced metric \(h_{ab}\) in Eq.~\eqref{Inducedmetric}.

\textbf{B. Pull-back}

\begin{itemize}
    \item[] \(\rm Def:\) Let \(\Psi:\mathcal{N}\to\mathcal{M}\) be a smooth map. The pullback \(\Psi^*\) is the linear map
    \begin{equation}
        \Psi^* : T^*_{\Psi(p)}(\mathcal{M}) \rightarrow T^*_p(\mathcal{N}),
    \end{equation}
    which pulls a covector at \(\Psi(p)\in\mathcal{M}\) back to a covector at \(p\in\mathcal{N}\). Thus, for injective \(\Psi\),
    \begin{equation}
        \forall \ \boldsymbol{\omega}\in T^*_{\Psi(p)}(\mathcal{M}), \qquad \exists \ \Psi^*\boldsymbol{\omega} \in T^*_p(\mathcal{N}).
    \end{equation}
\end{itemize}

Since covectors are defined by their action on vectors, the pullback is characterised by the relation
\begin{equation}
    \Psi^*\omega(\mathbf{V}) = \omega\big(\Psi_*(\mathbf{V})\big) \in \mathbb{R},
\end{equation}
for every \(\mathbf{V}\in T_p(\mathcal{N})\). Let \(\omega=\omega_\mu dx^\mu\) and \(\mathbf{V}=V^a\partial_{\xi^a}\). Then
\begin{equation}
    \Psi^* \omega (V^a\partial_{\xi^a})
    = V^a\omega (\Psi_* \partial_{\xi^a})
    = V^a\omega(e^\mu_a\partial_\mu)
    = V^a\omega_\mu e^\mu_a.
\end{equation}
Hence the pullback of \(\omega_\mu\) is the covector \(\Omega_a=\omega_\mu e^\mu_a\) on \(\mathcal{N}\). Returning to our case \(\mathcal{N}=\Sigma\subset\mathcal{M}\), the ambient metric can now be pulled back to yield the induced metric,
\begin{equation}
    \Psi^*(g)=h
    \qquad \text{or} \qquad
    h_{ab}=g_{\mu\nu}e^\mu_ae^\nu_b.
\end{equation}
More generally, any rank-\((0,n)\) tensor field \(T_{\mu\cdots\nu}\) can be pulled back as
\begin{equation}
    \Psi^*(\mathbf{T}) = T_{a\cdots b}=T_{\mu\cdots\nu}e^\mu_a\cdots e^\nu_b.
\end{equation}

\begin{itemize}
    \item[] \textbf{Example 1b:} For the sphere in Eq.~\eqref{Eg1}, using \(g_{\mu\nu}=\mathrm{diag}(1,1,1)\) and the tangents from Eq.~\eqref{tangents}, one obtains the induced metric
    \begin{equation}
        h_{ab} =
        \begin{pmatrix}
            a^2&0\\
            0&a^2\sin^2\theta^1
        \end{pmatrix},
    \end{equation}
    so that the line element on \(\Sigma\) becomes
    \begin{equation}
        ds^2 =a^2\big[(d\theta^1)^2+\sin^2\theta^1 (d\theta^2)^2\big].
    \end{equation}
\end{itemize}

Using the push-forward and pullback mechanisms, intrinsic tensors can be obtained very efficiently. There is, however, another useful way to restrict ambient tensors to the hypersurface, namely through projectors.

\vspace{2mm}

\textbf{C. Projectors}

The ambient metric \(g_{\mu\nu}\) may be restricted to the hypersurface in such a way that the normal direction becomes null with respect to the restricted tensor. Let \(N_\mu\) be the unit normal covector, with \(N_\mu N^\mu=\epsilon=\pm1\). Then the restricted metric is
\begin{equation}\label{Therestrictedmetric}
   g_{\mu\nu}\big|_{\Sigma}= h_{\mu\nu}= g_{\mu\nu} -\epsilon N_{\mu}N_\nu.
\end{equation}
Acting on the normal vector, one finds
\begin{equation}
    h_{\mu\nu}N^\nu=g_{\mu\nu}N^\nu -\epsilon^2N_\mu =0.
\end{equation}
Thus \(h_{\mu\nu}\) is a degenerate rank-\((0,2)\) ambient tensor, with \(N^\mu\) as a null vector.\footnote{Equivalently, \(h(N,N)=h_{\mu\nu}N^\mu N^\nu=0\).} For vectors \(V^\mu\) tangent to \(\Sigma\), one has
\begin{equation}
    h_{\mu\nu}V^\nu =g_{\mu\nu}V^\nu.
\end{equation}
Because \(h_{\mu\nu}\) is degenerate, it has no inverse. Nevertheless, one may define\footnote{Here \(h^{\mu\nu}\) is the push-forward of \(h^{ab}\), namely \(h^{\mu\nu}= h^{ab}e^\mu_ae^\nu_b\).}
\begin{equation}
    h^{\mu\nu} = g^{\alpha\mu}g^{\beta\nu}h_{\alpha\beta}.
\end{equation}
Contracting this with \(h_{\mu\nu}\) yields the tangential projector,
\begin{equation}\label{Projector}
      P^\mu_\rho=h^{\mu\nu}h_{\nu\rho} = \delta^{\mu}_\rho-\epsilon N^{\mu}N_\rho.
\end{equation}
Any vector \(V^\mu\) can then be decomposed into tangential and normal components via
\begin{equation}
    P^{\mu}_{\rho}V^\rho = V^\mu-\epsilon N^\mu (V^\rho N_\rho).
\end{equation}
The second term in Eq.~\eqref{Projector} is the normal projector,
\begin{equation}
    Q^{\mu}_{\rho} = \epsilon N^\mu N_\rho.
\end{equation}
Thus one has the completeness relation
\begin{equation}\label{decomposition}
    P^\mu_\nu +Q^{\mu}_{\nu}= \delta^\mu_\nu
    \qquad\text{or equivalently}\qquad
    \mathbf{P}\otimes\mathbf{V} +\mathbf{Q}\otimes\mathbf{V} =\mathbf{V}.
\end{equation}

The projector \(\mathbf{P}\) restricts rank-\((1,0)\) and \((0,1)\) objects to the hypersurface. It can also be used to restrict arbitrary ambient tensors. Such a restriction does not by itself produce an intrinsic tensor, since the result remains an ambient tensor; one must then pull it back to obtain an actual \(\Sigma\)-tensor. For example, given a rank-\((0,2)\) ambient tensor \(B_{\mu\nu}\), its tangential restriction is
\begin{equation}
    b_{\mu\nu} = P^\alpha_\mu P^\beta_\nu B_{\alpha\beta}.
\end{equation}
Pulling this back to \(\Sigma\), one obtains
\begin{equation}
    b_{ab} = e^\mu_ae^\nu_b b_{\mu\nu}= e^\mu_ae^\nu_b  P^\alpha_\mu P^\beta_\nu B_{\alpha\beta}  =e^\alpha_ae^\beta_b B_{\alpha\beta}.
\end{equation}
Since \(e^\mu_a N_\mu =0\), the process of restricting and then pulling back reduces to the direct pullback of \(B_{\mu\nu}\).

\vspace{2mm}

\begin{center}
\fbox{
\parbox{1.0\textwidth}{
\textbf{Lemma 1:} Let \((\mathcal{M},g)\) be an ambient manifold with metric \(g\), and let \(\mathbf{T}\in T^*(\mathcal{M})\times\cdots\times T^*(\mathcal{M})\) be a rank-\((0,n)\) tensor at some point \(\Psi(p)\in\mathcal{M}\). Let \((\Sigma,h)\) be an embedded submanifold with induced metric \(h\). Then the pullback of \(\mathbf{T}\) is equal to the pullback of its tangential projection:
\begin{equation}\label{Lemma1}
\Psi^*(\mathbf{T})=\Psi^*(\mathbf{P}^{\bigotimes n}\mathbf{T}),
\end{equation}
where \(\mathbf{P}^{\bigotimes n}=\mathbf{P}\otimes\mathbf{P}\cdots\otimes\mathbf{P}\) (\(n\) times), with
\[
\mathbf{P}:= P^\mu_\nu (\partial_\mu\otimes dx^\nu)= [\delta^\mu_\nu-\epsilon N^\mu N_\nu]\partial_\mu\otimes dx^\nu,
\]
and
\[
\mathbf{T}:= T_{\mu\cdots\nu}(dx^\mu \otimes\cdots \otimes dx^\nu).
\]
}
}
\end{center}

\vspace{2mm}

\textbf{D. Intrinsic Covariant Derivatives and Curvature Tensors}

The existence of the induced metric \(h_{ab}\) guarantees the existence of a unique torsion-free, metric-compatible connection on \(\Sigma\). This defines the intrinsic covariant derivative \(\hat{\nabla}_a\):
\begin{align}\label{intrinsic covariant deivatives}
    &\hat\nabla_a U_b = \partial_aU_b -\hat{\Gamma}_{ab}^cU_c, \\
    &\hat\nabla_a U^b = \partial_aU^b +\hat{\Gamma}_{ac}^bU^c,
\end{align}
where
\begin{equation}
    \hat{\Gamma}_{ab}^c =\frac{1}{2}h^{cd}\big[\partial_ah_{db}+\partial_bh_{da}-\partial_dh_{ab}\big],
\end{equation}
and metric compatibility implies \(\hat{\nabla}_ah_{bc}=0\). Using Lemma \eqref{Lemma1}, it is straightforward to show that the intrinsic covariant derivative is simply the pullback of the ambient covariant derivative acting on tangent covectors. More precisely, if \(V^\mu N_\mu=0\) and \(U_b=e^\mu_bV_\mu\), then
\begin{equation}\label{thepullbackofcovderiv}
    \hat\nabla_a U_b  =e^\mu_ae^\nu_b \nabla_\mu V_\nu.
\end{equation}
Similarly, the intrinsic Riemann and Ricci tensors are defined by
\begin{equation}
    \hat R_{abcd} = \partial_c \hat \Gamma_{abd}- \partial_d \hat \Gamma_{abc} +\hat \Gamma_{acm}\hat \Gamma^m_{bd}-\hat\Gamma_{adm}\hat \Gamma^m_{bc},
    \qquad
    \hat R_{ab} =h^{cd}\hat R_{cadb}.
\end{equation}

This concludes the discussion of the intrinsic geometry of hypersurfaces, governed by the first fundamental form \(h_{ab}\) induced from the ambient metric \(g_{\mu\nu}\). The geometry of an embedded hypersurface, however, is not exhausted by its intrinsic metric. There is another important question: how does the hypersurface bend inside the ambient spacetime? This is the content of the extrinsic geometry, which is not captured by \(h_{ab}\) and its derivatives alone.

\newpage

\subsection{Extrinsic Geometry}

The intrinsic geometry of a hypersurface characterises distances, areas, and volumes measured entirely within the hypersurface itself. It does not, however, fully capture how the hypersurface bends inside the ambient spacetime. To make this distinction concrete, let us consider a simple example.

\begin{itemize}
    \item[]\textbf{Example 3:} Consider the one-dimensional manifold \(S^1\) equipped with the intrinsic metric
    \begin{equation}
        ds^2=a^2d\phi^2, \qquad \phi\in[0,2\pi].
    \end{equation}
    Since the metric component is constant, the Christoffel symbols vanish, and consequently the Ricci curvature also vanishes, \(\rm R=0\). At the intrinsic level, this appears to contain the full geometric information about the circle. However, once \(S^1\) is embedded into a higher-dimensional space, such as \(\mathbb{R}^3\) or \(T^2\subset\mathbb{R}^3\), the intrinsic data no longer captures the full geometry. For example, consider the embedding
    \begin{align}
        &x^1(\phi)=[L_1+L_2\cos(q\phi)]\cos(p\phi),\\
        &x^2(\phi)=[L_1+L_2\cos(q\phi)]\sin(p\phi),\\
        &x^3(\phi)=L_2\sin(q\phi),
    \end{align}
    where \(L_1>L_2\) and \((p,q)\) are integers. The induced metric on \(S^1\subset\mathbb{R}^3\) is
    \begin{equation}
        ds^2 = \left[q^2L_2^2+p^2(L_1+L_2\cos(q\phi))^2\right]d\phi^2.
    \end{equation}
    Even though the Ricci curvature still vanishes, \(\rm R=0\), the curve is highly bent in the ambient space \(\mathbb{R}^3\), as illustrated in the three-dimensional plot below.
    \begin{figure}[htbp]
        \centering
        \includegraphics[width=0.5\linewidth]{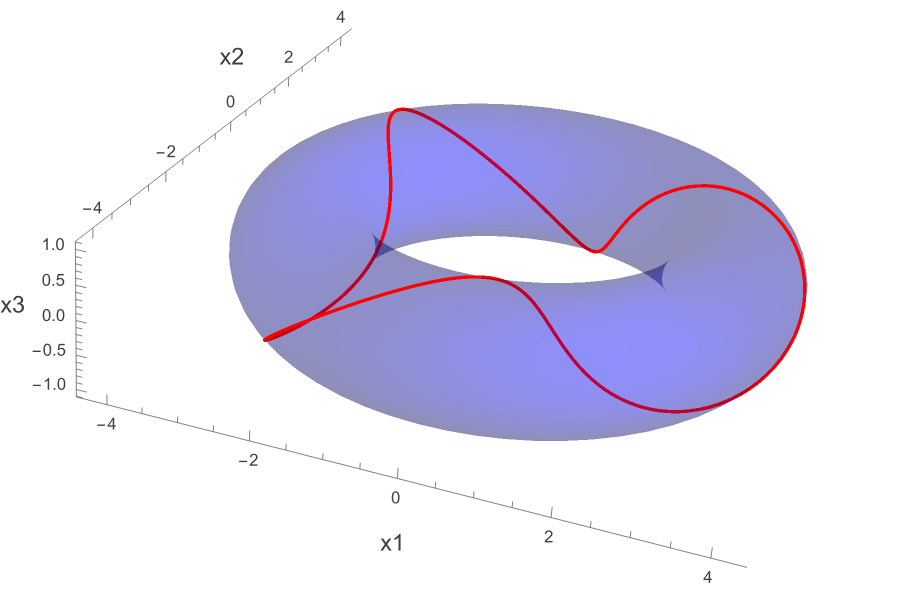}
        \caption{\(S^1\) embedded in \(\mathbb{R}^3\) as a torus knot with winding numbers \(p=1\), \(q=3\), major radius \(L_1=3\), and minor radius \(L_2=1\). The red curve has codimension \(2\) with respect to \(\mathbb{R}^3\), and codimension \(1\) with respect to \(T^2\subset\mathbb{R}^3\).}
        \label{fig:2}
    \end{figure}
\end{itemize}

This example makes clear that there is a geometric aspect hidden from intrinsic data whenever an embedding is involved. This additional aspect is called the \textit{extrinsic geometry}. In this subsection, we briefly discuss how extrinsic geometry is quantified in terms of extrinsic curvature tensors and their associated scalars. We will also study their relation to intrinsic curvature through the Gauss--Codazzi relations.

\vspace{2mm}

\textbf{Extrinsic curvature:}

A natural way to quantify extrinsic geometry is to measure how the normal covector changes as one moves along \(\Sigma\). Let \(N_\mu\) be the normal covector obtained from the level-set construction, \(N_\mu\sim\nabla_\mu\Phi\). Then the extrinsic curvature tensor on \(\mathcal{M}\) is defined as the tangential projection
\begin{equation}
    K_{\mu\nu}=P^\alpha_\mu P^\beta_\nu\nabla_\alpha N_\beta.
\end{equation}
Equivalently, \(K_{\mu\nu}\) may be defined as the tangential projection of the Lie derivative of the ambient metric along the normal direction,
\begin{equation}
    K_{\mu\nu} = \frac{1}{2}P^\alpha_\mu P^\beta_\nu \mathcal{L}_{N}g_{\alpha\beta}
    = \frac{1}{2}P^\alpha_\mu P^\beta_\nu \left[\nabla_\alpha N_\beta +\nabla_\beta N_\alpha\right].
\end{equation}
The Frobenius integrability condition,
\begin{equation}
    \nabla_\alpha N_\beta-\nabla_\beta N_\alpha=0,
\end{equation}
ensures that these two definitions coincide. Pulling this tensor back to the hypersurface yields
\begin{equation}\label{extrinsiccurvature}
    \hat K_{ab} =e^\mu_a e^\nu_b K_{\mu\nu} \xrightarrow{\text{Lemma 1}} e^\mu_ae^\nu_b \nabla_\mu N_\nu.
\end{equation}
While \(h_{ab}\) is the first fundamental form, \(\hat K_{ab}\) is the symmetric second fundamental form on \(\Sigma\). In particular, \(\hat K_{ab}\) is symmetric under exchange of the indices \(a,b\).

Two important scalar quantities can be constructed from the first and second fundamental forms.

\begin{itemize}
    \item[]\textbf{Mean curvature scalar:} Contracting Eq.~\eqref{extrinsiccurvature} with \(h^{ab}\), one obtains the mean curvature scalar,
    \begin{align}
        \hat K =h^{ab}\hat K_{ab} = h^{ab}e^\mu_ae^\nu_b \nabla_\mu N_\nu  =\nabla_\mu N^\mu,
    \end{align}
    which is the same as \(K=g^{\mu\nu}\nabla_\mu N_\nu\).

    \item[]\textbf{Squared extrinsic curvature:} This is obtained by contracting \(\hat K_{ab}\) with \(\hat K^{ab}\),
    \begin{equation}
        \hat K_{ab}\hat K^{ab} = \hat K^{ab}e^\mu_a e^\nu_bK_{\mu\nu} =h_{\mu\nu}h_{\alpha\beta}K^{\mu\alpha}K^{\nu\beta}.
    \end{equation}
    Using Eq.~\eqref{Therestrictedmetric}, the right-hand side simplifies to
    \begin{equation}
         \hat K_{ab}\hat K^{ab} =K_{\mu\nu} K^{\mu\nu} = \nabla_\mu N_\nu \nabla^\mu N^\nu-\epsilon A^\mu A_\mu,
         \qquad \text{for} \qquad A^\mu=N^\nu\nabla_\nu N^\mu.
    \end{equation}
    If the normal vector field \(N^\mu\) is tangent to a spacelike geodesic congruence, then the acceleration term vanishes, since \(A^\mu=0\).
\end{itemize}

The extrinsic curvature tensor also appears as an additional contribution when one pulls back the covariant derivative of a vector field that is not necessarily tangent to \(\Sigma\). Any vector field \(\mathbf{V}\) can be decomposed into tangential and normal parts using Eq.~\eqref{decomposition},
\begin{align}
    &\mathbf{V} =\mathbf{P}\otimes\mathbf{V}+\mathbf{Q}\otimes\mathbf{V},\\
    &V^\mu =P^\mu_\rho V^\rho+Q^\mu_\rho V^\rho.
\end{align}
The tangential part may be written as
\begin{equation}
    P^\mu_\rho V^\rho =e^\mu_a U^a=v^\mu,
\end{equation}
where \(U_a =e^\mu_a V_\mu\). Lowering the index of the decomposed vector using the ambient metric \(g_{\mu\nu}\), and then pulling back the resulting rank-\((0,2)\) tensor after acting with the \(g\)-compatible covariant derivative, one finds
\begin{equation}
    e^\mu_ae^\nu_b \nabla_\mu V_\nu = \hat \nabla_a U_b +\epsilon g(\mathbf{N},\mathbf{V})\hat{K}_{ab},
\end{equation}
where
\begin{equation}
    g(\mathbf{N},\mathbf{V}) =g_{\mu\nu} N^\mu V^\nu.
\end{equation}

Now consider the quantity \(e^\mu_a \nabla_\mu V_\beta\), which is not the full pullback of \(\nabla_\mu V_\nu\). In the case \(g(\mathbf{N},\mathbf{V})=0\), contraction with the normal vector gives
\begin{equation}
    N^\nu e^\mu_a \nabla_\mu V_\nu =-e^\mu_a V^\nu\nabla_\mu N_\nu=-\hat K_{ab} U^b,
\end{equation}
and hence the decomposition
\begin{equation}
    e^\mu_a \nabla_\mu V^\nu = e^\nu_b\hat \nabla_a U^b -\epsilon \hat K_{ab} U^b N^\nu.
\end{equation}
Contracting this expression with \(U^a\) leads to the useful relation
\begin{equation}\label{Geodesicbending}
    V^\mu\nabla_\mu V^\nu = e^\nu_b [U^a\hat \nabla_a U^b]-\epsilon \hat K_{ab} U^a U^b N^\nu.
\end{equation}

This relation has an immediate geometric interpretation. If \(K_{\mu\nu}=0\), then every geodesic of the ambient manifold \(\mathcal{M}\), satisfying \(V^\mu \nabla_\mu V^\nu=0\), is also a geodesic of the hypersurface, satisfying \(U^a\hat \nabla_a U^b=0\). Such hypersurfaces are called \textit{totally geodesic}. If, on the other hand, the extrinsic curvature is non-vanishing, then an observer confined to \(\Sigma\) and freely falling within \(\Sigma\), so that \(U^a\hat \nabla_a U^b=0\), follows a geodesic of the hypersurface but an accelerated curve in the ambient manifold. Conversely, if the curve is geodesic in the ambient manifold, \(V^\mu\nabla_\mu V^\nu=0\), then Eq.~\eqref{Geodesicbending} implies that both its tangential and normal parts must vanish separately. In that case, for non-vanishing extrinsic curvature, one must have
\begin{equation}
    \hat K_{ab} U^a U^b=0.
\end{equation}
Thus, a geodesic of the ambient spacetime can remain confined to the hypersurface while also being geodesic in \(\Sigma\) only along directions for which \(\hat K_{ab}U^aU^b\) vanishes.

These consequences of Eq.~\eqref{Geodesicbending} will be important in the later chapters. The remaining ingredient of this subsection is the set of Gauss relations.

\vspace{2mm}

\textbf{Gauss relations:}

These relations connect the intrinsic Riemann tensor \(\hat R_{abcd}\) of the hypersurface with the projected components of the ambient Riemann tensor \(R_{\mu\nu\rho\sigma}\) and the extrinsic curvature \(K_{ab}\). Consider the commutators
\begin{align}\label{commutators}
    &[\nabla_\mu , \nabla_\nu] V^\rho = {\rm R^{\rho}_{\sigma\mu\nu}}V^\sigma
    \quad \text{or} \quad
    [\nabla_\mu , \nabla_\nu] V_\rho = {\rm R_{\mu\nu\rho\sigma}}V^\sigma, \\
    &[\hat\nabla_a, \hat\nabla_b]U^c ={\rm\hat R^c_{dab}}U^d
    \quad \text{or} \quad
    [\hat\nabla_a, \hat\nabla_b]U_c ={\rm \hat R_{abcd}}U^d.
\end{align}
To derive the desired relations, one may either project the ambient commutators onto \(\Sigma\) using \(P^\mu_\nu\), or write the intrinsic commutators in terms of ambient quantities through \(\hat \nabla_a = e^\mu_a \nabla_\mu\) and \(U_c = e^\mu_c V_\mu\). After a careful calculation, one arrives at
\begin{align}
    & {\rm R_{\mu\nu\sigma\rho}\big|_\Sigma} =P^\alpha_\mu P^\beta_\nu P^\gamma_\sigma P^{\delta}_\rho{\rm R_{\alpha\beta\gamma\delta}}+\epsilon [K_{\mu\sigma} K_{\nu\rho}-K_{\mu\rho} K_{\nu\sigma}],\label{Restrictedriemann}\\
   & {\rm \hat R_{abcd}} =e^\mu_ae^\nu_be^\rho_c e^\sigma_d{ \rm R_{\mu\nu\rho\sigma}} +\epsilon [\hat K_{ac} \hat K_{bd}-\hat K_{ad}\hat K_{bc}].\label{pullbackriemann}
\end{align}
Equation \eqref{pullbackriemann} is simply the pullback of Eq.~\eqref{Restrictedriemann}. These relations explicitly show that the pullback of the ambient Riemann tensor is non-trivial in the sense of Lemma 1. The second term in Eq.~\eqref{pullbackriemann} encodes the bending of the hypersurface in the ambient manifold. For example, for \(S^n\subset\mathbb{R}^{n+1}\), the ambient Riemann tensor vanishes, yet the intrinsic curvature of the sphere is non-zero because the extrinsic curvature does not vanish, with \(K_{ab}(S^n)=g_{ab}/L\), where \(L\) is the radius of the sphere.

Contracting Eq.~\eqref{pullbackriemann} with \(h^{ac}\), one obtains the intrinsic Ricci tensor and scalar,
\begin{align}
    &{\rm \hat R_{bd}}= e^\nu_be^\sigma_d {\rm R_{\nu\sigma}} +\epsilon e^\nu_b e^\sigma_dN^\mu N^\rho{ \rm R_{\mu\nu\rho\sigma}}+\epsilon [\hat K \hat K_{bd}-\hat K_{d}^c\hat K_{bc}],\label{intrinsicricci0}\\
    & {\rm \hat R} = {\rm R}+2\epsilon {\rm R_{\mu\nu}}N^\mu N^\nu -[\hat K ^2-\hat K^{bc}\hat K_{bc}].\label{intrinsicricci}
\end{align}
The scalar \(\hat R\) is often referred to as the Gaussian curvature of \(\Sigma\), and Eq.~\eqref{intrinsicricci} is the scalar Gauss relation. Similar manipulations yield the intrinsic Einstein tensor,
\begin{equation}\label{pulledbackeinsteintensor}
\hat G_{ab} = e^\mu_ae^\nu_b G_{\mu\nu}+\epsilon e^\nu_b e^\sigma_dN^\mu N^\rho{ \rm R_{\mu\nu\rho\sigma}}+\epsilon [\hat K \hat K_{bd}-\hat K_{d}^c\hat K_{bc}]-\frac{1}{2}h_{ab}\bigg[2\epsilon {\rm R_{\mu\nu}}N^\mu N^\nu -[\hat K ^2-\hat K^{bc}\hat K_{bc}]\bigg].
\end{equation}
Once again, even in a flat ambient manifold, the intrinsic Einstein tensor need not vanish. The additional terms in Eqs.~\eqref{pullbackriemann}, \eqref{pulledbackeinsteintensor}, \eqref{intrinsicricci}, and \eqref{intrinsicricci0} are precisely what distinguish extrinsic geometry from intrinsic geometry.

The hypersurfaces discussed so far are all codimension-one objects. Embedded submanifolds of general codimension possess a much richer geometric structure. We will briefly discuss such higher-codimension surfaces in the context of generalised Raychaudhuri equations and the dynamics of their congruences in the next sections.

\newpage

\section{Foliations}\label{2.2Foliations}

So far, we have analysed a single hypersurface embedded in an ambient manifold, together with its intrinsic and extrinsic geometry. In many physical and geometrical situations, however, one must consider a continuous family of hypersurfaces filling a region of spacetime, rather than a single isolated hypersurface. A standard example is the Hamiltonian formulation of General Relativity, which requires the ADM decomposition of spacetime \cite{Misner:1973prb}. Under global hyperbolicity, the spacetime manifold admits a Cauchy surface, and the manifold is diffeomorphic to the product \(\mathcal{M}\simeq \mathbb{R}\times \Sigma\), with leaves \(\{\Sigma_t\}\) given by non-intersecting spacelike hypersurfaces belonging to the same continuous family.

The construction of such a family is also essential for our own purpose, namely the construction of a tube around a concentrated source of codimension greater than one, in our case a timelike curve \(\gamma\). This motivates the introduction of a continuously parametrised family of hypersurfaces and, more generally, of a foliation. Our discussion here follows Ref.~\cite{gourgoulhon200731formalismbasesnumerical}. That reference focuses on the \(3+1\) formalism based on spacelike foliations,\footnote{By a spacelike foliation we mean that all leaves are spacelike hypersurfaces.} whereas our eventual goal is to construct timelike foliations. Nevertheless, we adopt from it the general method of construction, the associated kinematics, and the use of adapted coordinates, which we will later extend to tubular timelike foliations in Chapter 3.

\subsection{Construction and Geometry}

Before discussing the specific case of spacelike Cauchy foliations, it is useful to recall the general definition of a foliation. Following Ref.~\cite{Lawson1974}, one defines:

\begin{center}
\fbox{
\parbox{1.0\textwidth}{
\textbf{Foliation:} A \(p\)-dimensional, class \(C^r\) foliation of a \(D\)-dimensional manifold \(\mathcal{M}\) is a decomposition of \(\mathcal{M}\) into a union of disjoint connected subsets \(\{L_{\alpha}\}\), called the leaves of the foliation, with the following property:

Every point in \(\mathcal{M}\) has a neighbourhood \(U\) and a system of local coordinates \(\mathbf{x}:U\to \mathbb{R}^D\) of class \(C^r\), such that for each leaf \(L_\alpha\), the connected components of \(U\cap L_\alpha\) are described by the equations \(\{x^{p+1}=c^1, \cdots , x^D=c^D\}\), where each \(c^i\) is constant. Such a foliation is denoted by
\begin{equation}\label{Foliations}
    \mathcal{F} =\{L_{\alpha}\},
\end{equation}
and every leaf is an embedded \(p\)-dimensional submanifold.
}
}
\end{center}

For foliations to exist, the relevant integrability conditions must be satisfied, in particular the Frobenius integrability condition for \(p>1\). Nevertheless, every open manifold admits a \(p=D-1\) dimensional, or codimension-one, foliation; see Ref.~\cite{Lawson1974}. This definition will be used later in Chapter 3 for our construction.

Under global hyperbolicity, spacetime admits a Cauchy hypersurface \(\Sigma\), which is spacelike. In that case the global topology takes the product form
\begin{equation}
    \mathcal{M}\simeq \Sigma\times \mathbb{R}.
\end{equation}
The spacetime may then be foliated by constant-time spacelike leaves, with the foliation written as
\begin{equation}
    \mathcal{F}= \{\Sigma_s\} = \bigg\{x\in\mathcal{M}\bigg| \quad \Phi(x)=s, \quad \nabla\Phi\big|_{\Phi=s}\neq0,\quad \forall s\in\mathbb{R}\bigg\}.
\end{equation}

\begin{figure}[htbp]
    \centering
    \includegraphics[width=0.5\linewidth]{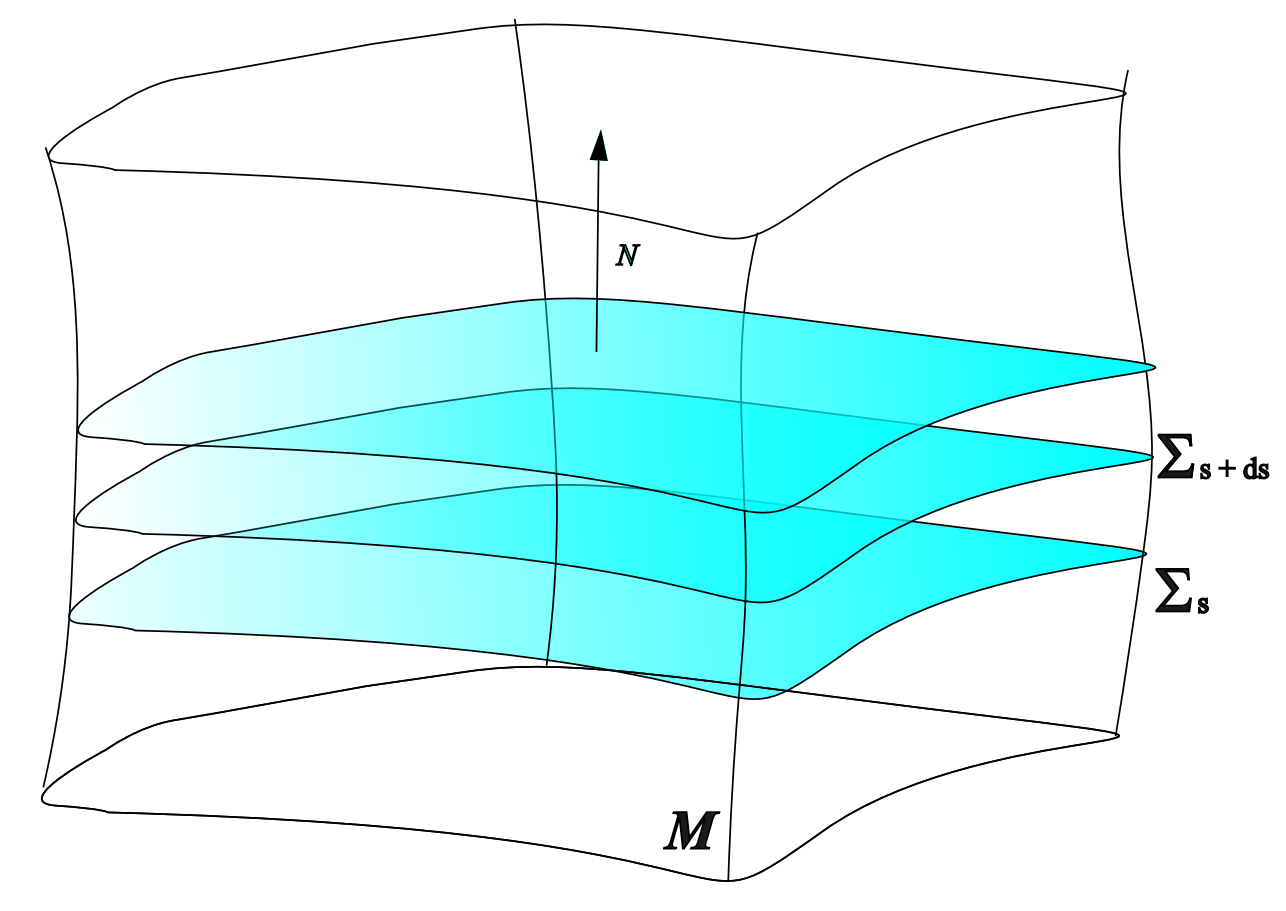}
    \caption{A continuous family of hypersurfaces in the region \(M\). For Cauchy foliations, \(M=\mathcal{M}\) and \(t=s\), whereas in general \(M\subset\mathcal{M}\).}
    \label{fig:3}
\end{figure}

\noindent\textbf{A. Lapse function:}

The unit normal vector to the leaf \(\Sigma_s\) is collinear with \(\nabla\Phi\), with normalisation factor \(\mathcal{N}\):
\begin{equation}
    N^\mu = -\mathcal{N} \nabla^\mu \Phi=-\mathcal{N} \nabla^\mu t.
\end{equation}
The minus sign is chosen so that the normal remains future-directed when the scalar field increases toward the future. The normalisation factor is
\begin{equation}
    \mathcal{N} =(-g^{\mu\nu}\nabla_\mu \Phi\nabla_\nu\Phi)^{-1/2},
\end{equation}
and is called the \textit{lapse function}. In compact notation, the unit normal vector \(\mathbf{N}\) and the corresponding covector \(\underline{\mathbf{N}}\) may be written as
\begin{equation}\label{newnormal}
    \mathbf{N}=-\mathcal{N}\vec \nabla \Phi
    \quad \text{and} \quad
    \underline{\mathbf{N}} =-\mathcal{N}\mathbf d\Phi,
\end{equation}
where \(\mathbf{d}\Phi\) is the gradient one-form. By construction, the lapse function is always positive.

\vspace{2mm}

\noindent\textbf{B. Normal evolution vector:}

One now defines the normal evolution vector as the timelike vector normal to \(\Sigma_s\),
\begin{equation}
    \mathbf{m}=\mathcal{N}\mathbf{N},
    \qquad \text{with} \qquad
    \mathbf{g}(\mathbf{m},\mathbf{m})=-\mathcal{N}^2.
\end{equation}
This vector is especially useful because it satisfies
\begin{equation}
    \mathbf{g}(\mathbf{m},\vec{\nabla}\Phi)=1
    \qquad \text{or equivalently} \qquad
    g_{\mu\nu}m^\mu \nabla^\nu \Phi=1.
\end{equation}
Because of this property, the nearby hypersurface \(\Sigma_{s+\delta s}\) may be obtained by Lie-dragging the points of \(\Sigma_s\) along \(\mathbf{m}\). If \(p\in\Sigma_s\), then
\begin{equation}
    p' = p+\mathbf{m}\,\delta\Phi,
    \qquad \Phi(p')=\Phi(p)+\delta\Phi.
\end{equation}

\vspace{2mm}

\noindent\textbf{C. Acceleration:}

The acceleration of an Eulerian observer moving along \(\mathbf{N}\) is defined by
\begin{equation}
    \mathbf{a} =\nabla_{\mathbf {N}}\mathbf{N}.
\end{equation}
Using Eq.~\eqref{newnormal}, this can be written as
\begin{equation}
    a_\mu =P^\alpha_\mu\nabla_\alpha \ln\mathcal{N}
    \quad \text{or} \quad
    a_\mu =D_\mu\ln \mathcal{N}.
\end{equation}
Since \(g_{\mu\nu}a^\mu N^\nu=0\), the acceleration is tangent to the leaf. It also appears in the decomposition of derivatives of the normal and of the normal evolution vector:
\begin{align}
    &\nabla_{\beta} N_\alpha = -K_{\alpha\beta} -a_\alpha N_\beta,\\
    &\nabla_\beta m^\alpha= -\mathcal{N}K_{\beta}^\alpha-N_\beta D^\alpha\ln \mathcal{N}+N^\alpha \nabla_\beta\mathcal{N}.
\end{align}
In the presence of non-zero acceleration, the projected Ricci tensor and the intrinsic Ricci scalar acquire additional terms,
\begin{align}
   &{\rm  R_{\mu\nu}} \bigg|_{\Sigma_s}= P^\alpha_\mu P^\beta_\nu {\rm R_{\alpha\beta}} - P^\alpha_\mu P^\beta_\nu N^\gamma N^\delta{ \rm R_{\gamma\alpha\delta\beta}}-[ K  K_{\mu\nu}- K_{\nu}^\rho K_{\mu\rho}]+ \frac{1}{\mathcal{N}}D_{\mu}D_{\nu}\mathcal{N},\\
   &{\rm \hat R} = {\rm R} -2{\rm R}_{\mu\nu}N^\mu N^\nu -[\hat K ^2-\hat K^{bc}\hat K_{bc}]+\frac{1}{\mathcal{N}}D_\mu D^\mu\mathcal{N}+\frac{1}{\mathcal{N}}N^\mu N^\nu D_\mu D_\nu\mathcal{N}.
\end{align}
The intrinsic Einstein tensor is similarly modified:
\begin{align}
       &\hat G_{ab} =e^\mu_a e^\nu_b G_{\mu\nu} - e^\nu_a e^\sigma_bN^\mu N^\rho{ \rm R_{\mu\nu\rho\sigma}}- [\hat K \hat K_{ab}-\hat K_{a}^c\hat K_{bc}]-\frac{1}{2}h_{ab}\bigg[2\epsilon {\rm R_{\mu\nu}}N^\mu N^\nu -[\hat K ^2-\hat K^{bc}\hat K_{bc}]\bigg]\nonumber\\
       &\quad \quad \quad +\frac{1}{\mathcal{N}}\bigg[e^\mu_ae^\nu_b D_\mu D_\nu\mathcal{N}-\frac{1}{2}h_{ab}\bigg[D_\mu D^\mu\mathcal{N}+N^\mu N^\nu D_\mu D_\nu\mathcal{N}\bigg]\bigg].
\end{align}
Here we are working with a timelike normal field, so \(\epsilon=N^\mu N_\mu=-1\). These expressions show that the lapse function \(\mathcal{N}\) plays an important role in the geometry of the foliation.

To interpret it geometrically, regard the unit timelike normal vector as the velocity vector \(N^\mu=U^\mu\) of a future-directed timelike observer following a curve \(\gamma(\tau)\) that intersects each hypersurface at a single point. Then an infinitesimal displacement \(\delta t\,\mathbf{m}\) along that curve measures the proper time interval \(\delta\tau\) in terms of the coordinate time interval \(\delta t\):
\begin{equation}
    \delta\tau=\sqrt{-\mathbf{g}(\delta t\mathbf{m},\delta t\mathbf{m})}=\mathcal{N}\delta t.
\end{equation}
For comoving observers, the lapse is constant, \(\mathcal{N}=1\). At this stage, however, we do not restrict ourselves to any particular class of observers. It is now natural to discuss the adapted coordinate system associated with such a foliation.

\subsection{Adapted Metric and the Coordinates}\label{ADMadaptedmetricandcoordinates}

Let the coordinates on a leaf \(\Sigma_s\) be \(\xi^a\), with \(a\in[1,2,3]\). Provided they vary smoothly from one leaf to the next, one may define coordinates on the entire foliated region by setting
\begin{equation}
    x^\mu=[\Phi=t,\xi^a],
\end{equation}
that is, \(x^0=\Phi=t\) and \(x^a=\xi^a\). The coordinates \(\xi^a\) may naturally be interpreted as spatial coordinates. The corresponding basis for the tangent space \(T_p(\mathcal{M})\) is
\begin{equation}
    \boldsymbol{\partial_\mu}=[\boldsymbol{\partial_t},\boldsymbol{\partial_a}].
\end{equation}

\begin{figure}[htbp]
    \centering
    \includegraphics[width=0.5\linewidth]{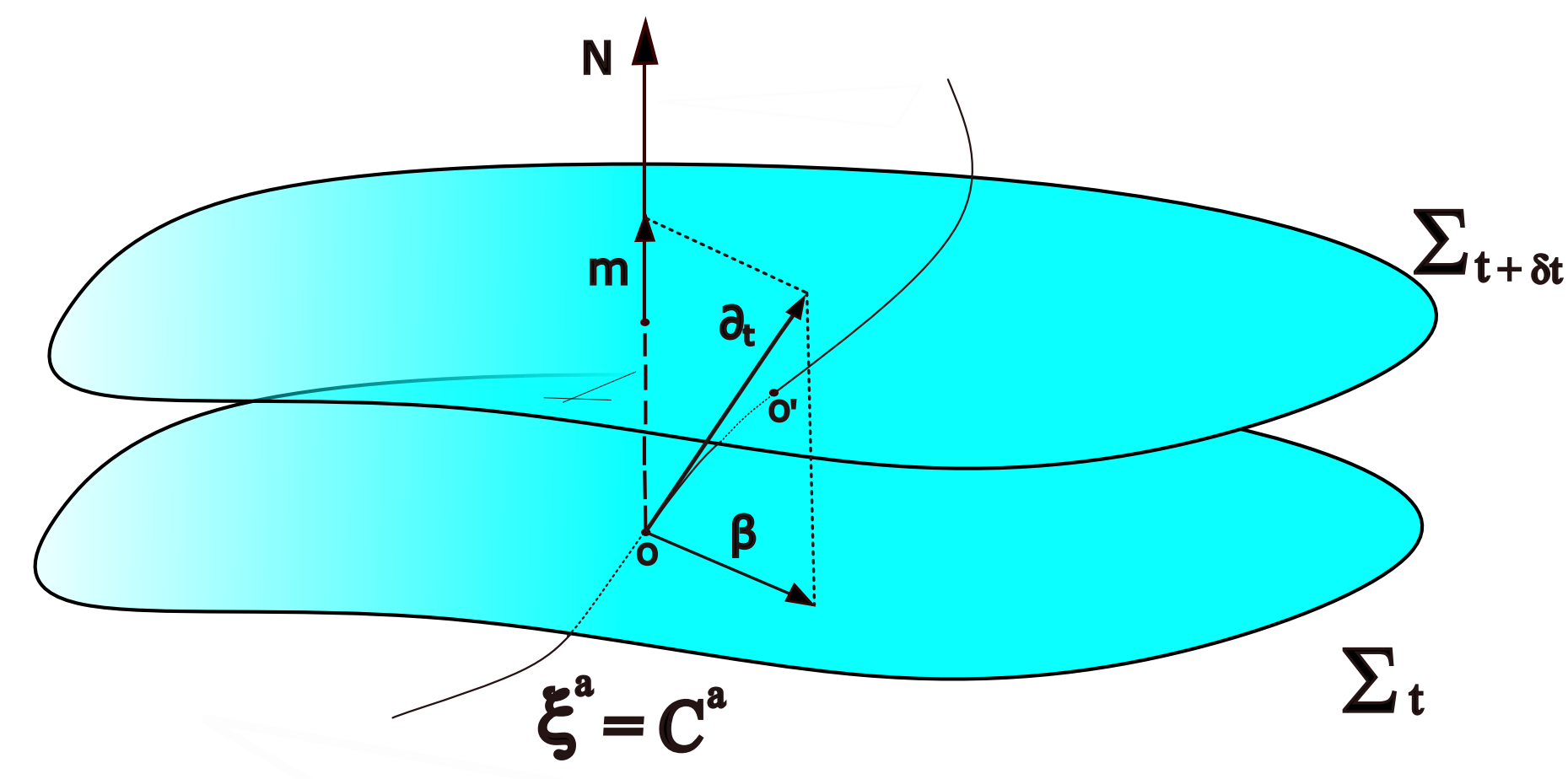}
    \caption{Each point \(\xi^a=C^a\) on the hypersurface \(\Sigma_t\) evolves along the vector \(\boldsymbol{\partial}_t\) and traces a curve in \(\mathcal{M}\).}
    \label{fig:4}
\end{figure}

Each point on the leaf \(\Sigma_t\) is the intersection of the coordinate surfaces \(\xi^a=C^a\), where \(C^a\in\mathbb{R}\) are constants. These coordinate points are not required to evolve along the unit normal vector \(\mathbf{N}\) or along the normal evolution vector \(\mathbf{m}\). Instead, they evolve along \(\boldsymbol{\partial}_t\), which need not be timelike a priori. The difference between \(\mathbf{m}\) and \(\boldsymbol{\partial}_t\) is encoded in the shift vector \(\boldsymbol{\beta}\):
\begin{equation}\label{Shiftdecompo}
     \boldsymbol{\partial_t}= \mathbf{m}+\boldsymbol{\beta}.
\end{equation}
Here \(\boldsymbol{\partial}_t\) is tangent to the curve traced by the point \(O:\xi^a=C^a\), while the vectors \(\boldsymbol{\partial}_a\) are tangent to the leaf \(\Sigma_t\), so that \(\boldsymbol{\partial}_a\in T_O(\Sigma_t)\).

To show that the shift vector is tangential to the leaf, consider the dual basis \(\mathbf{d}x^\mu\in T^*_O(\mathcal{M})\), satisfying
\begin{equation}\label{veccovecmapping}
    \langle \mathbf{d}x^\mu,\boldsymbol{\partial}_\nu\rangle=\delta^\mu_\nu.
\end{equation}
Since \(\langle\mathbf{d}t,\boldsymbol{\partial}_t\rangle=1\), and
\begin{equation}\label{suport}
    \langle\mathbf{d}t ,\boldsymbol{m}\rangle
    =\langle\mathbf{d}t ,\mathcal{N}\mathbf{N}\rangle
    =\mathcal{N}\langle\mathbf{d}t ,\mathbf{N}\rangle
    =-\langle\underline{\mathbf{N}} ,\mathbf{N}\rangle
    =1,
\end{equation}
it follows from Eq.~\eqref{Shiftdecompo} that
\begin{equation}
   \langle \mathbf{d}t,\boldsymbol{\beta} \rangle
   =\langle\mathbf{d}t, \boldsymbol{\partial}_t\rangle-\langle\mathbf{d}t ,\boldsymbol{m}\rangle=0.
\end{equation}
Hence \(\boldsymbol{\beta}\) indeed lies in the tangent space of the leaf. The evolution vector therefore decomposes into a normal and a tangential part:
\begin{equation}
     \boldsymbol{\partial_t}= \mathcal{N}\mathbf{N}+\boldsymbol{\beta}.
\end{equation}
It follows immediately that
\begin{equation}\label{00component}
    \mathbf{g}(\boldsymbol{\partial}_t,\boldsymbol{\partial}_t)=g_{tt} = -\mathcal{N}^2+\beta_a \beta^a.
\end{equation}
Therefore the causal character of \(\boldsymbol{\partial}_t\) is determined by
\begin{align}
   \text{\(\boldsymbol{\partial}_t\) is }
\begin{cases}
\text{timelike}, & \text{if} -\mathcal{N}^2+\beta_a \beta^a   < 0, \\[6pt]
\text{spacelike}, & \text{if } -\mathcal{N}^2+\beta_a \beta^a > 0, \\[6pt]
\text{null}, & \text{if } -\mathcal{N}^2+\beta_a \beta^a = 0.
\end{cases}
\end{align}
If the shift vanishes, \(\beta^a=0\), then the evolution is timelike and proceeds entirely along the normal evolution vector \(\mathbf{m}\), provided \(\mathcal{N}\neq0\). Since \(\boldsymbol{\beta}\) is tangent to \(\Sigma\), it may be written as
\begin{equation}
    \boldsymbol{\beta} =\beta^a \boldsymbol{\partial}_a
    \quad \text{and} \quad
    \underline{\boldsymbol\beta}= \beta_a \mathbf{d}\xi^a=\beta_a\mathbf{d}x^a.
\end{equation}
Using Eq.~\eqref{Shiftdecompo}, the components of the unit normal vector in the basis \(\{\boldsymbol{\partial}_\mu\}\) are
\begin{equation}\label{normalinbeta}
    N^\mu =\bigg[\frac{1}{\mathcal{N}},\frac{-\beta^a}{\mathcal{N}}\bigg],
\end{equation}
while the normal covector follows from Eq.~\eqref{newnormal}:
\begin{equation}\label{conormalinbeta}
     N_\mu =\bigg[-{\mathcal{N}}, 0\bigg].
\end{equation}

\vspace{2mm}

\noindent\textbf{A. The metric tensor}

The ambient metric and the induced metric are given by
\begin{equation}
    \mathbf{g} = g_{\mu\nu}\mathbf{d}x^\mu\otimes\mathbf{d}x^\nu
    \quad \text{and} \quad
    \mathbf{h}= h_{ab}\mathbf{d}\xi^a\otimes\mathbf{d}\xi^b,
\end{equation}
respectively, with components
\begin{equation}
    g_{\mu\nu} = \mathbf{g}(\boldsymbol{\partial}_\mu,\boldsymbol{\partial}_\nu)
    \quad \text{and} \quad
    h_{ab} =\mathbf{h}(\boldsymbol{\partial}_a,\boldsymbol{\partial}_b ).
\end{equation}
Since \(h_{ab}\) is the pullback of the ambient metric, we now compute the ambient metric components in these adapted coordinates. The \(g_{00}=g_{tt}\) component has already been obtained in Eq.~\eqref{00component}:
\begin{equation}
    g_{00}=g_{tt}= -\mathcal{N}^2+\beta_a\beta^a.
\end{equation}
The mixed components are
\begin{equation}
    g_{a0} =g_{at}=\mathbf{g}(\boldsymbol{\partial}_a,\boldsymbol{\partial}_t).
\end{equation}
Using Eqs.~\eqref{normalinbeta} and \eqref{conormalinbeta}, one has \(g_{a\mu}N^\mu=0\), and hence
\begin{equation}
    g_{a0}=g_{ab} \beta^b.
\end{equation}
The purely spatial components are simply
\begin{equation}
    g_{ab} =\mathbf{g}(\boldsymbol{\partial}_a,\boldsymbol{\partial}_b) = h_{ab}.
\end{equation}
Therefore, the full ambient metric and its inverse in the foliated region are
\begin{equation}\label{theadaptedmetric}
    g_{\mu\nu}=
            \begin{pmatrix}
                -\mathcal{N}^2+\beta_a\beta^a&\beta_b\\
                \beta_a& h_{ab}
            \end{pmatrix},\quad 
    g^{\mu\nu}=
            \begin{pmatrix}
                -\frac{1}{\mathcal{N}^2}&\frac{\beta^b}{\mathcal{N}^2}\\
                \frac{\beta^a}{\mathcal{N}^2}&h^{ab} -\frac{\beta^a\beta^b}{\mathcal{N}^2}
            \end{pmatrix}.
\end{equation}
In general, \(g_{ab}=h_{ab}\), but \(g^{ab}\neq h^{ab}\). This is the standard adapted metric for global spacelike, or Cauchy, foliations, commonly recognised as the \(3+1\) metric. In Chapter 3, we will follow an analogous procedure to obtain the metric corresponding to a different type of foliation, namely timelike foliations in the tubular neighbourhood of a timelike curve.

\newpage
\section{Congruences}\label{2.3 congruence}

In the previous section, we studied foliations of spacetime, in particular Cauchy foliations. At first sight, the construction appears globally regular, but foliations may break down at certain points, which we shall refer to as critical points. The goal of this section is to study the conditions under which such breakdown occurs. It turns out that the theory of geodesic congruences is especially useful for this purpose. For geodesic congruences, we will frequently refer to Ref.~\cite{Poisson:2009pwt}, adapting its presentation to the language needed for our construction. However, the notion of geodesic congruence is only the starting point. One may go beyond curves to congruences of hypersurfaces and, more generally, to congruences of submanifolds of arbitrary codimension. For that broader setting, we draw on Refs.~\cite{Capovilla,Capovilla_1995}, which study the dynamics of relativistic membranes and generalise congruence theory in both perturbative and non-perturbative settings.

\subsection{Geodesic Congruence}

A congruence of curves is a family of one-dimensional curves whose members never intersect. Since we work in a pseudo-Riemannian setting, congruences of geodesics may be timelike, spacelike, or null. The null case is not relevant for our present purposes, whereas the timelike and spacelike cases admit closely analogous constructions.

A geodesic congruence may be regarded as a one-dimensional foliation of an open region \(\mathcal{U}\subset\mathcal{M}\), with the leaves \(L_\alpha\) being non-null geodesics, in the sense of the general foliation definition \eqref{Foliations}. We will return to this foliation viewpoint shortly. As an intuitive picture, imagine a sphere of dust in four-dimensional spacetime, with each particle following a timelike geodesic. Because of ambient curvature, the dust cloud may in general expand or contract, shear, and rotate. In the language of continuum kinematics, these effects are encoded in a deformation tensor.

Let \(\gamma(\tau)\) be a fiducial timelike geodesic with unit tangent vector \(U^\mu\), satisfying \(U^\mu U_\mu=-1\). Let \(U^\mu\) be normal to a spacelike hypersurface \(\Sigma_t\), whose induced metric is
\begin{equation}
     h_{\mu\nu} = g_{\mu\nu}+U^\mu U^\nu.
\end{equation}
The corresponding tangential projector is
\begin{equation}
     P^\mu_\nu =\delta^\mu_\nu+U^\mu U_\nu.
\end{equation}
Now consider the rank-\((0,2)\) tensor built from the tangent field,
\begin{equation}
     B_{\mu\nu}= \nabla_{\mu}U_\nu,
     \qquad
     U^\mu B_{\mu\nu}=U^\mu \nabla_\mu U_\nu=0.
\end{equation}
The first equality follows because the curve is geodesic and hence unaccelerated. Moreover, \(U^\nu B_{\mu\nu}=0\) also holds because \(U^\nu U_\nu=-1\). Projecting this tensor onto \(\Sigma_t\) does not yield anything new, but one may identify the projected form with the extrinsic curvature tensor of \(\Sigma_t\):
\begin{equation}\label{ExtrinsictensorofSigma}
    K_{\mu\nu} =P^\alpha_\mu P^\beta_\nu \nabla_\alpha U_\beta = B_{\mu\nu}.
\end{equation}
When viewed as a matrix, this tensor admits a decomposition into expansion, shear, and rotation:
\begin{equation}\label{distortiontensor}
    K_{\mu\nu} = \frac{1}{3}\theta h_{\mu\nu}+\sigma_{\mu\nu}+\omega_{\mu\nu},
\end{equation}
where
\begin{align}
    &\sigma_{\mu\nu}= \frac{1}{2}[B_{\mu\nu}+B_{\nu\mu}]-\frac{1}{3}\theta h_{\mu\nu},\\
    &\omega_{\mu\nu} = \frac{1}{2}[B_{\mu\nu}-B_{\nu\mu}],\label{rotationtensor}
\end{align}
and
\begin{equation}
    \theta=g^{\mu\nu}B_{\mu\nu}
\end{equation}
is the expansion scalar. Congruences with vanishing rotation tensor, \(\omega_{\mu\nu}=0\), are said to be hypersurface orthogonal, meaning that all members of the congruence are orthogonal to a family of hypersurfaces \(\{\Sigma_t\}\).\footnote{See Ref.~\cite{Poisson:2009pwt} for a proof.}

To derive the evolution equation for \(\theta\), consider
\begin{align*}
    U^\rho \nabla _\rho B_{\mu\nu}
    &= U^\rho \nabla_\rho \nabla_\mu U_\nu\\
    &= U^\rho\Big[\nabla_{\mu}\nabla_\rho U_\nu- {\rm R}_{\mu\sigma\nu\rho}U^\sigma \Big]\\
    &= -B_{\mu\sigma}B^{\sigma}_{\ \nu} - {\rm R}_{\mu\rho\nu\sigma} U^\rho U^\sigma.
\end{align*}
Taking the trace gives
\begin{equation}
    \frac{d\theta}{d\tau} = -B_{\mu\nu}B^{\mu\nu}-{\rm R}_{\mu\nu}U^\mu U^\nu.
\end{equation}
Substituting the decomposition \eqref{distortiontensor}, one obtains the Raychaudhuri equation:
\begin{equation}\label{RaychaudhuriEq}
    \frac{d\theta}{d\tau} =-\frac{1}{3}\theta^2-\sigma_{\mu\nu}\sigma^{\mu\nu}+\omega_{\mu\nu}\omega^{\mu \nu}- {\rm R}_{\mu\nu} U^\mu U^\nu.
\end{equation}
Since the shear and rotation tensors are purely spatial, one has
\begin{equation}
    \sigma_{\mu\nu}\sigma^{\mu\nu}\geq0,
    \qquad
    \omega_{\mu\nu}\omega^{\mu\nu}\geq0.
\end{equation}

This equation is of central importance in General Relativity. For a hypersurface-orthogonal congruence, \(\omega_{\mu\nu}=0\), so the first two terms on the right-hand side are non-positive. The sign of the last term becomes clear once Einstein's equation is invoked:
\begin{align}
    &{\rm R}_{\mu\nu}-\frac{1}{2}g_{\mu\nu}{\rm R}= M_p^2{\rm T}_{\mu\nu}
    \quad \Rightarrow \quad
    {\rm R}_{\mu\nu} =M_p^2\Big[{\rm T}_{\mu\nu} -\frac{1}{2}g_{\mu\nu} {\rm T}\Big],\\
    &{\rm R}_{\mu\nu}U^\mu U^\nu=M_p^2\Big[{\rm T_{\mu\nu}}U^\mu U^\nu+\frac{1}{2}{\rm T}\Big].\label{Energycondition}
\end{align}
The requirement that \({\rm R}_{\mu\nu}U^\mu U^\nu\geq0\) is the strong energy condition. If this condition holds, then Raychaudhuri's equation implies
\begin{equation}
    \frac{d\theta}{d\tau}\leq0.
\end{equation}
Thus the expansion decreases along the congruence. An initially diverging congruence, \(\theta>0\), diverges less rapidly with evolution, while an initially converging congruence, \(\theta<0\), converges more rapidly. In this sense, gravity behaves attractively when the strong energy condition holds. In particular, an initially converging congruence is guaranteed to develop a caustic.

\begin{figure}[htbp]
    \centering
    \includegraphics[width=0.5\linewidth]{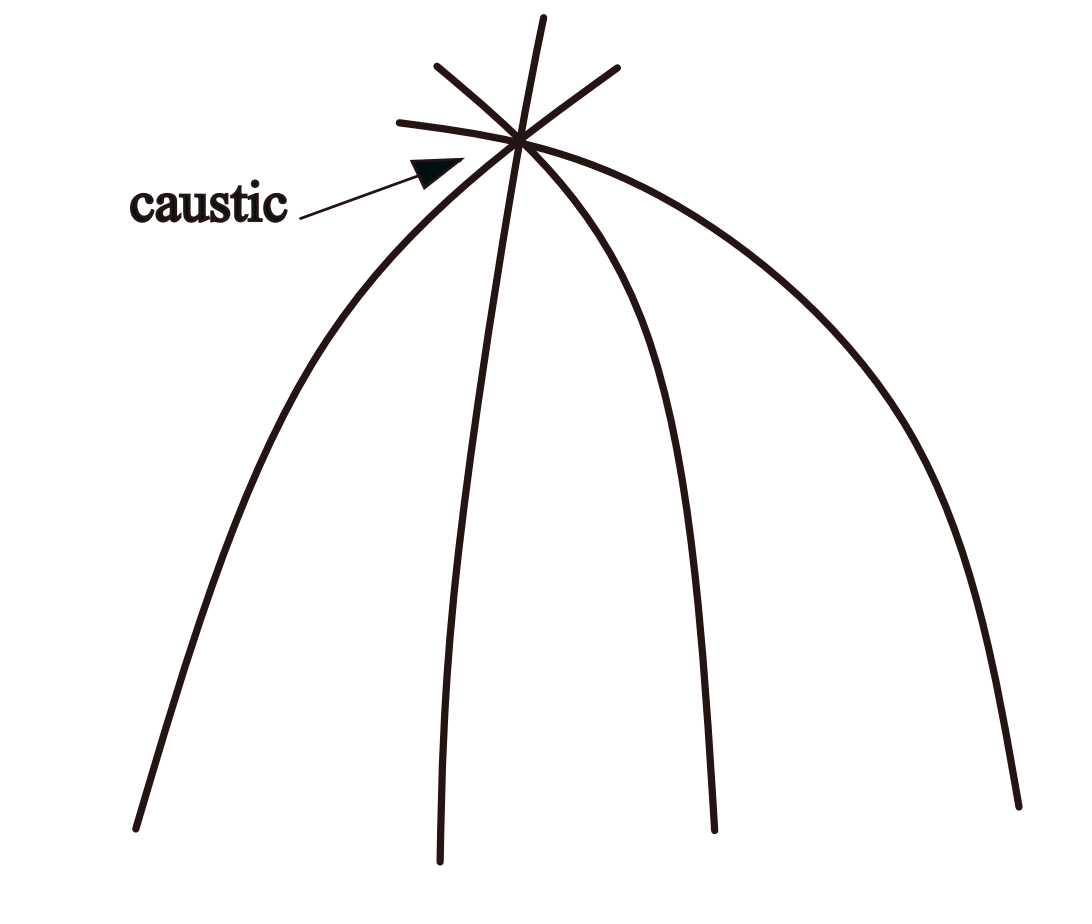}
    \caption{A converging congruence forming a caustic.}
    \label{fig:6}
\end{figure}

One may say that the one-dimensional timelike foliation of the open region \(\mathcal{U}\subset\mathcal{M}\) degenerates at the caustic.\footnote{A more precise quantification of degeneracy will be given in Chapter 3.} If the evolution parameter is proper time, then one may verify that this degeneration occurs within finite proper time by integrating
\begin{equation}
    \frac{d\theta}{d\tau}\leq-\frac{1}{3}\theta^2
    \quad \xRightarrow{\text{integrating}} \quad
    \theta^{-1}(\tau)= \theta^{-1}_0+\frac{\tau}{3},
\end{equation}
where \(\theta_0=\theta(0)\). Since \(\theta_0<0\), the expansion scalar blows up to \(\theta=-\infty\) within
\begin{equation}
    \tau\leq \frac{3}{|\theta_0|}.
\end{equation}
A similar analysis applies to spacelike congruences, where the parameter is proper distance rather than proper time. In that case, a caustic forms at a finite proper distance if the spacelike geodesics are initially converging and if
\begin{equation}
    {\rm R_{\mu\nu}}N^\mu N^\nu\geq0,
\end{equation}
where \(N^\mu\) now denotes the tangent to the spacelike geodesics. The blow-up of the expansion scalar may also be viewed as the blow-up of the extrinsic curvature scalar \(K\) of the spacelike hypersurface \(\Sigma_t\), since \(K=\theta\). Geometrically, this means that the hypersurfaces become extremely bent in the ambient spacetime at the caustic.

\subsection{Congruence of Submanifolds}\label{congruenceofsubmanifolds}

A congruence of non-null geodesics is a one-dimensional foliation of an open region \(\mathcal{U}\subset\mathcal{M}\). More generally, however, a spacetime, or a region of spacetime, may be foliated by leaves of any dimension \(d<D\). In that case, the analysis of caustic formation is considerably more involved than in the one-dimensional case described by the Raychaudhuri equation \eqref{RaychaudhuriEq}. There, one benefits from the facts that the shear and rotation scalars are non-negative,
\begin{equation}
    \sigma_{\mu\nu}\sigma^{\mu\nu}\geq0,
    \qquad
    \omega_{\mu\nu}\omega^{\mu\nu}\geq0,
\end{equation}
and that the governing equation is first order. These features make it relatively straightforward to determine the onset of caustics once Einstein's equations and the energy conditions are imposed.

For foliations by \(d\)-dimensional submanifolds, the situation is more intricate. Ref.~\cite{Capovilla_1995} develops a generalised form of the Raychaudhuri equation for congruences of \(d\)-dimensional membranes, or equivalently \(d\)-dimensional foliations. In the next subsection we import the necessary geometric machinery from that framework, which will later be used to study the formation of caustics, and hence the bounds of the tubular neighbourhood of a timelike curve, in the next chapter. 

\vspace{2mm}

\noindent\textbf{\underline{Geometry of the Submanifold:}}

\vspace{2mm}

Let \(\Sigma\) be a \(d\)-dimensional embedded timelike submanifold of a \(D\)-dimensional spacetime manifold \(\mathcal{M}\). Let the coordinates on \(\Sigma\) be \(\xi^a\), with \(a\in[0,d-1]\), and let the embedding maps be
\begin{equation}\label{submanifoldembedding}
    x^\mu =X^\mu(\xi^a),
\end{equation}
for \(\mu\in[0,D-1]\). \textit{Throughout this subsection, Greek indices \((\mu,\nu,\rho,\dots)\) refer to tensors on the ambient manifold, whereas lower-case Latin indices \((a,b,\dots)\) refer to tensors intrinsic to \(\Sigma\).}

The metric induced on \(\Sigma\) is
\begin{equation}
    \mathbf{g}(\mathbf e_a,\mathbf e_b)=h_{ab} =g_{\mu\nu}e^\mu_a e^\nu_b,
    \qquad \text{where} \qquad
    e^\mu_a=\frac{\partial X^\mu}{\partial\xi^a},
    \qquad
    \mathbf{e}_a =e^\mu_a\boldsymbol{\partial}_{\mu}.
\end{equation}
One may equivalently work in an orthonormal basis, for which
\begin{equation}
    \mathbf{g}(\mathbf{E}_a, \mathbf{E}_b) = \eta_{ab}.
\end{equation}

The tangent space along \(\Sigma\) splits into tangential and normal parts; see Eq.~\eqref{splittingoftangentspace}. The normal space is spanned by \(D-d\) independent normals, denoted by \(\{\mathbf{N}^A\}\), with \(A\in[1,D-d]\). \textit{Capital Latin indices are used to label the normal basis.} The Gram matrix of the normals is positive definite, since the normal space has index zero:
\begin{equation}
    G^{AB}= \mathbf{g}(\mathbf{N}^A,\mathbf{N}^B)= g_{\mu\nu} N^{\mu A}N^{\nu B}.
\end{equation}
In practice, one usually chooses an orthonormal normal frame such that
\begin{equation}
    G^{AB}=\delta^{AB}.
\end{equation}
The capital indices are then raised and lowered with \(\delta^{AB}\) and \(\delta_{AB}\). All normals are orthogonal to the tangents to \(\Sigma\):
\begin{equation}
    \mathbf{g}(\mathbf{N}^A,\mathbf{E}_a)=g_{\mu\nu}N^{\mu A}E^\nu_a=0.
\end{equation}

The restricted degenerate metric on \(\Sigma\), together with its inverse-like counterpart, is defined by
\begin{equation}
    h_{\mu\nu} =g_{\mu\nu}-G_{AB}N_{\mu}^AN_\nu ^B,
    \qquad \text{and} \qquad
    h^{\mu\nu}=g^{\mu\nu}-G_{AB}N^{\mu A}N^{\nu B}.
\end{equation}
It is then natural to define the tangential projector that restricts ambient tensors to \(\Sigma\):
\begin{equation}
    P^{\mu}_\nu= \delta^{\mu}_\nu-G_{AB}N^{\mu A}N_{\nu}^B.
\end{equation}

This projector may now be used to define the extrinsic curvature tensor, whose pullback to \(\Sigma\) is
\begin{equation}
    K_{\mu\nu}^A = P^\alpha _\mu P^\beta_\nu \nabla_{\alpha}N_\beta^A
    \quad \xRightarrow{\text{pullback}} \quad
    \hat K^{A}_{ab} =E^\mu_a E^\nu_b \nabla_\mu N^A_\nu.
\end{equation}
The surface is said to be \textit{extremal} if
\begin{equation}
    h^{\mu\nu}K^A_{\mu\nu}=h^{ab} \hat K^A_{ab}=0.
\end{equation}
In this way, the notion of a geodesic is generalised from curves to higher-dimensional surfaces, and arises naturally from Nambu--Goto-type actions describing relativistic membranes.

The discussion of action-based surface dynamics will be taken up shortly. Before that, let us introduce the intrinsic covariant derivative
\begin{equation}
    \hat{\nabla}_a =E^\mu_a \nabla_\mu,
\end{equation}
viewed as the pullback of the ambient, \(g_{\mu\nu}\)-compatible covariant derivative \(\nabla_\mu\). Let the surface \(\Sigma\) be spanned by the orthonormal frame \(\{\mathbf{E}_a,\mathbf{N}^A\}\). The gradients of these basis vectors satisfy the generalised Gauss--Weingarten equations,
\begin{align}\label{Generalized Gauss-Weingarten}
    &\hat{\nabla}_a\mathbf{E}_b = \hat{\Gamma}^c_{ab}\mathbf{E}_c-G_{AB}\hat K_{ab}^A \mathbf{N}^B,\\
    &\hat{\nabla}_a \mathbf{N}^A= \hat {K}_{ab}^A \mathbf{E}^b+G_{BC}\omega_a^{AB}\mathbf{N}^C.
\end{align}
These equations give the orthonormal decomposition of the gradients of the adapted basis. The coefficients \(\hat{\Gamma}_{abc}=-\hat{\Gamma}_{acb}\) are the Ricci rotation coefficients intrinsic to \(\Sigma\), while \(\omega_a^{AB}=-\omega^{BA}_a\) are the twist potentials of \(\Sigma\).

A standard computation involving the commutators of covariant derivatives in Eq.~\eqref{commutators}, together with the generalised Gauss--Weingarten equations, leads to relations between the intrinsic and ambient curvatures, namely the Gauss--Codazzi relations:
\begin{equation}
    {\rm \hat R}_{abcd}=E^\mu_a E^\nu_b E^\rho_c E^\sigma_d {\rm R}_{\mu\nu\rho\sigma} +G_{AB}[K^{A}_{ac}K^{B}_{bd}-K^{A}_{ad}K^{B}_{bc}].
\end{equation}
The intrinsic and extrinsic geometry must also satisfy the Codazzi--Mainardi and Ricci integrability conditions, respectively:
\begin{align}
    &\bar{\nabla}_a \hat K^A_{bc}-\bar{\nabla}_b \hat K^A_{ac} = E^\mu_a E^\nu_bE^\rho_cN^{\sigma A}{\rm R}_{\mu\nu\rho\sigma},\\
    &\Omega^{AB}_{ab}-h^{cd}\hat K^A_{ac}\hat K^B_{db}+h^{cd}\hat K^A_{bc}\hat K^B_{da}=E^\mu_a E^\nu_b N^{\rho A}N^{\sigma B}{\rm R}_{\mu\nu\rho\sigma}.
\end{align}
Here \(\bar \nabla_a\) is the covariant derivative acting on fields that transform as tensors under rotations of the normal frame:
\begin{equation}\label{normalfield covairant derivative}
    \bar \nabla_a \mathbf{T}^{A} =\hat\nabla_a \mathbf{ T}^A-G_{CD}\omega^{AC}_{a}\mathbf{T}^D,
\end{equation}
and \(\Omega\) is the curvature associated with the twist potential,
\begin{equation}\label{twist potential}
   \Omega^{AB}_{ab} =\hat \nabla_a \omega_{b}^{AB}-\hat \nabla_b\omega_{a}^{AB}  + G_{CD}\omega^{AC}_a\omega^{DB}_b- G_{CD}\omega^{AC}_b\omega^{DB}_a.
\end{equation}
In particular, the twist potential may be gauged away whenever
\begin{equation}
    \Omega^{AB}_{ab}=0.
\end{equation}

Equations \eqref{submanifoldembedding}--\eqref{twist potential} capture the intrinsic as well as the extrinsic geometry of an isolated submanifold of codimension \(D-d\geq1\). We will say that if the Gram metric \(G^{AB}\) degenerates on a set of points, then that set is a \textit{singular set}, and the geometry is said to degenerate there. Such degeneracy may arise either through null directions or through a trivial, vanishing-gradient type of degeneration. In most of the standard literature, however, the normal bundle \(N(\Sigma)\) is treated locally as being isomorphic to \(\mathbb{R}^{D-d}\), so that the Gram matrix is positive definite, non-degenerate, and equivalent to \(\delta^{AB}\).

\vspace{2mm}

\noindent\textbf{\underline{Family of Submanifolds:}}

\vspace{2mm}

A family of submanifolds, a congruence of submanifolds, and a \(d\)-dimensional foliation are, in essence, three equivalent ways of describing a region of spacetime as a union of disjoint connected sets. For the moment, we follow the terminology of congruences or families, as used in the parent literature. Later, for our own purposes, it will be more convenient to adopt the language of foliations.

Consider a one-parameter family of \(d\)-dimensional surfaces \(\{\Sigma_s\}\), parametrised by
\begin{equation}
    x^\mu =X^\mu(\xi^a,s).
\end{equation}
To measure relative displacement in the transverse, that is, normal, directions, one needs the gradients of the orthonormal basis along those transverse directions. Let
\begin{equation}
    \boldsymbol{\delta}=\boldsymbol{\partial}_s
\end{equation}
be the evolution vector. This vector need not be normal to \(\Sigma\), so we project it along the normal directions using the transverse projector
\begin{equation}
    Q^\mu_\nu =G_{AB}N^{\mu A}N_{\nu}^B,
\end{equation}
and define
\begin{equation}
    \bar\delta^\mu= Q^\mu_\nu\delta^\nu.
\end{equation}
Let \(\boldsymbol{\bar\delta}=\mathbf{N}^A\). The transverse covariant derivatives are then defined by
\begin{equation}
    D^A = N^{\mu A}\nabla_\mu.
\end{equation}
The gradients of the basis \(\{\mathbf{E}_a,\mathbf{N}^A\}\) along the transverse directions satisfy
\begin{align}\label{transversegaussweingarten}
    &D_A\mathbf{E}_a = S_{ab A}\mathbf{E}^b + J_{aAB}\mathbf{N}^B,\\
    &D_A \mathbf{N}_B=-J_{aAB}\mathbf{E}^a +\gamma^C_{AB}\mathbf N_C.
\end{align}
These are the generalised Gauss--Weingarten equations in the transverse directions, complementing the tangential relations in Eq.~\eqref{Generalized Gauss-Weingarten}. Here \(\gamma^C_{AB}\) plays the role of Ricci rotation coefficients in the transverse sector, while \(S_{abA}\) is the analogue of the tangential twist potential. Most importantly, the object \(J^{a}_{AB}\) is the analogue of the extrinsic curvature tensor \(\hat K^A_{ab}\), but for the transverse surfaces to which the family \(\{\Sigma_s\}\) is normal.

\textit{\textbf{Remark:}} For geodesic congruences, the expansion scalar \(\theta\) was related to the extrinsic curvature tensor of the spacetime hypersurfaces normal to the congruence through \(\theta=h^{ab}\hat K_{ab}\). By analogy, the scalar constructed from \(J\) may be regarded as the quantity measuring the composite expansion along all transverse directions to \(\Sigma_s\).

\vspace{8mm}

To obtain the generalised Raychaudhuri equation, one decomposes the analogue of \(\hat K^A_{ab}\) into trace, symmetric traceless, and antisymmetric parts:
\begin{align}\label{generaldecomposition}
    &J^{AB}_a =\frac{1}{N-D}G^{AB}\Theta_a +\Sigma^{AB}_a +\Lambda^{AB}_a.
\end{align}
In the one-dimensional case \(d=1\), these correspond, respectively, to the analogues of expansion, shear, and rotation. For \(d>1\), the interpretation is less direct.

To proceed further, one introduces the quantity
\begin{equation}\label{intermediate}
    \bar\nabla_b J_a^{AB} = -\bar\nabla^A \hat K_{ab}^B-G_{CD}J^{AC}_{b}J^{DB}_a-h^{cd}\hat K^A_{bc}\hat K^B_{da}+N^{\mu A}E^\nu_bE^\rho_aN^{\sigma B}{\rm R}_{\mu\nu\rho\sigma},
\end{equation}
where \(\bar \nabla_A\) and \(\nabla_A\) are defined by
\begin{align}
    &\bar\nabla_A \mathbf T_{abcd\cdots mn} =\nabla_A\mathbf T_{abcd\cdots mn} -S_{akA}\mathbf{T}^k_{bcd\cdots mn} -\cdots - S_{nkA}\mathbf{T}^k_{abcd\cdots m},\\
    & \nabla_A \mathbf T_B =D_A \mathbf{T}_B-\gamma_{ABC}\mathbf{T}^C.
\end{align}
Tracing Eq.~\eqref{intermediate} with \(h^{ab}\), one obtains
\begin{equation}\label{tracedraychaudhuri}
    h^{ab}\bar\nabla_b J_a^{AB}=-h^{ab}\bar\nabla^A \hat K_{ab}^B -h^{ab}G_{CD}J^{AC}_{b}J^{DB}_a-h^{ab}h^{cd}\hat K^A_{bc}\hat K^B_{da}+h^{ab}N^{\mu A}E^\nu_bE^\rho_aN^{\sigma B}{\rm R}_{\mu\nu\rho\sigma}.
\end{equation}

The first term on the right-hand side is often regarded in the literature as a problematic source term. Suppose one prescribes initial values of \(J^{AB}\) on a spacelike hypersurface \(\Sigma_t\); the question then arises as to how these quantities evolve at later times according to Eq.~\eqref{intermediate}. The obstruction lies in the fact that the source term remains undetermined.

The standard way to eliminate this source is to restrict attention to a particular class of surfaces whose dynamics are governed by the classical Nambu--Goto action. Such surfaces are extremal and satisfy the equations of motion
\begin{equation}
    h^{ab}\hat K^A_{ab}=0.
\end{equation}
Equation \eqref{tracedraychaudhuri} is then the traced Raychaudhuri equation for the congruence of submanifolds. Substituting the decomposition \eqref{generaldecomposition} into Eq.~\eqref{tracedraychaudhuri}, one obtains the key equation
\begin{equation}\label{generalizedraychaudhuri}
    \hat\nabla_a \Theta^a =-\frac{1}{N-D}\Theta^a \Theta_a-\Sigma^{aAB}\Sigma_{aAB} + \Lambda^{aAB}\Lambda_{aAB} -G_{AB}M^{AB},
\end{equation}
where
\begin{equation}\label{effective mass}
    M^{AB} = \hat K^A_{ab} \hat K^{abB}+{\rm R}_{\mu\nu} N^{\mu A} N^{\nu B}.
\end{equation}
For the complete set of partial differential equations,\footnote{The generalised shear and generalised twist each satisfy their own evolution equations, and there is also an antisymmetric Raychaudhuri equation, which we do not discuss here.} see Ref.~\cite{Capovilla_1995}.

Unlike the ordinary geodesic case, one cannot analyse caustic formation here with the same degree of simplicity. However, for \(d=D-1\), corresponding to a congruence of hypersurfaces, the generalised shear \(\Sigma^{aAB}\), the generalised twist \(\Lambda^{aAB}\), and the twist potential \(\omega_a^{AB}\) all vanish, which considerably simplifies the structure. One still lacks a definite sign for \(\hat K^A_{ab}\hat K^{abB}\), since \(\hat K_{ab}\) is Lorentzian rather than purely spatial. Nevertheless, the energy conditions can still be used to constrain the sign of the remaining term in Eq.~\eqref{effective mass}, namely \({\rm R}_{\mu\nu}N^{\mu A}N^{\nu B}\).

\noindent\textbf{\underline{Jacobi Equations}}

To obtain the perturbative version of the generalised Raychaudhuri equation, one must invoke the dynamical equation for the surface, namely the Nambu--Goto dynamics:
\begin{equation}\label{Nambugoto}
    S[X^\mu] = -\mathcal{T}\int d^d\xi \sqrt{-h}
    \quad \xRightarrow{\text{E.O.M.}} \quad
    \hat K^A =0.
\end{equation}
The development leading to Eq.~\eqref{generalizedraychaudhuri} is non-perturbative and describes general deformations of relativistic membranes. By linearising Eq.~\eqref{tracedraychaudhuri}, one obtains the perturbative Jacobi equation for small deformations. Since the perturbation of the antisymmetric Raychaudhuri equation is vacuous, the traced equation carries the relevant perturbative content.

Consider a deformation of the embedding,
\begin{equation}
     \mathbf x =\mathbf{ X+\delta \mathbf{X}}
     \qquad \Rightarrow\qquad
     x^\mu=X^{\mu}(\xi^a)+\delta X^{\mu}(\xi^a).
 \end{equation}
This may be decomposed into tangential\footnote{Ref.~\cite{Capovilla} uses \(\mathbf{e}_a\) as the tangential basis rather than the orthonormal basis \(\mathbf{E}_a\) employed above.} and normal parts:
 \begin{equation}
     \delta \mathbf {X} = {\varepsilon}^a\mathbf{e}_a+G_{AB}\varepsilon^A \mathbf{N}^B.
 \end{equation}
The tangential deformations correspond to diffeomorphisms of \(\Sigma\) and are therefore gauge in nature. The true physical deformations are the normal ones, \(G_{AB}\varepsilon^A \mathbf{N}^B\). The Jacobi equation governing these normal deformations is
 \begin{equation}\label{generalizedjacobieqn}
     \bar \nabla^a\bar \nabla_a \varepsilon^A -\bigg[ G_{BC}M^{AB}\bigg] \varepsilon^C=0.
 \end{equation}
For \(d=D-1\), there is only one scalar deformation mode \(\varepsilon^1\), which we denote simply by \(\varepsilon\). In this case, the covariant derivative \(\bar \nabla_a\) reduces to the intrinsic covariant derivative \(\hat \nabla_a\), since the twist potential vanishes; see Eq.~\eqref{normalfield covairant derivative}. The Jacobi equation then becomes
\begin{equation}\label{codim1jacobiequation}
    \hat \nabla^a \hat\nabla_a \varepsilon + \bigg[\hat K_{ab}\hat K^{ab} +{\rm R}_{\mu\nu}N^\mu N^\nu\bigg]\varepsilon=0.
\end{equation}
This resembles the equation of motion for a massive scalar field with effective mass
\begin{equation}
    M^2 =\hat K_{ab}\hat K^{ab} + {\rm R}_{\mu\nu}N^\mu N^\nu,
\end{equation}
namely,
\begin{equation}
    \big[\square  +M^2\big]\varepsilon=0.
\end{equation}
The stability conditions of such congruences, based on the equation above, will be discussed in later chapters, where the analysis must be coupled to Einstein's equations with a well-behaved source. The methodology developed in Ref.~\cite{Capovilla_1995} differs from that of Ref.~\cite{Capovilla}, but both approaches lead to the same Jacobi equations \eqref{generalizedjacobieqn}. A codimension-one congruence of hypersurfaces possesses only one breathing mode, namely \(\varepsilon\), which lies along the normal direction. In that sense, one may say that a \((D-1)\)-foliation \textit{breathes} along its normal direction.

\newpage

\section{Tubular Neighborhood}\label{2.4tubularneighbourhood}

In the preceding discussion of embedded hypersurfaces, we characterised both the intrinsic and extrinsic geometry of codimension-one submanifolds in terms of tensorial data, such as the induced metric, the extrinsic curvature, and the pullback of ambient curvature tensors. While this framework provides a complete differential-geometric description of how a submanifold sits inside the ambient space, it does not answer a further structural question: what does the ambient manifold itself look like in a neighbourhood of the submanifold? The tubular-neighbourhood theorem, originally introduced by Whitney \cite{Whitney1}, provides a precise description of such neighbourhoods.

The tubular-neighbourhood theorem appears in several different formulations in the literature, depending on the additional structure imposed on the ambient manifold. In the smooth category, it is often established using abstract differential-topological arguments. In the presence of a Riemannian metric, it may also be constructed explicitly via the exponential map along normal directions. Although these approaches are equivalent in their conclusions, they differ in both notation and method. In this work, we adopt the formulation of Ref.~\cite{oliva2002geometric}, which is particularly well suited to our purposes. For completeness, we will occasionally refer to alternative perspectives from \cite{Lee2013,spivak1975comprehensive,mukherjee2015differential} where they provide additional geometric insight. Before stating the tubular-neighbourhood theorem itself, we must first review the normal bundle and a few related notions.

\vspace{2mm}

\textbf{Normal Bundle:}

Let \(\mathcal{M}\) be a \(D\)-dimensional ambient manifold equipped with a pseudo-Riemannian metric \(g\), and let \(\Sigma\) be a \(d\)-dimensional submanifold embedded in \(\mathcal{M}\) via \(\Psi:\Sigma\to \mathcal{M}\). At each point \(p\in \Sigma\), one defines the normal space by
\begin{equation}\label{Normalspace}
   N_p(\Sigma):= \{ \mathbf{V}\in T_p(\mathcal{M})\mid g(\mathbf{V},\mathbf{W})=0,\ \forall \mathbf{W}\in T_p(\Sigma)\}.
\end{equation}
The tangent space of the ambient manifold at \(p\) then splits as
\begin{equation}\label{splittingoftangentspace}
    T_p(\mathcal{M}) = N_p(\Sigma)\bigoplus T_p(\Sigma).
\end{equation}
Taking the union of the normal spaces over all \(p\in\Sigma\), one obtains the total normal space,\footnote{The total normal space has dimension \(\dim(N(\Sigma))=\dim(\mathcal{M})=D\).}
\begin{equation}\label{normalspacefiber}
    N(\Sigma) = \bigcup_{p\in \Sigma} N_p(\Sigma).
\end{equation}
At each point \(p\), the vector space \(N_p(\Sigma)\) is the fibre \(\rm F\) of the normal bundle. Its dimension is
\begin{equation}
    \dim(\rm F)=D-d,
\end{equation}
which is precisely the codimension of \(\Sigma\). The total normal space, together with the projection map \(\pi\) and the base space \(\Sigma\), defines the normal bundle \((N,\pi,\Sigma)\). The projection map is
\begin{equation}
    \pi: N(\Sigma)\to \Sigma,
\end{equation}
and acts as
\begin{equation}
    \pi(\mathbf{V}_p)=p,
\end{equation}
for \(\mathbf{V}_p\in N_p(\Sigma)\) and \(p\in\Sigma\).

The existence of the normal bundle is not, by itself, sufficient to define a tubular neighbourhood of \(\Sigma\), since the normal bundle is still merely a submanifold of \(T(\mathcal{M})\). For that purpose, one also needs the notions of a section and the zero section.

\vspace{2mm}

\textbf{Section:}

Let \((E,\pi,M)\) be a normal vector bundle. A map \(\sigma:M\to E\) is called a \textit{section} if
\begin{equation}
    \pi \circ \sigma =\rm Id_\mathcal{M}.
\end{equation}
Intuitively, a section assigns to each point \(p\in M\) a point in the fibre attached to \(p\). Here \(E\) and \(M\) are both manifolds. The \textit{zero section} is the map sending each point of the submanifold to the zero vector in the corresponding fibre:
\begin{equation}
    \mathbf{0}:\Sigma\to N(\Sigma),
\end{equation}
with
\begin{equation}
    \mathbf{0}(p)=0_p.
\end{equation}
With this machinery in place, we may now state the notion of a tubular neighbourhood, following Ref.~\cite{oliva2002geometric}.

\begin{center}
\fbox{
\parbox{1.0\textwidth}{
\textbf{Tubular Neighbourhood:} Let \((\mathcal{M},g)\) be a pseudo-Riemannian manifold, with \(\mathcal{M}\in C^\infty\) and \(g\in C^k\), \(k\geq2\), and let \(\Sigma \subset \mathcal{M}\) be an embedded submanifold of dimension \(d=\dim(\Sigma)<\dim(\mathcal{M})=D\). Then there exists\footnotemark\ a tubular neighbourhood \(f:Z\to\Omega\) of class \(C^{k-1}\) of \(\Sigma\) in \(\mathcal{M}\).

\vspace{1mm}

Here \(f:Z\to\Omega\) is a diffeomorphism from an open neighbourhood \(Z\) of the zero section in \(N(\Sigma)\) onto an open set \(\Omega\subset\mathcal{M}\) containing \(\Sigma\), such that \(f(0_p)=p\) for every zero vector \(0_p\in N(\Sigma)\), \(p\in \Sigma\). The neighbourhood \(Z\) is a tube in \(N(\Sigma)\), whereas the neighbourhood \(f(Z)\)\footnotemark\ is a tube in \(\mathcal{M}\).
}
}
\footnotetext{See Ref.~\cite{oliva2002geometric}, Proposition 9.4, p.~45, for a proof of existence.}
\footnotetext{For further geometric intuition, see also \cite{Lee2013,spivak1975comprehensive,mukherjee2015differential}.}
\end{center}

The diffeomorphism \(f\) associates to a vector in the normal bundle a point in \(\mathcal{M}\) lying in a neighbourhood of \(\Sigma\), in the direction specified by that vector at the corresponding base point. Before proceeding, however, one must address the region of validity of the map \(f\). In other words, for how far does \(f\) remain a diffeomorphism? To answer this, we need additional structure, namely the exponential map and the behaviour of geodesics.

\subsection{Exponential map}

Let \((\mathcal{M},g)\) be a pseudo-Riemannian manifold endowed with its unique Levi--Civita connection. Suppose that for given initial data \(\gamma(0)=p\) and \(\dot\gamma(0)=\mathbf{V}\), there exists a unique geodesic \(\gamma:[0,1]\to\mathcal{M}\). Then the exponential map at \(p\in\mathcal{M}\) is defined by
\begin{equation}\label{exponentialmap}
    \exp_p:T_p(\mathcal{M})\to\mathcal{M},
\end{equation}
with
\begin{equation}
    \exp_p(\mathbf{V})=\gamma(1),
\end{equation}
for \(\mathbf{V}\in T_p(\mathcal{M})\). By construction, this map is defined wherever the corresponding geodesics exist up to the required affine-parameter value. The exponential map is locally a diffeomorphism, and it remains so until geodesics intersect or, equivalently, until caustics form.

In the Riemannian case, the exponential map is valid up to the injectivity radius; see, for example, \cite{petersen2006riemannian,jost2013riemannian}. These definitions rely crucially on the positivity of the metric norm. In the pseudo-Riemannian setting, however, the norm is not positive definite in general, owing to the distinction between spacelike \((V^\mu V_\mu>0)\), timelike \((V^\mu V_\mu<0)\), and null \((V^\mu V_\mu=0)\) directions. Accordingly, the region of validity of the exponential map is no longer characterised in exactly the same way. Instead, the relevant notion is that of a normal neighbourhood; see, for example, Ref.~\cite{o1983semi}.

Restricting the domain of the exponential map \eqref{exponentialmap} to the subspace \(N_p(\Sigma)\subset T_p(\mathcal{M})\), one sees immediately that the tubular-neighbourhood map \(f\) may be identified locally with the exponential map. In this sense, the normal neighbourhood is precisely the neighbourhood \(Z\subset N(\Sigma)\), with image \(\exp(Z)\subset\mathcal{M}\). In the pseudo-Riemannian case relevant to us, the actual region of validity is the tubular neighbourhood of \(\Sigma\) along the spacelike normal geodesics, assuming that all \(D-d\) normals to \(\Sigma\) are spacelike, and only up to the point at which spacelike caustics form.

Thus, one obtains a geometrically precise notion of a tubular neighbourhood around an embedded submanifold \(\Sigma\subset\mathcal{M}\).

\newpage

\section{Submersions}\label{2.5 submersions}

This section is important for our later construction, since we will use the notion of a submersion to avoid the degeneracies that appear when foliating spacetime in the tubular neighbourhood of a higher-codimension manifold. The occurrence of such degeneracies will be discussed in detail in the next chapter. For the standard definition of a Riemannian submersion, we refer to Ref.~\cite{Lee2013}, and for the pseudo-Riemannian case to Ref.~\cite{o1983semi}.

\vspace{2mm}

A key feature that distinguishes different classes of smooth maps is their rank.

\begin{center}
\fbox{
\parbox{1.0\textwidth}{
\textbf{Rank:} Let \(f:M\to N\) be a smooth map between smooth manifolds \(M\) and \(N\). The rank of \(f\) at a point \(p\in M\) is defined as the rank of the Jacobian matrix of the differential
\[
d_pf:T_p(M)\to T_{f(p)}(N).
\]
}
}
\end{center}

If the rank is the same at every point, then \(f\) is said to have constant rank. The rank of \(f\) is bounded above by
\begin{equation}
    \min(\dim(M),\dim(N)).
\end{equation}
Maps whose rank attains this upper bound everywhere are said to have full rank. A submersion is one such map.

\begin{center}
\fbox{
\parbox{1.0\textwidth}{
\textbf{Submersion:} A smooth map \(f:M\to N\) is called a smooth submersion if its differential is surjective at every point, or equivalently if
\[
{\rm Rank}(f)=\dim(N),
\]
with \(\dim(M)>\dim(N)\).
}
}
\end{center}

A submersion fails whenever the rank of the differential drops below its maximal value at some point of \(M\). Such a point is called a \textit{singular point}. If this happens on a set of points, one speaks of a \textit{singular set}.

\newpage
\subsection{Pseudo-Riemannian Submersion}

For pseudo-Riemannian manifolds, the tangent space naturally splits into vertical and horizontal subspaces. In that setting, a pseudo-Riemannian submersion is defined as follows:

\begin{center}
\fbox{
\parbox{1.0\textwidth}{
\textbf{Pseudo-Riemannian Submersion:} Let \(f:M\to N\) be a smooth map between pseudo-Riemannian manifolds \(M\) and \(N\). Then \(f\) is a pseudo-Riemannian submersion if:

\textbf{A.} The fibres \(f^{-1}(b)\), for \(b\in N\), are semi-Riemannian submanifolds of \(M\).

\textbf{B.} The differential of \(f\) preserves the scalar products of vectors normal to the fibres.

\vspace{2mm}

Vectors tangent to the fibres are called \textit{vertical}, while vectors normal to the fibres are called \textit{horizontal}; these belong to mutually orthogonal subspaces. At a point \(p\in f^{-1}(b)\subset M\), the tangent space therefore splits as
\begin{equation}
    T_{p}(M)=  \mathcal{H}_p(f^{-1}(b))\bigoplus\mathcal{V}_p(f^{-1}(b)),
\end{equation}
where \(\mathcal{V}_p\) is the vertical space and \(\mathcal{H}_p\) is the horizontal space.
}
}
\end{center}

Translating this into the language used in the tubular-neighbourhood section, the fibres \(f^{-1}(b)\) are precisely the smooth submanifolds \(\Sigma\) of the ambient pseudo-Riemannian manifold \(M=\mathcal{M}\). Let \(\Sigma\) be a smooth \(d\)-dimensional submanifold of \(\mathcal{M}\). Then \(\Sigma\) serves as the fibre of a submersion \(f:\mathcal{M}\to N\). The tangent space at a point \(p\in\Sigma\) splits as
\begin{equation}
    T_p(\mathcal{M}) = N_p(\Sigma)\bigoplus T_p(\Sigma).
\end{equation}
In this interpretation, the normal space \(N_p(\Sigma)\) is the required horizontal space, while the tangent space \(T_p(\Sigma)\) plays the role of the vertical space. Thus statement \(\mathbf{A}\) is automatically satisfied. To understand statement \(\mathbf{B}\), however, some additional remarks are needed.

Let \(\Sigma\) be a \(d\)-dimensional submanifold of a \(D\)-dimensional manifold \(\mathcal{M}\). Then \(\Sigma\) may be described locally as the common level set of scalar fields \(\varphi=\{\varphi^A\}\), with \(A\in[1,D-d]\):
\begin{equation}
    \Sigma=\bigg\{x\in \mathcal{M}\bigg| \quad \varphi^A(x) =0,\quad \nabla\varphi^A\neq0\bigg\}.
\end{equation}
One may regard the collection of fields \(\varphi\) as a submersion in some region \(\mathcal{U}\subset \mathcal{M}\),
\begin{equation}
    \varphi: \mathcal{U}\subset \mathcal {M}\to \mathbb{R}^{D-d}.
\end{equation}
Then statement \(\mathbf{B}\) translates into the requirement that the scalar products of the normal gradients \(\vec\nabla\varphi^A\in N_p(\Sigma)\),
\begin{equation}
    \mathbf{g}(\vec\nabla\varphi^A,\vec\nabla\varphi^B),
\end{equation}
be preserved in character. In particular, the normals must remain either spacelike for all \(A,B\in[1,D-d]\), or timelike for all \(A,B\in[1,D-d]\). Their causal character must not change. Degeneracy occurs if any normal vector becomes null, that is, if
\begin{equation}
    g_{\mu\nu}\nabla^\mu \varphi^A\nabla^\nu\varphi^A=0,
\end{equation}
or if the gradient vanishes,
\begin{equation}
    \nabla^\mu \varphi^A=0,
\end{equation}
for some \(A\).

Thus statement \(\mathbf{B}\) amounts to the requirement that the gradient of the map \(f\) preserve its causal type and not degenerate. One way to diagnose such degeneracy is through the loss of rank of the submersion. In our case, the rank of \(\varphi\) is
\begin{equation}
    {\rm rank}(\varphi)=D-d,
\end{equation}
which is constant. However, if at any point of \(\mathcal{M}\) the rank drops below this value, that is,
\begin{equation}
    {\rm rank}(\varphi)<D-d,
\end{equation}
then the submersion breaks down, and one may say that degeneracy has occurred.

\newpage
\section{Summary and Bridges}

In Section \ref{2.1geometryofhypersurface}, we discussed the intrinsic and extrinsic geometry of a single isolated hypersurface, and showed how \(\Sigma\)-tensors may be constructed from ambient tensors using pullback and tangential projection. The discussion there was restricted to non-null hypersurfaces. The purpose of that section was to understand codimension-one surfaces in detail, so that they could later serve as the basic geometric building blocks for the codimension-one foliations studied in Chapter 3.

Our need to foliate spacetime then led naturally to Section \ref{2.2Foliations}, where the concept of foliation was reviewed in its general form through Definition \eqref{Foliations}, before turning to the specific example of Cauchy, or ADM-type, foliations. That discussion was important not because we ultimately intend to work with standard Cauchy foliations, but because it provided the necessary machinery for analysing foliation geometry and for constructing adapted coordinate systems and adapted metrics. As already noted in Section \ref{2.2Foliations}, Cauchy foliations are only one example among many possible foliations of spacetime.

For our purposes, however, the relevant region is not an arbitrary open subset of spacetime, but rather the tubular neighbourhood of an auxiliary timelike curve, which was constructed in Section \ref{2.4tubularneighbourhood}. This tubular neighbourhood is not merely any open neighbourhood: locally, it resembles the normal bundle, and its extent is controlled by the validity of the exponential map. Consequently, the foliations to be constructed in Chapter 3 are to be studied inside a geometrically well-defined open neighbourhood.

Since exponential maps remain diffeomorphic only up to the onset of caustics, it was necessary to include Section \ref{2.3 congruence}, where we studied the conditions for caustic formation in a geodesic congruence. In order to determine the region of validity of the tubular neighbourhood, one must understand where caustics form along the normal geodesics emanating from the auxiliary timelike curve. This, in turn, led us to the broader viewpoint that a \(d\)-dimensional foliation may equally well be regarded as a congruence of \(d\)-dimensional surfaces. For this reason, we included a subsection on the geometry and kinematics of congruences of \(d\)-dimensional submanifolds. That framework will be essential in Chapter 3 when studying the kinematics of codimension-one foliations inside the tubular neighbourhood of the timelike source.

Finally, it is equally important to understand the limitations of foliations not only through caustic formation, whether along normal directions to the source or during the physical evolution of the congruence, but also through the degeneracy of the underlying submersion maps. This was the motivation for developing Section \ref{2.5 submersions}. The concepts introduced there will play an important role in avoiding degeneracies in the foliation-based constructions of the next chapter.
\chapter{Tubular Foliation and Geometry} \label{chapter3}

The Geroch--Traschen obstruction in general relativity imposes a strong restriction on which concentrated sources are compatible with Einstein's equations. In particular, sources of higher codimension, \(\mathrm{co\text{-}dim}\geq 2\), produce distributional stress-energy tensors with delta-function support on the source, and in that case no sufficiently regular metric solves Einstein's equations in the usual distributional sense \cite{GEROCHandtraschen}. This naturally raises the question: \textit{how should one model point particles in general relativity?}

One possible approach is to smear the source so that it is effectively replaced by an object of codimension less than \(2\), thereby respecting the Geroch--Traschen compatibility criterion. Israel's thin-shell formalism provides a canonical realisation of this idea \cite{Israel:1966rt,Mansouri:1996ps}. There, the source is supported on a codimension-one hypersurface, the metric remains continuous across the hypersurface, its first derivatives are allowed to jump, and the second derivatives generate the delta-function contribution in the curvature tensors. In this way, Einstein's equations remain meaningful for a distributional stress-energy tensor supported on a hypersurface. Thus codimension-one sources occupy a distinguished position in the distributional formulation of general relativity.

A different strategy is to give up localisation altogether, so that one never encounters a singular stress-energy tensor concentrated on a lower-dimensional source. An isolated hypersurface \(\Sigma_\Phi\) still carries delta-function support \(\delta(\Phi)\) in the stress-energy tensor, although in a controlled manner. In the present work, however, our intention is to introduce a tube geometrically by means of a space-filling codimension-one foliation in the tubular neighbourhood of a timelike curve, thereby removing the delta support altogether. In this picture, the notion of a worldline is replaced by the notion of a tube, and each leaf of the foliation carries a portion of the total stress-energy.

In the previous chapter, we assembled the mathematical tools needed for this construction. We now use that machinery to construct tubular foliations in a controlled way and to study their geometry, thereby preparing the ground for the dynamical analysis of the next chapter.

\section{Construction and Geometry}\label{constructionandgeometry}

Let \(\gamma(\tau)\) be a one-dimensional timelike auxiliary curve in the ambient manifold \((\mathcal{M},g)\), with velocity tangent vector \(U^\mu\). At this stage, the curve is not assumed to be geodesic. Every point \(p\) on the curve admits a normal space, defined by
\begin{equation}
    N_p(\gamma)=\bigg\{\mathbf{V}\in T_p(\mathcal{M})\bigg| \quad \mathbf{g}(\mathbf{V},\mathbf{W})=0,\ \forall\mathbf{W}\in T_p(\gamma)\bigg\}.
\end{equation}
By orthogonality, the tangent space \(T_p(\mathcal{M})\) splits as
\begin{equation}\label{orthogonalsplitting}
    T_p(\mathcal{M}) =N_p(\gamma)\bigoplus T_p(\gamma).
\end{equation}

It is important to note that the metric \(g\) has signature \((-,+,+,+,\dots)\). Hence the ambient tangent space has index one,
\begin{equation}
    \mathrm{Ind}\big(T_p(\mathcal{M})\big)=1.
\end{equation}
Since \(\gamma(\tau)\) is timelike, \(U^\mu U_\mu<0\), the tangent space \(T_p(\gamma)\) also has index one.\footnote{Equivalently, the induced metric on the curve has signature \((-)\), so that \(g^\gamma_{00}=\mathbf{g}(\mathbf{U},\mathbf{U})=-1\).} Therefore, by the orthogonal splitting \eqref{orthogonalsplitting}, the normal space must have vanishing index:
\begin{equation}
    \mathrm{Ind}\big(T_p(\mathcal{M})\big)
    =
    \mathrm{Ind}\big(N_p(\gamma)\big)
    +
    \mathrm{Ind}\big(T_p(\gamma)\big)
    \qquad \Rightarrow \qquad
    \mathrm{Ind}\big(N_p(\gamma)\big)=0.
\end{equation}
Thus all \(D-1\) normals to the curve are spacelike.\footnote{A submanifold of codimension \(D-d\) admits \(D-d\) independent normals; see, for example, \cite{Capovilla}.} In other words,
\begin{equation}
    \mathbf{g}(\mathbf{N}^i,\mathbf{N}^i)>0,
    \qquad
    \mathbf{N}^i\in N_p(\gamma),
    \qquad
    i\in[1,\dots,D-1].
\end{equation}
This is geometrically important: the Lorentzian structure of spacetime admits only one timelike direction, which is already occupied by the tangent to the curve. It is also crucial for our construction, since we wish to foliate a neighbourhood of \(\gamma(\tau)\) by timelike hypersurfaces whose normals are therefore spacelike.

The total normal space is the union of all fibres \(N_p(\gamma)\) along the curve:
\begin{equation}
    N(\gamma)=\bigcup_{p\in\gamma(\tau)}N_p(\gamma).
\end{equation}
By the tubular-neighbourhood theorem, there exists a tubular neighbourhood of the curve \(\gamma(\tau)\), which we state as follows.

\begin{center}
\fbox{
\parbox{1.0\textwidth}{
\textbf{Definition:} Let \((\mathcal{M},g)\) be a pseudo-Riemannian spacetime manifold, with \(\mathcal{M}\in C^\infty\), \(g\in C^k\), \(k\geq2\), and let \(\gamma\subset\mathcal{M}\) be an embedded submanifold of dimension \(1=\dim(\gamma)<\dim(\mathcal{M})=D\). Then there exists a tubular neighbourhood \(f=\exp:Z\to\Omega\) of class \(C^{k-1}\) of \(\gamma\) in \(\mathcal{M}\).

\vspace{1mm}

Here \(\exp:Z\to\Omega\) is a diffeomorphism from an open neighbourhood \(Z\) of the zero section in \(N(\gamma)\) onto an open set \(\Omega\subset\mathcal{M}\) containing \(\gamma\), such that \(\exp(0_p)=p\) for every zero vector \(0_p\in N(\gamma)\), \(p\in\gamma\). The neighbourhood \(Z\) is a tube in \(N(\gamma)\), whereas the neighbourhood \(\exp(Z)\) is a tube in \(\mathcal{M}\).
}
}
\end{center}

This tubular neighbourhood \(\exp(Z)\) is an open subset of \(\mathcal{M}\), valid within the region where the exponential map remains a diffeomorphism. As discussed in Section \eqref{2.4tubularneighbourhood}, this diffeomorphism breaks down at the proper distance at which caustics form. That issue will be studied separately in detail. For the present discussion, it is sufficient that a smooth tubular neighbourhood exists around the timelike curve, within which the desired foliation may be constructed.

Having established such a neighbourhood, we now proceed to foliate it smoothly and study the geometry of the resulting leaves.

\subsection{Tubular Foliation and its Geometry}

In Section \eqref{2.2Foliations}, we reviewed the general notion of foliation and discussed Cauchy foliations as a familiar example, where spacetime decomposes globally as \(\mathcal{M}\simeq \Sigma\times\mathbb{R}\). However, a Cauchy foliation is by no means the only foliation a pseudo-Riemannian manifold may admit. Depending on the problem at hand, other foliations may be more natural. In the present case, our aim is to foliate only a finite region of spacetime, namely the tubular neighbourhood of an auxiliary timelike curve, for the reasons explained at the beginning of this chapter.

We define the tubular foliation as follows.

\begin{center}
\fbox{
\parbox{1.0\textwidth}{
\textbf{Definition:} A \(p\)-dimensional, class \(C^r\), \(r\geq1\), foliation of a \(D\)-dimensional pseudo-Riemannian manifold \((\mathcal{M},g)\) is a decomposition of a region of \(\mathcal{M}\) into a union of disjoint connected subsets \(\Sigma_\Phi\), called leaves, with the following property:

The timelike curve \(\gamma(\tau)\subset\mathcal{M}\) admits a tubular neighbourhood \(\exp(Z)\), together with a local coordinate system of class \(C^r\),
\[
\mathbf{x}:\exp(Z)\to\mathbb{R}^D,
\]
such that each timelike leaf \(\Sigma_\Phi\) is locally described by equations of the form
\[
x^{p+1}=c^1,\dots,x^D=c^D,
\]
with \(c^i\in\mathbb{R}\) constants. The foliation is then denoted by
\begin{equation}\label{tubularfoliation}
    \mathcal{F}=\{\Sigma_\Phi\}
    =
    \bigg\{
    x\in\mathcal{W}\subset \exp(Z)\subset\mathcal{M}
    \bigg|
    \quad
    \Phi(x)=s,\ 
    \nabla\Phi\big|_{\Phi=s}\neq0,\ 
    \forall s\in\mathbb{R},
    \ \Phi_{\max}=\mathcal{E}
    \bigg\}.
\end{equation}
Here \(\mathcal{F}\) is a codimension-one foliation, so that \(p=D-1\), and \(\mathcal{W}\cap\Sigma_\Phi\) has a single component described locally by one equation, \(x^D=\Phi(x)=s\). In particular, the leaves may be interpreted as the level sets of the scalar field \(\Phi(x)\).
}
}
\end{center}

It is important to notice that the central curve around which the spacetime is foliated is not itself a leaf of \(\mathcal{F}\). Indeed, if one tries to identify the scalar field \(\Phi(x)\) with the proper distance \(r(x)\) from the curve \(\gamma(\tau)\), then the foliation degenerates at \(r=0\), because \(\nabla r(x)\) vanishes there. Hence the curve does not belong to the family of leaves:
\begin{equation}
    \gamma(\tau)\not\subset \{\Sigma_\Phi\}.
\end{equation}

\begin{figure}[htbp]
    \centering
    \includegraphics[width=0.5\linewidth]{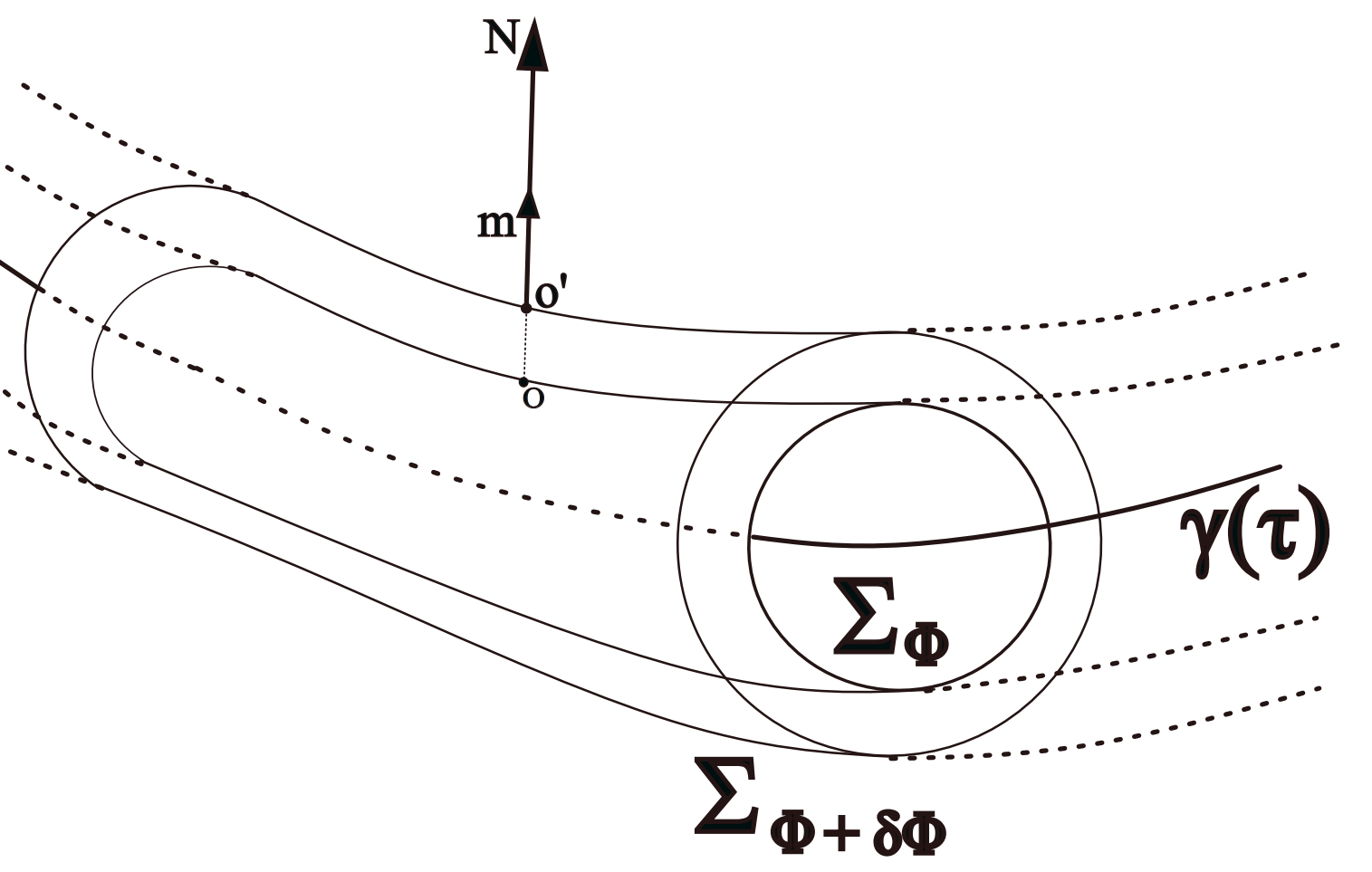}
    \caption{Tubular foliation around the timelike curve \(\gamma(\tau)\).}
    \label{fig:5}
\end{figure}

This central degeneracy is expected, since a single scalar field cannot define a codimension-\((D-1)\) curve; the level set \(\Phi=0\) is still a hypersurface. Thus the degeneracy of the foliation \eqref{tubularfoliation} at the central curve must be avoided if one wishes to construct a consistent foliation in the tubular neighbourhood of \(\gamma(\tau)\). We postpone the detailed analysis of this issue to the next section. For now, we focus on the geometry of the leaves themselves.

We choose the negative sign convention for the normal in the direction of increasing \(\Phi\), that is, away from the central curve \(\gamma\). The spacelike unit normal vector is defined by
\begin{equation}\label{ch3normal}
    \mathbf{N}=-\mathcal{N}\vec{\nabla}\Phi
    \qquad \text{or} \qquad
    N^\mu =-\mathcal{N}\nabla^\mu\Phi,
\end{equation}
with normalisation factor
\begin{equation}\label{Normallapse}
    \mathcal{N}=(g^{\mu\nu}\nabla_\mu\Phi\nabla_\nu\Phi)^{-1/2},
\end{equation}
so that
\begin{equation}
    N^\mu N_\mu =1.
\end{equation}
In analogy with the ADM construction discussed in Section \eqref{2.2Foliations}, the factor \(\mathcal{N}\) may again be interpreted as a lapse-type function. Since it is defined here for a timelike foliation rather than a spacelike one, we shall refer to \eqref{Normallapse} as the \textit{normal-lapse function}.

The unit normal covector \(\underline{\mathbf{N}}\) is collinear with the gradient one-form \(\mathbf{d}\Phi\), and satisfies
\begin{equation}\label{conormal}
    \mathbf{d}\Phi=\mathcal{N}^{-1}\underline{\mathbf{N}}.
\end{equation}
The next important vector is the normal evolution vector \(\mathbf{m}\), defined by
\begin{equation}
    \mathbf{m} =\mathcal{N}\mathbf{N},
    \qquad \text{with} \qquad
    \mathbf{g}(\mathbf{m},\mathbf{m})=\mathcal{N}^2.
\end{equation}
This vector Lie-drags points from the leaf \(\Sigma_\Phi\) to the neighbouring leaf \(\Sigma_{\Phi+\delta\Phi}\), so that for \(p\in\Sigma_\Phi\) and \(p'\in\Sigma_{\Phi+\delta\Phi}\),
\begin{equation}
    \Phi(p')=\Phi(p)+\delta\Phi,
    \qquad
    p'=p+\mathbf{m}\,\delta\Phi.
\end{equation}
As in the Cauchy-foliation case, \(\mathbf{m}\) plays an important role in constructing an adapted metric on the foliated region.

Before doing so, observe that if the normal-lapse function is not constant, then the acceleration of the normal flow is non-vanishing:
\begin{equation}
    \mathbf{a}=\nabla_\mathbf{N} \mathbf{N}.
\end{equation}
Substituting Eq.~\eqref{ch3normal} and simplifying, one finds
\begin{equation}\label{Acceleration}
    a^\nu =-D^\nu \ln\mathcal{N},
    \qquad \text{equivalently} \qquad
    \mathbf{a}=-\vec{\boldsymbol{D}}\ln\mathcal{N},
\end{equation}
where
\begin{equation}
    P^\nu_\mu=\delta^\nu_\mu-N^\nu N_\mu
\end{equation}
is the tangential projector. In particular, the acceleration is tangent to the leaves, since
\begin{equation}
    \mathbf{g}(\mathbf{a},\mathbf{N})=0.
\end{equation}
This acceleration describes the failure of the spacelike normal curves emanating from the timelike curve \(\gamma(\tau)\) to be geodesics.

The non-vanishing acceleration contributes both to the extrinsic curvature and to the gradient of the normal evolution vector:
\begin{align}\label{accelaratedextrinsic curvature}
    &K_{\mu\nu} =P^\alpha_\mu P^\beta_\nu \nabla_\alpha N_\beta
    = \nabla_\mu N_\nu-N_\mu N^\alpha \nabla_\alpha N_\nu \notag\\
    &\qquad =\nabla_\mu N_\nu-N_\mu a_\nu
    =\nabla_\mu N_\nu+N_\mu D_\nu\ln \mathcal{N},
\end{align}
and
\begin{align}
    \nabla_\mu m^\nu
    &= N^\nu\nabla_\mu\mathcal{N}+\mathcal{N}\nabla_\mu N^\nu \notag\\
    &= N^\nu \nabla_\mu \mathcal{N}+\mathcal{N}K_\mu {}^{\nu} - N_\mu D^\nu \mathcal{N}.
\end{align}

The presence of acceleration also modifies the projected Ricci tensor and hence the scalar curvature relation. Consider
\begin{equation}\label{midRicciequation}
    P^\mu_{\alpha}N^\sigma P^\nu_\beta \big[\nabla_\nu,\nabla_\sigma\big]N_\mu
    =
    P^\mu_{\alpha}N^\sigma N^\rho P^\nu_\beta{\rm R}_{\mu\rho\nu\sigma}.
\end{equation}
Using Eq.~\eqref{accelaratedextrinsic curvature}, the left-hand side becomes
\begin{align}
    &P^{\mu}_{\alpha} P^\nu_{\beta} N^\sigma \bigg[\nabla_\nu  \Big(K_{\sigma\mu}+N_{\sigma}a_\mu\Big)-\nabla_\sigma  \Big(K_{\nu\mu}+N_{\nu}a_\mu\Big)\bigg] \notag\\
    &=P^{\mu}_{\alpha} P^\nu_{\beta}\bigg[-K_{\sigma\mu}K^\sigma_\nu-N^\sigma\nabla_\sigma K_{\nu\mu}+D_\nu a_\mu-a_\nu a_\mu\bigg] \notag\\
    &= -K_{\sigma\alpha}K^\sigma_\beta-P^{\mu}_{\alpha} P^\nu_{\beta}N^\sigma\nabla_\sigma K_{\nu\mu}+D_\alpha a_\beta -a_\alpha a_\beta.
\end{align}
On the other hand, the projected Ricci relation gives
\begin{equation}
     {\rm R}_{\alpha\beta}\bigg|_\Sigma
     =
     P_\alpha^\mu P_\beta^\nu {\rm R}_{\mu\nu}
     +P_\alpha^\mu P_\beta^\nu N^\rho N^\sigma {\rm R}_{\mu\rho\nu\sigma}
     +\big[KK_{\alpha\beta}-K^\delta_\alpha K_{\delta\beta}\big].
\end{equation}
Substituting the second term from this equation using \eqref{midRicciequation}, one obtains
\begin{equation}
   {\rm R}_{\alpha\beta}\bigg|_\Sigma
   =
   P_\alpha^\mu P_\beta^\nu {\rm R}_{\mu\nu}
   +\big[KK_{\alpha\beta}-K^\delta_\alpha K_{\delta\beta}\big]
   -K_{\sigma\alpha}K^\sigma_\beta
   -P^{\mu}_{\alpha} P^\nu_{\beta}N^\sigma\nabla_\sigma K_{\nu\mu}
   +D_\alpha a_\beta -a_\alpha a_\beta.
\end{equation}
Pulling this back to the leaf gives
\begin{equation}\label{Gausscoddazirelationfornormals}
    \hat {\rm R} _{ab}
    =
    e_a^\mu e_b^\nu {\rm R}_{\mu\nu}
    +\big[\hat K\hat K_{ab}-\hat K^c_a \hat K_{cb}\big]
    -\hat K_{ca}\hat K^c_b
    -e^{\mu}_{a} e^\nu_{b}N^\sigma\nabla_\sigma K_{\nu\mu}
    +\hat D_a a_b -a_a a_b.
\end{equation}
Contracting with \(h^{ab}\), one further obtains the scalar relation
\begin{equation}\label{Scalargaussfornormal}
    \hat {\rm R}
    =
    {\rm R} -{\rm R}_{\mu\nu} N^\mu N^\nu
    +\Big[\hat K^2-2\hat K_{ab} \hat K^{ab}\Big]
    -N^\sigma \nabla_\sigma K
    +\hat{D}_a a^a-a_a a^a.
\end{equation}
Here \(\hat D_a=\hat\nabla_a\) in the sense of Lemma \eqref{Lemma1}. For extremal surfaces obeying the Nambu--Goto condition \eqref{Nambugoto}, the mean extrinsic curvature vanishes,
\begin{equation}
    \hat K=K=0.
\end{equation}
Equivalently, the expansion scalar of the spacelike congruence normal to the family \(\{\Sigma_\Phi\}\) vanishes. The above equations therefore relate the intrinsic and extrinsic geometry of the timelike leaves in the presence of non-zero normal acceleration.

\subsection{Integrability and the bounds of foliation}\label{secintegrabilityandbound}

Let us now interpret the foliation \eqref{tubularfoliation} more concretely by taking \(\Phi=s\) to represent proper distance from the central curve \(\gamma(\tau)\), so that the normal curves emanate from the timelike curve \(\gamma(\tau)\).

\begin{figure}[htbp]
    \centering
    \includegraphics[width=0.8\linewidth]{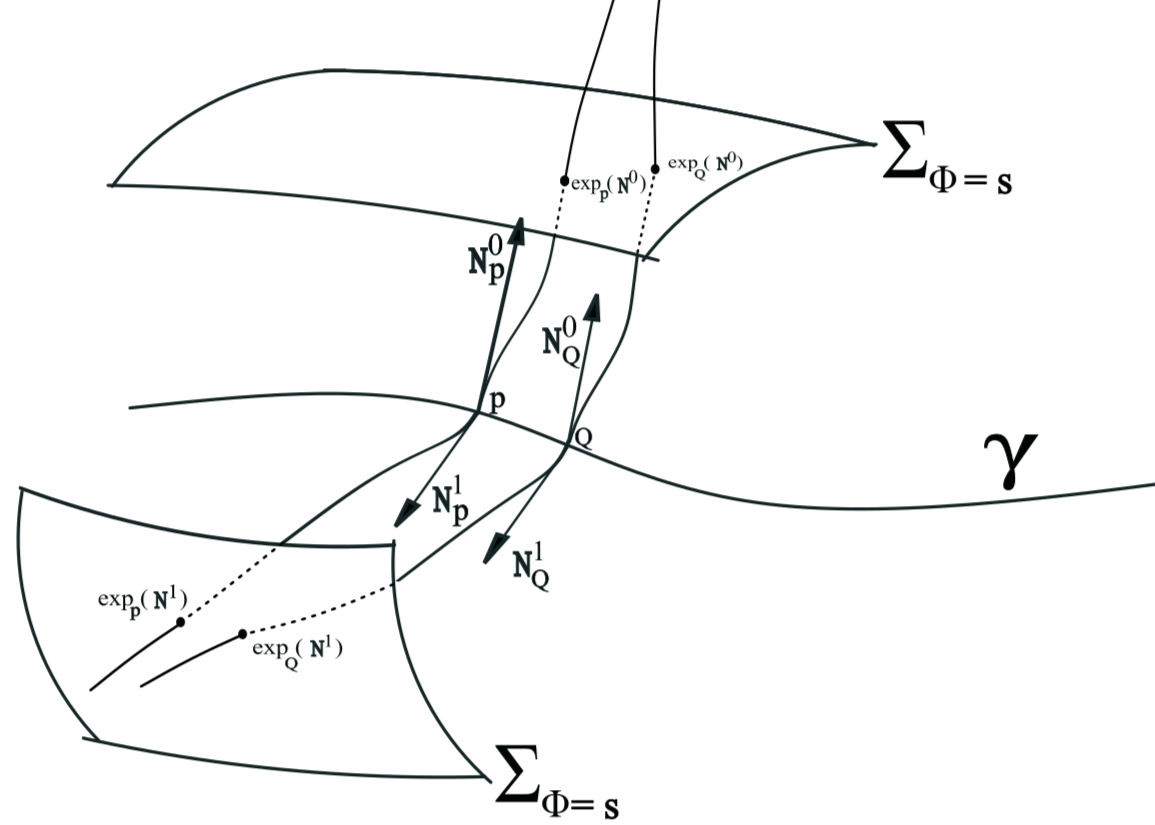}
    \caption{A normal flow emanating from the timelike curve \(\gamma\) along the basis \(\{\mathbf{N}^A\}\) of the normal space \(N_p(\gamma)\), intersecting a leaf \(\Sigma_{\Phi=s}\) at proper distance \(s\) from the central curve \(\gamma\).}
    \label{fig:7}
\end{figure}

Since \(\dim N_p(\gamma)=D-1\), the normal space admits a basis \(\{\mathbf{N}^A\}\). Let \(\mathbf{U}\) be any vector in \(N_p(\gamma)\), expanded as
\begin{equation}\label{flowgenerator}
    \mathbf{U} =G_{AB} C^A \mathbf N^B,
    \qquad \text{or in components} \qquad
    U^\mu = G_{AB}C^A N^{\mu A}.
\end{equation}
For the moment, we assume that the Gram metric \(G^{AB}\) is non-degenerate, so that it is invertible and the index operations on \(A,B,C,\dots\in[1,D-1]\) are well defined. Thus
\begin{equation}
    (G^{AB})^{-1}=G_{AB}.
\end{equation}
The coefficients \(C_A\) are scalar functions with respect to the evolution parameter \(s\).

Let \(\eta(s)\) be a fiducial curve in the spacelike congruence inside an open region \(\mathcal{O}\subset\mathcal{M}\), with \(\mathcal{O}\cap\gamma\neq\emptyset\). Let \(U^\mu\) be its tangent vector, written in the transported basis as
\begin{equation}
    U^\mu(\tau,s) =C_A(\tau) N^{\mu A}(\tau,s).
\end{equation}
Then
\begin{equation}
    U^\mu \nabla_\mu U^\nu = C_A(\tau)C_B(\tau)N^{\mu A}\nabla_\mu N^{\nu B}.
\end{equation}
The norm of \(\mathbf{U}\) is
\begin{equation}
    \mathbf{g}(\mathbf{U},\mathbf{U})
    =
    \mathbf{g}(G_{AB} C^A \mathbf N^B,G_{ED} C^E \mathbf N^D)
    =
    G_{AB}G_{ED}G^{BD} C^AC^E
    =
    C^AC_A.
\end{equation}
Hence the associated unit tangent vector is
\begin{equation}
    \mathbf{U}= \frac{1}{\sqrt{C_BC^B}}C_A\mathbf N^A.
\end{equation}

We now follow the same strategy used in the derivation of the usual Raychaudhuri equation. Define the restricted metric and projector by
\begin{equation}
    h_{\mu\nu} =g_{\mu\nu}-U_\mu U_\nu,
    \qquad
    h^{\mu\nu} =g^{\mu\nu}-U^\mu U^\nu,
\end{equation}
with
\begin{equation}
    h_{\mu\nu}h^{\mu\nu}=3,
\end{equation}
and
\begin{equation}
    P^{\mu}_\nu =\delta^\mu_\nu -U^\mu U_\nu,
    \qquad
    P^\mu_\nu P^\nu_\alpha =P^\mu_\alpha,
\end{equation}
respectively. Consider now the rank-\((0,2)\) tensor \(\nabla_\mu U_\nu\), projected onto the hypersurfaces transverse to the flow:
\begin{equation}\label{anotherrank02}
    b_{\mu\nu } = P^{\alpha}_\mu P^{\beta}_\nu \nabla_\alpha U_\beta
    = \nabla _\mu U_\nu -U_\mu a_\nu.
\end{equation}
This tensor decomposes into trace, traceless symmetric, and antisymmetric parts:
\begin{equation}\label{accelerated tensor}
    b_{\mu\nu} = \frac{1}{3} \hat\theta h_{\mu\nu} +\hat \sigma_{\mu\nu}+\hat\omega_{\mu\nu}.
\end{equation}
Here the expansion scalar is
\begin{equation}
    \hat \theta =h^{\mu\nu}\nabla_\mu U_\nu.
\end{equation}
Differentiating Eq.~\eqref{anotherrank02} along the flow and contracting with \(h^{\mu\nu}\), one obtains
\begin{equation}
    U^\rho\nabla_\rho \hat\theta =- b_{\mu\nu}b^{\mu\nu} -{\rm R}_{\mu\nu}U^\mu U^\nu+\nabla_\mu a^\mu.
\end{equation}
Substituting the decomposition \eqref{accelerated tensor}, we arrive at
\begin{equation}\label{normalraychaudhuri}
    \frac{d\hat\theta}{ds}
    =-\frac{1}{3}\hat\theta^2-\hat\sigma^2 +\hat \omega^2 -{\rm R}_{\mu\nu}U^\mu U^\nu+\nabla_\mu a^\mu.
\end{equation}

This has the same structure as the usual Raychaudhuri equation \eqref{RaychaudhuriEq}, except that \(\hat \sigma_{\mu\nu}\) and \(\hat \omega_{\mu\nu}\) are now Lorentzian rather than purely spatial tensors. Their quadratic contractions therefore do not have a definite sign in general. However, if the hypersurfaces are extremal, so that \(\hat\theta=0\), and if the flow is hypersurface orthogonal, so that \(\hat\omega_{\mu\nu}=0\), then in the absence of Lorentzian shear the equation simplifies to
\begin{equation}
     \frac{d\hat\theta}{ds}=-{\rm R}_{\mu\nu}U^\mu U^\nu+\nabla_\mu a^\mu.
\end{equation}
The conditions for normal caustics to form away from \(\gamma\) will be analysed only after the stress-energy tensor has been specified and Einstein's equations have been used to constrain the sign of \({\rm R}_{\mu\nu}U^\mu U^\nu\). For safety, we shall assume that the normal caustic occurs beyond \(\Phi=\Lambda^{-1}\).

\noindent\textbf{\textit{Remark:}} \textit{Since \(\mathbf{U}\) is constructed as a linear combination of the normal basis vectors, the curve \(\eta(s)\) is necessarily spacelike.}

\vspace{2mm}

To guarantee that the flow generated by \(\mathbf{U}\) is hypersurface orthogonal, one must impose the vanishing of the Frobenius three-form,
\begin{equation}
    U_{[\mu}\nabla_\nu U_{\rho ]}=0.
\end{equation}
See Appendix \eqref{A.3INTEGRABILITY}. For the general definition \eqref{flowgenerator}, this is not automatic, and instead imposes constraints on the normal-bundle coefficients \(C_A\). Let us compute these conditions explicitly.

The gradient of the unnormalised covector \(\underline{\mathbf{U}}\) is
\begin{equation}
    \nabla_\mu U_\nu = N^A_\nu\nabla_\mu C_A +C_A \nabla_\mu N_\nu^A.
\end{equation}
Substituting into the antisymmetrised Frobenius bracket,
\begin{equation}
    U_{[\mu}\nabla_\nu U_{\rho ]}
    =
    \frac{1}{3!}
    \big[
    U_\mu \nabla_\rho U_\nu
    + U_\rho \nabla_\nu U_\mu
    + U_\nu \nabla_\mu U_\rho
    - U_\mu \nabla_\nu U_\rho
    - U_\rho \nabla_\mu U_\nu
    - U_\nu \nabla_\rho U_\mu
    \big],
\end{equation}
one finds the decomposition
\begin{equation*}
     U_{[\mu}\nabla_\nu U_{\rho ]} = T_1 +T_2,
\end{equation*}
where
\begin{align}
    & T_1 =\frac{1}{6}C_AC_B N^B_{[\mu}\nabla_\rho N^A_{\nu]},\\
    & T_2=\frac{1}{6}\Bigg[C_B\nabla_\rho C_A\big(N^B_\mu N^A_\nu-N^B_\nu N^A_\mu\big)+C_B\nabla_\nu C_A\big(N^B_\rho N^A_\mu-N^B_\mu N^A_\rho\big)+C_B\nabla_\mu C_A\big(N^B_\nu N^A_\rho-N^B_\rho N^A_\nu\big)\Bigg].
\end{align}
Therefore,
\begin{align}
    U_{[\mu}\nabla_\nu U_{\rho ]}=0
    \quad \text{if} \quad
\begin{cases}
C_{A}\nabla_\mu C_{B} -C_{B}\nabla_\mu C_{A}=0,\\[6pt]
C_AC_B N^B_{[\mu}\nabla_\rho N^A_{\nu]}=0.
\end{cases}
\end{align}
These are the integrability conditions for the flow.

The first condition may be rewritten as
\begin{equation}
   \frac{\nabla_\mu C_A}{C_A} = \frac{\nabla_\mu C_B}{C_B},
   \qquad \forall A,B,
\end{equation}
which integrates to
\begin{equation}
    C_A(x) =\mathcal{C}_A g(x),
    \qquad \text{with} \qquad
    \mathcal{C}_A\in\mathbb{R}
    \ \text{constants}.
\end{equation}
Then Eq.~\eqref{flowgenerator} becomes
\begin{equation}
    U^\mu =g(x) \mathcal{C}_A N^{\mu A} =g(x) \bar N^\mu.
\end{equation}
Thus the first integrability condition selects a unique direction in the normal space, making the velocity vector collinear to a preferred normal direction. When the second condition is imposed simultaneously, it becomes the standard Frobenius condition for the flow generated by \(\bar N^\mu\):
\begin{equation}
    g^2(x)\,\bar N_{[\mu}\nabla_\rho \bar N_{\nu]} =0,
    \qquad \text{for} \qquad
    g(x)\neq0.
\end{equation}
Dropping the bar from the chosen normal and using the scalar field \(\Phi\), we may write, as in Eq.~\eqref{ch3normal},\footnote{Recall that the normals \(\mathbf{N}^A\) used above were not assumed to be normalised, and the integrability analysis was performed for the unnormalised vector \(\mathbf{U}\).}
\begin{equation}
    U^\mu = -g(x)\nabla^\mu \Phi.
\end{equation}
It follows that the function \(g(x)\) is precisely the normal-lapse function \(\mathcal{N}\):
\begin{equation}
    g(x) = \mathcal{N} =(g_{\mu\nu}\nabla^\mu \Phi\nabla^\nu \Phi)^{-1/2}.
\end{equation}
Hence \(U^\mu\) is hypersurface orthogonal for \(s>0\).

In general, the normal-lapse function is not constant, and therefore the acceleration remains non-zero:
\begin{equation}
    a^\mu =D^\mu \ln \mathcal{N}.
\end{equation}
Its divergence contributes to the right-hand side of Eq.~\eqref{normalraychaudhuri}. If \(\nabla_\mu a^\mu\) contributes positively, it can oppose focusing of the normal congruence \(\eta(s)\); if not, it enhances focusing, in accordance with the general Raychaudhuri mechanism. The same acceleration terms also appear in the scalar Gauss relation \eqref{Scalargaussfornormal} and in the Ricci-form Gauss--Codazzi relation \eqref{Gausscoddazirelationfornormals} for the timelike leaves.

\newpage
\section{Adapted metric and coordinates}

Let the coordinates on the leaves \(\Sigma_s\) be \(\xi^a\), with \(a\in[0,D-2]\). One may then define adapted coordinates on the foliated region \(\mathcal{W}\subset\mathcal{M}\) around the curve \(\gamma(\tau)\), and hence construct an adapted metric following the same general strategy as in the ADM foliation discussed in Section \ref{ADMadaptedmetricandcoordinates}. Let the natural basis of the tangent space \(T(\mathcal{W})\) be
\begin{equation}
    \boldsymbol{\partial}_\mu=[\boldsymbol{\partial}_s,\boldsymbol{\partial}_a].
\end{equation}
Here \(\boldsymbol{\partial}_s\) is tangent to the curve traced by the points \(\{O:\xi^a=C^a\}\), while the vectors \(\boldsymbol{\partial}_a\) are tangent to the leaf \(\Sigma_s\), so that
\begin{equation}
    \boldsymbol{\partial}_a\in T_O(\Sigma_s).
\end{equation}

The curves generated by \(\boldsymbol{\partial}_s\) need not evolve purely along the unit normal \(\mathbf{N}\) to the leaves. The difference is captured by the shift vector \(\boldsymbol{\beta}\), through the decomposition
\begin{equation}\label{shiftdecomposed}
    \boldsymbol{\partial}_s=\mathbf{m}+ \boldsymbol{\beta},
\end{equation}
where
\begin{equation}
    \mathbf{m}=\mathcal{N}\mathbf{N}.
\end{equation}

To show that the shift vector belongs to the tangent space of the leaf, consider the dual basis \(\mathbf{d}x^\mu\in T_O^*(\mathcal{W})\), satisfying
\begin{equation}
    \langle \mathbf d x^\mu, \boldsymbol{\partial}_\nu\rangle=\delta^\mu_\nu.
\end{equation}
In particular,
\begin{equation}
    \langle \mathbf d s, \boldsymbol{\partial}_s\rangle=1.
\end{equation}
Using Eq.~\eqref{conormal}, one also finds
\begin{equation}
     \langle \mathbf d s,\mathbf{m}\rangle
     =
     \langle \mathbf d s,\mathcal N\mathbf N\rangle
     =
     \langle \mathbf d\Phi,\mathcal N\mathbf N\rangle
     =
     \langle \underline{\mathbf{N}},\mathbf{N}\rangle
     =1.
\end{equation}
Therefore,
\begin{equation}
    \langle \mathbf{d}s, \boldsymbol{\beta}\rangle
    =
    \langle \mathbf d s, \boldsymbol{\partial}_s\rangle
    -
    \langle \mathbf d s,\mathbf{m}\rangle
    =0,
\end{equation}
and hence
\begin{equation}
    \boldsymbol{\beta}\in T_O(\Sigma_s).
\end{equation}

The vector \(\boldsymbol{\partial}_s\) is not guaranteed a priori to be spacelike. Indeed,
\begin{equation}\label{sscomponent}
    g_{ss}
    =
    \mathbf{g}(\boldsymbol{\partial}_s,\boldsymbol{\partial}_s)
    =
    \mathcal{N}^2+\beta_a\beta^a.
\end{equation}
Since \(\boldsymbol{\beta}\) is tangent to a timelike leaf, its norm need not have a definite sign in general. Therefore, the causal character of \(\boldsymbol{\partial}_s\) is determined by the sign of \(g_{ss}\):
\begin{align}
   \text{ \(\boldsymbol{\partial}_s\) is }
\begin{cases}
\text{timelike}, & \text{if} \quad \mathcal{N}^2+\beta_a\beta^a<0, \\[6pt]
\text{spacelike}, & \text{if} \quad \mathcal{N}^2+\beta_a\beta^a>0, \\[6pt]
\text{null}, & \text{if} \quad \mathcal{N}^2+\beta_a\beta^a=0.
\end{cases}
\end{align}
For the present discussion, we restrict attention to the non-null cases, so as to avoid the null degeneration of the adapted metric.

The shift and normal vectors may be written in component form as
\begin{equation}
    \boldsymbol{\beta}= \beta^a\boldsymbol{\partial}_a,
    \qquad
    \mathbf{N}=N^\mu\boldsymbol{\partial}_\mu.
\end{equation}
Using Eqs.~\eqref{shiftdecomposed} and \eqref{conormal}, the components of the normal and conormal vectors are
\begin{equation}
    N^\mu = \bigg[\frac{1}{\mathcal{N}}, \frac{-\beta^a}{\mathcal{N}}\bigg],
    \qquad
    N_\mu =\bigg[{\mathcal{N}},0\bigg],
\end{equation}
respectively.

From Eq.~\eqref{sscomponent}, we already have the \(ss\)-component of the adapted metric. The remaining components follow similarly:
\begin{align}
    &g_{sa}
    =
    \mathbf{g}(\boldsymbol{\partial}_s,\boldsymbol{\partial}_a)
    =
    \mathbf{g}(\mathcal N\mathbf N+\boldsymbol\beta,\boldsymbol{\partial}_a)
    =
    \beta_a,\\
    &g_{ab}
    =
    \mathbf{g}(\boldsymbol{\partial}_a,\boldsymbol{\partial}_b)
    =
    h_{ab}.
\end{align}
Hence the full metric tensor in adapted coordinates takes the form
\begin{equation}
    g_{\mu\nu}=
    \begin{pmatrix}
        \mathcal{N}^2+\beta_c\beta^c & \beta_b\\
        \beta_a & h_{ab}
    \end{pmatrix},
\end{equation}
or equivalently,
\begin{equation}
    dS^2=\mathcal{N}^2 ds^2+h_{ab}(s,\xi)\,[d\xi^a+\beta^a ds][d\xi^b+\beta^b ds].
\end{equation}
The corresponding inverse metric is\footnote{It may be verified directly from \(g^{\mu\nu}g_{\nu\rho}=\delta^\mu_\rho\).}
\begin{equation}\label{thenormaladaptedmetric}
    g^{\mu\nu}=
            \begin{pmatrix}
                \frac{1}{\mathcal{N}^2}&-\frac{\beta^b}{\mathcal{N}^2}\\
                -\frac{\beta^a}{\mathcal{N}^2}&h^{ab} +\frac{\beta^a\beta^b}{\mathcal{N}^2}
            \end{pmatrix}.
\end{equation}
This adapted coordinate system is valid locally within the tubular neighbourhood \(\exp(Z)\), away from the singular behaviour at the core that will be discussed next.

\newpage
\section{The Central Degeneracy}

Our construction of a timelike codimension-one foliation in the tubular neighbourhood encounters two distinct kinds of central degeneracy. In Section \ref{secintegrabilityandbound}, we considered the affinely parametrised spacelike geodesics \(\eta(s)\) emanating from points on the central curve \(\gamma(\tau)\), in order to determine the bounds of the tubular neighbourhood. Each point of \(\gamma\) behaves as the origin of a congruence of spacelike geodesics radiating outward and filling a local neighbourhood in spacetime. However, the source point \(s=0\) is precisely the common intersection point of all such geodesics emanating from the same point of the curve. This is the first kind of degeneracy, and it is useful to formulate it more carefully.

\begin{center}
\fbox{
\parbox{1.0\textwidth}{
\textbf{Proposition (trivial central degeneracy):} Let \(\Sigma\) be a timelike submanifold of \(\mathcal{M}\). The polar normal-coordinate parametrisation of the normal exponential map degenerates at \(\Sigma\): all angular data in the normal directions collapse at \(s=0\), so the source is a set of trivial caustic points for the radial normal flow.
}
}
\end{center}

\textbf{Proof:} Let \(\Sigma\) be a \(d\)-dimensional timelike source, so that at each point \(p\in\Sigma\),
\begin{equation}
    T_p(\mathcal{M})= N_p(\Sigma)\oplus T_p(\Sigma),
\end{equation}
with
\begin{equation}
    \dim N_p(\Sigma)=D-d,
    \qquad
    \mathrm{Ind}(N_p(\Sigma))=0.
\end{equation}
Let \(\mathbf{n}\) denote a unit vector in the normal space \(N_p(\Sigma)\), and consider the radial normal-coordinate map
\begin{equation}
    F:\Sigma\times S^{D-d-1}\times[0,\lambda)\to\mathcal{M},
    \qquad
    F(p,\mathbf n,s)=\exp_p^*(s\mathbf n),
\end{equation}
where \(\exp_p^*\) is the normal exponential map at \(p\). For \(s>0\), the parameter \(s\) measures proper distance along the corresponding spacelike geodesic. At \(s=0\), however,
\begin{equation}
    F(p,\mathbf n,0)=p,
\end{equation}
for every unit normal direction \(\mathbf n\in S^{D-d-1}\). Thus all angular directions collapse to the same point on the source. In particular, the angular variables cease to distinguish points at \(s=0\), and the polar normal-coordinate description becomes degenerate there.

Therefore, the source \(\Sigma\) is a set of trivial caustic points for the radial normal flow. Equivalently, the adapted coordinates based on \((s,\mathbf n)\) are valid only for
\begin{equation}
    0<s<\lambda,
\end{equation}
and break down at the source itself. This is directly analogous to the degeneracy of cylindrical or spherical coordinates at their centre. \(\square\)

The trivial central degeneracy is therefore structural. Any adapted metric written using hypersurfaces of constant \(s\) is naturally defined only in the region \(0<s<\lambda\), where \(\lambda\) denotes the bound of the tubular neighbourhood. Likewise, the Raychaudhuri equation for the emanating normal geodesics must be interpreted within the same range of affine parameter.

This is the first kind of degeneracy encountered at the source. There is, however, a second and more subtle degeneracy tied directly to the foliation itself. In constructing the tubular foliation \eqref{tubularfoliation}, one finds that the leaves cannot be continuously deformed all the way to the central source without some loss of regularity. A simple model illustrating this behaviour is the cylinder
\begin{equation}
    C=S^1\times \mathbb R,
\end{equation}
embedded in \(\mathbb R^3\), with intrinsic coordinates \(\xi^a=[\theta,z]\) and metric
\begin{equation}
    dS^2 =dz^2+R^2 d\theta^2,
    \qquad \text{or equivalently} \qquad
    h_{ab}=\begin{pmatrix}
        1&0\\
        0&R^2
    \end{pmatrix}.
\end{equation}
At \(R=0\), the metric degenerates because
\begin{equation}
    \lim_{R\to0}\det(h_{ab})=0.
\end{equation}
Thus the continuous deformation of the intrinsic geometry of the surface encounters degeneracy at the centre. In our case, the leaves \(\Sigma_\Phi\) have dimension \(D-1\), whereas the central source has dimension \(d<D-1\), and the same basic phenomenon occurs.

This behaviour is naturally described by the theory of singular foliations \cite{laurentgengoux2024invitationsingularfoliations}. Although several definitions exist in the literature, for our purposes the weakest notion is sufficient.

\begin{center}
\fbox{
\parbox{1.0\textwidth}{
\textbf{Singular foliation:} A foliation \(\mathcal{F}=\{\Sigma_\Phi\}\) is said to be singular at a leaf \(\Sigma_{\Phi^*}\) if
\[
\dim(\Sigma_\Phi)\neq \dim(\Sigma_{\Phi^*})
\qquad \text{for} \qquad
\Phi\neq\Phi^*.
\]
}
}
\end{center}

Thus the open region of the manifold is regarded as a union of connected, non-intersecting leaves that may have different dimensions. In particular, a foliation in the tubular neighbourhood of an auxiliary source is singular unless
\begin{equation}
    \dim(\text{source})=\dim(\text{leaves}).
\end{equation}
In our case the source is the timelike auxiliary curve \(\gamma\), while the leaves are \((D-1)\)-dimensional timelike hypersurfaces generated by the level sets of a single scalar field \(\Phi(x)\). This singular behaviour is expected, since a curve requires \(D-1\) scalar equations to be specified, whereas a hypersurface requires only one. Equivalently, the mismatch arises from the dimension of the normal space,
\begin{equation}
    \dim N_p(\Sigma)=D-d,
\end{equation}
which for \(\Sigma=\gamma\) reduces to \(D-1\).

Therefore, if one attempts to interpret the foliation-generating scalar field \(\Phi(x)\) simply as the proper distance \(s\) from the source, the construction necessarily breaks down at \(\Phi=0\). At that locus,
\begin{equation}
    \nabla\Phi=0,
\end{equation}
and the level set ceases to be regular. Consequently, \(\Phi=0\) does not define a smooth hypersurface and cannot be regarded as an ordinary leaf of the foliation. The regular foliation introduced earlier,
\begin{equation}
    \mathcal{F} = \{\Sigma_\Phi\}
    =
    \bigg\{
    x \in\mathcal{W}\subset \exp(Z) \subset \mathcal{M}
    \ \bigg| \
    \Phi(x)=s,\ 
    \nabla \Phi\big|_{\Phi=s} \neq 0,\ 
    \forall s \in \mathbb{R},\ 
    \Phi_{\rm max}=\mathcal{E}
    \bigg\},
\end{equation}
already excludes such points. Hence the central curve cannot be included in this regular foliation.

To incorporate the core properly, one must enlarge the framework from a regular foliation to a singular one. For this purpose, the scalar field \(\Phi\) is no longer regarded as an arbitrary defining function, but rather as a Morse--Bott function whose critical set is identified with the central curve \(\gamma\). For the relevant definitions of critical set, Morse--Bott function, and index, see Appendix \eqref{A.5Morsebottfunctions}.

\begin{center}
\fbox{
\parbox{1.0\textwidth}{
\textbf{Proposition:} Let \(\Phi\) be the scalar field generating the tubular foliation \eqref{tubularfoliation} in a finite region \(\mathcal W\) of the tubular neighbourhood \(\Omega\) of the auxiliary curve \(\gamma\). Then the foliation is singular at \(\gamma\), and the curve may be represented as
\begin{equation}\label{morsebottcriticalset}
    \gamma=\{x\in\mathcal{M}\ |\ \nabla \Phi=0\},
\end{equation}
provided \(\Phi\) is reinterpreted as a Morse--Bott function.
}
}
\end{center}

\textbf{Proof:} Let
\begin{equation}
    \boldsymbol{\varphi}:\mathcal{U}\subset \mathcal{M}\to\mathbb {R}^{D-1}
\end{equation}
be a smooth pseudo-Riemannian submersion with non-vanishing gradients \(\nabla\varphi^A\neq0\) for all components \(\varphi^A\). Let the timelike curve \(\gamma\) be defined locally by
\begin{equation}\label{Core1}
    \gamma=\{x\in \mathcal{M}\ |\ \varphi^A=0,\ \nabla\varphi^A\neq0\},
    \qquad
    A\in [1,D-1].
\end{equation}
Then for any \(p\in \gamma\), there exists a local coordinate chart \((\mathcal{U},\mathbf{x})\) with adapted coordinates
\begin{equation}
    \mathbf x=(t,\boldsymbol{\varphi}).
\end{equation}

Now let
\begin{equation}
    \Phi:\mathcal{W}\subset\mathcal{M}\to \mathbb R
\end{equation}
be the foliation-generating scalar field, defined in the tubular neighbourhood \(\Omega\) of \(\gamma\). At critical points of \(\Phi\), the differential vanishes:
\begin{equation}
    d_p\Phi=0.
\end{equation}
The critical set of \(\Phi\) is therefore
\begin{equation}\label{singularset}
    C_\Phi=\{x\in\mathcal M\ |\ \nabla \Phi=0\}.
\end{equation}
Let \(C\subset C_\Phi\) be a connected component that is a non-degenerate critical submanifold of \(\mathcal{M}\). We propose that \(C=\gamma\).

At each \(p\in\gamma\), the Hessian \(\mathbf H_p\Phi\) is degenerate along the tangent space \(T_p(\gamma)\). Equivalently,
\begin{equation}
    \mathbf H_p(\mathbf U, \mathbf V) =0,
    \qquad
    \forall \mathbf U\in T_p(\gamma),\ \forall \mathbf V\in T_p(\mathcal{M}).
\end{equation}
In components, with \(H_{\mu\nu}=\nabla_\mu\nabla_\nu\Phi\), one has for any \(\mathbf U\in T_p(\gamma)\),
\begin{equation}
    V^\nu U^\mu H_{\mu\nu}
    =
    V^\nu U^\mu \nabla_\mu \nabla_\nu\Phi
    =
    V^\nu \nabla_\nu(U^\mu\nabla_\mu\Phi)-V^\nu(\nabla_\nu U^\mu)\nabla_\mu\Phi
    =0,
\end{equation}
since \(\nabla_\mu\Phi=0\) on the critical set.

On the other hand, when restricted to the normal space \(N_p(\gamma)\), the Hessian is assumed to be non-degenerate and positive definite. Define the projected Hessian by
\begin{equation}
    h_{\mu\nu} = P^{\alpha}_\mu P^\beta _\nu H_{\alpha\beta},
    \qquad
    P^\alpha_\mu =\delta^\alpha_\mu +U^\alpha U_\mu,
\end{equation}
where \(U^\mu\) is the timelike unit tangent to \(\gamma\). Then, for \(\mathbf N^A,\mathbf N^B\in N_p(\gamma)\), impose
\begin{equation}
    \mathbf h_p(\mathbf N^A, \mathbf N^B)
    =
    h_{\mu\nu} N^{\mu A} N^{\nu B}>0,
    \qquad
    \mathbf N^A\neq 0.
\end{equation}
Thus the index of the Hessian restricted to the normal directions is zero. By the Morse--Bott lemma, one may therefore state the following local result.

\begin{center}
\fbox{
\parbox{1.0\textwidth}{
\textbf{Lemma 2:} Let \(\Phi:\mathcal{M}\to \mathbb R\) be a Morse--Bott function, and let \(\gamma\) be a connected component of \(C_{\Phi}\) of dimension \(d=1\). Then for every \(p\in \gamma\), there exists a locally adapted coordinate system
\[
(\mathcal{U}\subset\Omega,\mathbf x=\{t,\boldsymbol{\varphi}\}:\mathcal{U}\to V\subset\mathbb R\times \mathbb{R}^{D-1})
\]
containing \(p\), such that
\[
\mathbf x(\mathcal{U}\cap\gamma)=\{(t,\varphi^A)\in V:\varphi^A=0\},
\qquad
\mathbf x(p)=0,
\]
and
\begin{equation}\label{Lemma2}
    \Phi=\Phi(\gamma)+G_{AB}\varphi^A\varphi^B,
    \qquad
    A,B\in[1,D-1].
\end{equation}
In particular, the Hessian of \(\Phi\) is non-degenerate and positive definite in the normal directions.
}
}
\end{center}

Using Lemma 2, and setting \(\Phi(\gamma)=0\) for simplicity, we obtain near the core
\begin{equation}
    \nabla_\mu \Phi
    =
    2G_{AB}\varphi^B\nabla_\mu\varphi^A.
\end{equation}
If we further take \(G_{AB}=\delta_{AB}\), this becomes
\begin{equation}
    \nabla_\mu \Phi =2\varphi_A\nabla_\mu\varphi^A.
\end{equation}
By Eq.~\eqref{Core1}, the core \(\gamma\) is the set of points where \(\varphi^A=0\). Therefore,
\begin{equation}
    \nabla_\mu \Phi\big|_{\varphi^A=0}
    =
    2\varphi_A\nabla_\mu\varphi^A\big|_{\varphi^A=0}
    =0,
    \qquad \text{with} \qquad
    \nabla_\mu \varphi^A\neq0.
\end{equation}
Conversely, in a sufficiently small neighbourhood of \(\gamma\), the independence of the gradients \(\nabla\varphi^A\) and the positive definiteness of \(G_{AB}\) imply that \(\nabla\Phi=0\) only when \(\varphi^A=0\) for all \(A\). Hence, locally,
\begin{equation}
    \gamma=\{x\in\mathcal{M} \ |\ \nabla \Phi=0\}.
\end{equation}
This proves the claim. \(\square\)

\newpage

The submersion variables \(\boldsymbol{\varphi}\), adapted so as to define the local coordinate charts, are taken to have mass dimension
\begin{equation}
    [\boldsymbol{\varphi}]=-1.
\end{equation}
Then, by Lemma 2, the Morse--Bott construction assigns to \(\Phi\) the mass dimension
\begin{equation}\label{Massdimensionnegative}
    [\Phi]=2[\boldsymbol{\varphi}]=-2.
\end{equation}
Since \(\Phi\) will enter the Lagrangian density constructed in the next chapter, this negative mass dimension will play an important role in fixing the dimensions of the remaining terms in that Lagrangian. The detailed implications of this observation will be discussed there.

It is worth emphasising that the Morse--Bott viewpoint adopted here differs from the standard use of Morse--Bott theory in the mathematical literature \cite{Codimmorsebott,scardua2006codimensionfoliationsbottmorsesingularities}. The usual mathematical focus is primarily on the topology of singular foliations in the Riemannian setting. By contrast, the present construction emphasises the geometry of pseudo-Riemannian singular foliations, motivated by a Lagrangian framework tailored to general relativity.

With these ingredients in place, we finally define the tube \(\mathcal{W}\) as the singularly foliated finite region
\begin{equation}\label{tube}
    \mathcal{W}
    =
    \mathrm{Sing}(\mathcal{F})
    =
    \bigg\{
    x\in \mathcal{M}
    \ \bigg|\
    0<\Phi(x)\le \mathcal{E},\ \nabla\Phi\neq0
    \bigg\}
    \cup
    \bigg\{
    x\in \mathcal{M}
    \ \bigg|\
    \nabla\Phi=0
    \bigg\},
\end{equation}
with \(\Phi\) understood as a Morse--Bott function through Eq.~\eqref{Lemma2}.

\section*{Summary}

In this chapter, we constructed a timelike foliation of spacetime in the tubular neighbourhood of the auxiliary timelike worldline \(\gamma\), and studied the geometry of its leaves. To ensure a smooth foliation structure away from \(\gamma\), we identified the bounds of the tubular neighbourhood through the normal flow. We then recognised that the foliation-generating scalar field \(\Phi\) cannot incorporate the central curve within an ordinary regular foliation.

This led us to identify the foliation as singular at the core and to relax the condition \(\nabla\Phi\neq0\) at the central set. The curve \(\gamma\) was then characterised as a connected component of the critical set of \(\Phi\), allowing \(\Phi\) to be reinterpreted as a Morse--Bott function. In this formulation, the central curve becomes part of a singular foliation, and the scalar field acquires the mass dimension
\begin{equation}
    [\Phi]=-2.
\end{equation}

 \chapter{Action Principle for the Foliation Field}\label{chapter4}

In the previous chapter, we constructed the tubular foliation of spacetime generated by a scalar field \(\Phi\), and showed that by relaxing the regularity condition on \(\nabla\Phi\), the foliation naturally develops a singular core. This core was identified with the auxiliary timelike curve \(\gamma(\tau)\), namely the auxiliary worldline, and interpreted as the source of the foliation configuration, though not yet as the source appearing in Einstein's equations. Away from the core, the level sets of \(\Phi\) provide a smooth foliation, while the near-core geometry is controlled by the degeneracy structure of \(\Phi\), Eq.~\eqref{Lemma2}. In this chapter, we will frequently use the terms \textbf{\textit{core}} and \textbf{\textit{source}} interchangeably, since both refer to the same singular set of the foliation.

So far, however, the construction is purely kinematical. In particular, it does not yet specify how the scalar field \(\Phi\) contributes to the stress-energy tensor, nor how it couples to gravity or to the Einstein--Hilbert action.

If the aim is to avoid the distributional contribution associated with the point-particle description, then the stress-energy must arise from the scalar field \(\Phi\) and remain well defined throughout the entire tube, including at the core. This requirement imposes a strong constraint on the admissible class of actions. Since the geometry is encoded in \(\nabla\Phi\) and in the foliation generated by \(\Phi\), the most natural way to construct a Lagrangian is through combinations of the field and its gradients. At the same time, the action must be chosen so that the resulting stress-energy tensor remains finite and integrable near the core, despite the vanishing of \(\nabla\Phi\).

In this chapter, we introduce an action functional for \(\Phi\), adapted to the tubular configuration and designed to meet the requirements stated above. We will derive the corresponding stress-energy tensor, and we will also show explicitly how worldlines re-emerge as effective objects in the ultraviolet limit of the tube.

\newpage

\section{The Action}

Before introducing the action relevant for our construction, it is useful to discuss the basic inspiration behind it. The form of the action we propose is motivated by viewing the configuration as a continuous stacking of codimension-one branes, ultimately leading to a space-filling structure compatible with the foliation developed in the previous chapter. Our goal is therefore to construct a canonical action for the tube that is suited to the geometric and physical requirements stated in the introduction of this thesis.

To begin, consider the minimal action for a single isolated brane.\footnote{A brane is a \(d\)-dimensional timelike surface carrying intrinsic tension, embedded in a \(D\)-dimensional ambient manifold \(\mathcal{M}\).} Let \(\xi^a\) denote the intrinsic coordinates on the brane, with \(a\in[0,d-1]\), and let the embedding be given by \(x^\mu=X^\mu(\xi^a)\). The induced metric on the brane is \(h_{ab}\). The minimal action is then
\begin{equation}\label{braneaction}
\mathcal{S}_{\rm Brane} = -\mathcal{T} \int d^{d}\xi \sqrt{-h}.
\end{equation}
The corresponding equations of motion are \(\mathcal{T}K^A=0\), where \(A\in[1,D-d]\) labels the normal directions. These equations describe extremal embeddings of the brane. Here \(K^A\) denotes the mean extrinsic curvature along each normal direction, and \(\mathcal{T}\) is the intrinsic tension, assumed to be constant along the brane.

Despite the simplicity of the action, the associated stress-energy tensor is distributional and localised on the brane:
\begin{equation}
{\rm T}^{\mu\nu}_{\rm Brane}(x)
=
-\mathcal{T}
\int d^d\xi \sqrt{-h}\, h^{ab} e^\mu_a e^\nu_b\, \delta^{(D)}(x-X(\xi)),
\end{equation}
where \(\delta^{(D)}=\delta^D/\sqrt{-g}\) is the covariant delta function. In the special case \(d=1\), this reduces to the familiar action and stress-energy tensor of a point particle.

We now specialise to the case of a codimension-one brane, for which \(d=D-1\). In this case, the brane may equivalently be described as a level set of a scalar field,
\begin{equation}
\Sigma = \{ x \in \mathcal{M} \mid \Phi(x) = 0 \}.
\end{equation}
In this representation, the action admits the equivalent spacetime form
\begin{equation}
\mathcal{S}_{\rm Brane}
= -\mathcal{T} \int d^{D-1}\xi \sqrt{-h}
= -\mathcal{T} \int d^D x \sqrt{-g}\,\delta(\Phi)\sqrt{|g_{\mu\nu} \nabla^\mu \Phi \nabla^\nu \Phi|},
\end{equation}
see Appendix \ref{A.4Equaivalentactions} for the proof. The corresponding stress-energy tensor takes the form
\begin{equation}
{\rm T}^{\mu\nu}_{\rm Brane}
= -\mathcal{T}\,\delta(\Phi)\Bigg[
\frac{\nabla^\mu \Phi \nabla^\nu \Phi}{\sqrt{|g_{\rho\sigma} \nabla^\rho \Phi \nabla^\sigma \Phi|}}
- g^{\mu\nu} \sqrt{|g_{\rho\sigma} \nabla^\rho \Phi \nabla^\sigma \Phi|}
\Bigg].
\end{equation}
This expression makes it explicit that the stress-energy tensor is localised at \(\Phi=0\), with support entirely restricted by the delta function.

A natural generalisation arises if one considers not a single isolated brane, but a continuous family of such hypersurfaces. In the codimension-one case, there is a single transverse direction, and the scalar field \(\Phi\) itself parametrises that direction. One may then imagine a configuration in which each level set \(\Phi=\text{const}\) carries a brane-like contribution.

To construct such a configuration, one integrates the localised action \eqref{braneaction} along the flow generated by the unit normal associated with \(\nabla\Phi\). Allowing the tension to vary as a function of \(\Phi\), one is led to
\begin{equation}
\mathcal{S}_{\rm Stack}
= -\int \mathcal{T}(\Phi)\,d\Phi \int d^{D-1}\xi\sqrt{-h}.
\end{equation}
Using the same spacetime representation as before, this may be rewritten as
\begin{equation}\label{Almostanaction}
\mathcal{S}_{\rm Stack}
= -\int d^D x \sqrt{-g}\,\mathcal{T}(\Phi)\sqrt{|g_{\mu\nu}\nabla^\mu \Phi \nabla^\nu \Phi|}.
\end{equation}
This action has a clear geometric interpretation: it describes a continuous stacking of codimension-one branes, with the scalar field \(\Phi\) organising spacetime into a foliation and the function \(\mathcal{T}(\Phi)\) determining how the tension is distributed across the leaves. In contrast to the single-brane case, the support of the action is no longer localised on one hypersurface, but instead fills the tubular region through the foliation.

However, the stacked-brane picture implicitly assumes that the foliation is smooth everywhere and that \(\nabla\Phi\neq0\). Our tubular construction is different: it is a singular foliation for which \(\nabla\Phi\) is intentionally allowed to vanish at the core, which is a one-dimensional timelike curve. In that case, the action \eqref{Almostanaction} yields a stress-energy tensor of the form
\begin{equation}
{\rm T}^{\mu\nu}_{\rm Stack}
= -\mathcal{T} (\Phi)\Bigg[
\frac{\nabla^\mu \Phi \nabla^\nu \Phi}{\sqrt{|g_{\rho\sigma} \nabla^\rho \Phi \nabla^\sigma \Phi|}}
- g^{\mu\nu} \sqrt{|g_{\rho\sigma} \nabla^\rho \Phi \nabla^\sigma \Phi|}
\Bigg],
\end{equation}
which is well defined only in the region where \(\nabla\Phi\neq0\). At the core, where \(\nabla\Phi=0\), the first term becomes indeterminate, of the form \(0/0\). The origin of this problem is the square-root structure of the Lagrangian density.

A standard way to eliminate this difficulty is to introduce a non-dynamical auxiliary field \(e\), in close analogy with Polyakov-type treatments of square-root actions. This removes the square root and naturally splits the theory into on-shell and off-shell versions with respect to the auxiliary field. As will be shown below, this reformulation removes the indeterminate \(0/0\) structure from the stress-energy tensor. At the same time, we drop the subscript “Stack” and include a smooth potential \(\mathcal V(\Phi)\) for the scalar field.

\begin{center}
\fbox{
\parbox{1.0\textwidth}{
\begin{equation}\label{Ouraction}
    \mathcal{S}[\Phi,e,g]
    =
    -\int_{\mathcal{W}} d^Dx\sqrt{-g}
    \bigg[
    \frac{1}{2}\mathcal{T}(\Phi)\bigg(e\, g_{\mu\nu} \nabla^\mu\Phi \nabla^\nu \Phi+e^{-1}\bigg)+\mathcal{V}(\Phi)
    \bigg]
\end{equation}
}
}
\end{center}
Here \(\mathcal{W}\) is the tube defined in Eq.~\eqref{tube}. This is the action we propose for the tubular configuration. The corresponding Lagrangian density is
\begin{equation}\label{ourlagrangian}
    \mathcal{L}
    =
    -\frac{1}{2}\mathcal{T}(\Phi)\bigg(e\, g_{\mu\nu} \nabla^\mu\Phi \nabla^\nu \Phi+e^{-1}\bigg)-\mathcal{V}(\Phi).
\end{equation}

One may verify that variation of the action \eqref{Ouraction} with respect to the auxiliary field \(e\) gives the constraint equation
\begin{align}
    \frac{\delta\mathcal{S}}{\delta e}
    =
    g_{\mu\nu} \nabla^\mu\Phi \nabla^\nu \Phi-\frac{1}{e^2}
    =0
    \qquad \Rightarrow \qquad
    e=\frac{1}{\sqrt{g_{\mu\nu}\nabla^\mu\Phi \nabla^\nu \Phi}},
\end{align}
where we have chosen the positive branch. Substituting this back into the action yields the on-shell form
\begin{equation}\label{Almostanaction2}
\mathcal{S}
=
-\int_{\mathcal{W}} d^D x \sqrt{-g}\bigg[
\mathcal{T}(\Phi)\sqrt{|g_{\mu\nu}\nabla^\mu \Phi \nabla^\nu \Phi|}
+\mathcal{V}(\Phi)
\bigg].
\end{equation}
Thus the auxiliary-field formulation is classically equivalent to the square-root action, now supplemented by the potential term.

Before proceeding to the computation of the stress-energy tensor, it is useful to analyse the mass dimensions of the Lagrangian density. In the previous chapter, after identifying the auxiliary timelike curve as a Morse--Bott critical submanifold and the scalar field \(\Phi\) as a Morse--Bott function, we found that the scalar field carries the negative mass dimension given in Eq.~\eqref{Massdimensionnegative}. This immediately imposes strong restrictions on the remaining quantities appearing in the Lagrangian density \eqref{ourlagrangian}.

The Lagrangian density must have mass dimension
\begin{equation}
    [\mathcal{L}]=D.
\end{equation}
Accordingly, the dimensions of the various quantities in the theory are
\begin{align}
    &[\mathcal{L}]=D,
    \qquad
    [g_{\mu\nu}]=0,
    \qquad
    [\nabla]=1,
    \qquad
    [\Phi]=-2, \notag\\
    &[e]=1,
    \qquad
    [\mathcal{T}]=D+1,
    \qquad
    [\mathcal{V}]=D,
    \qquad
    [\mathcal{S}]=0.
\end{align}

In standard field theories, the mass dimension of a field is usually determined by requiring the action to be dimensionless, or equivalently that \([\mathcal{L}]=D\). For example, in four dimensions, the free scalar and vector fields satisfy
\begin{equation}
    [\phi]=[A^\mu]=1,
\end{equation}
while a free Dirac field has\cite{Peskin:1995ev}
\begin{equation}
    [\psi]=\frac{3}{2}.
\end{equation}
In such cases, once the free-field dimensions are fixed, the dimensions of couplings are then determined by the condition \([\mathcal{L}]=4\). In contrast, there are theories in which the dimension of a field is fixed instead by its underlying physical interpretation. A familiar example is Brans--Dicke theory\cite{Brans:1961sx}, where the scalar field is identified schematically as
\begin{equation}
    \phi_{\rm BD}\sim G^{-1},
\end{equation}
so that in four dimensions
\begin{equation}
    [\phi_{\rm BD}]=2.
\end{equation}
Our construction belongs to this second class: the unusual mass dimension of \(\Phi\) is not imposed by a free kinetic term, but follows from the geometric role assigned to the field. This non-standard dimensionality compels us to examine the ultraviolet and infrared behaviour of the theory defined by \eqref{ourlagrangian}.

\subsection{Ultraviolet and Infrared Behaviour of the Tube}\label{4.1.1UVIR}

The non-standard mass dimension of the scalar field, together with the square-root kinetic structure, implies that the tension function must carry a high positive mass dimension, namely
\begin{equation}
    [\mathcal{T}]=D+1,
\end{equation}
so that the Lagrangian density satisfies \([\mathcal{L}]=D\). It is therefore natural to parametrise the tension function in terms of an energy scale \(\Lambda\) as
\begin{equation}
    \mathcal{T}(\Phi)=\Lambda^{D+1}\hat{\mathcal{T}}(\Lambda^2 \Phi),
    \qquad
    [\hat{\mathcal{T}}]=0,
    \qquad
    [\Lambda]=1.
\end{equation}

The dependence on \(\Lambda\) allows us to probe different regimes of the theory, in particular the ultraviolet and infrared regimes. The limit \(\Lambda\to\infty\) corresponds to short-distance or high-energy behaviour, namely the ultraviolet regime. Conversely, \(\Lambda\to0\) corresponds to long-distance or low-energy behaviour, namely the infrared regime. Studying these limits tells us which terms in the Lagrangian density dominate in each regime.

A natural question then arises: what physical quantity does the scale \(\Lambda\) correspond to in the tubular-foliation configuration? To answer this, recall the near-worldline behaviour of the scalar field given by Eq.~\eqref{Lemma2}, together with the definition of the tube \(\mathcal{W}\) in Eq.~\eqref{tube}. Let \(r\) denote the transverse proper distance, and assume spherical symmetry \(S^{D-2}\) in the transverse directions, so that
\begin{equation}
    \varphi_A\varphi^A=r^2.
\end{equation}
Then, under scaling by \(\Lambda\),
\begin{equation}
    \Lambda^2 \Phi =(\Lambda r)^2,
    \qquad
    [\Lambda^2 \Phi]=[(\Lambda r)^2]=0.
\end{equation}
Thus the dimensionless combination \(\Lambda^2\Phi\) is naturally controlled by the product \(\Lambda r\). If the Lagrangian density is effectively supported in the region
\begin{equation}
    0\leq \Lambda^2 \Phi \leq 1,
\end{equation}
with \(\Phi_{\max}=\mathcal{E}\), then \(\Lambda\) scales as the inverse transverse size of the tube:
\begin{equation}\label{scale}
    \Lambda\sim \frac{1}{r}.
\end{equation}
Accordingly, we identify
\begin{equation}
    \Lambda=\frac{1}{r_0}=\frac{1}{\sqrt{\mathcal{E}}},
\end{equation}
where \(r_0\) is the characteristic transverse radius of the tube.

Hence, the ultraviolet regime corresponds to a very thin tube,
\begin{equation}
    r_0\ll1
    \qquad \Leftrightarrow \qquad
    \Lambda\gg1,
\end{equation}
whereas the infrared regime corresponds to a broad tube,
\begin{equation}
    r_0\gg1
    \qquad \Leftrightarrow \qquad
    \Lambda\ll1.
\end{equation}

Let us now consider the scaled Lagrangian density and analyse its ultraviolet and infrared behaviour. We write
\begin{equation}\label{ScaledLagrangian}
    \mathcal{L}_{\Lambda}
    =
    -\Lambda ^{D+1}\hat{\mathcal{T}}(\Lambda^2\Phi)\sqrt{\nabla_\mu \Phi \nabla^\mu \Phi}
    -\Lambda^D\hat{\mathcal{V}}(\Lambda^2\Phi),
\end{equation}
with
\begin{align}
    &[\mathcal{L}_{\Lambda}]=D,
    \qquad
    [\Lambda]=1,
    \qquad
    [\Phi]=-2,\\
    &[\hat{\mathcal{T}}(\Lambda^2\Phi)]=0,
    \qquad
    [\hat{\mathcal{V}}(\Lambda^2\Phi)]=0.
\end{align}

To obtain a qualitative picture of which terms dominate in the ultraviolet and infrared regimes, it is useful to consider representative profiles for the tension and potential functions.

\begin{itemize}
    \item[1.] \textbf{Gaussian profiles:} Let \(\Phi=r^2\). Then the rescaled tension and potential may be chosen to have Gaussian profiles in \(r\), for \(r\neq0\):
    \begin{equation}\label{Gaussianprofile}
    \hat{\mathcal{T}}(\Lambda^2\Phi)=  \hat{\mathcal{T}}_0 e^{-\Lambda^2\Phi},
    \qquad
    \hat{\mathcal{V}}(\Lambda^2\Phi)=  \hat{\mathcal{V}}_0 e^{-\Lambda^2\Phi}.
    \end{equation}
    The kinetic and potential contributions then scale as
    \begin{equation}
        \mathcal{L}_{\Lambda}^{\rm Kin.}
        =
        -\hat{\mathcal{T}}_0 \Lambda^{D+1} e^{-(\Lambda r)^2}\sqrt{X},
        \qquad
        \mathcal{L}_{\Lambda}^{\rm Pot.}
        =
        -\hat{\mathcal{V}}_0 \Lambda^{D} e^{-(\Lambda r)^2},
    \end{equation}
    where
    \begin{equation}
        X=\nabla_\mu\Phi\nabla^\mu\Phi.
    \end{equation}
    Individually, both contributions vanish in the infrared limit \(\Lambda\to0\) and also in the ultraviolet limit \(\Lambda\to\infty\), provided \(r\neq0\). Hence no divergence appears in either asymptotic regime away from the core. However, if one considers the ratio
    \begin{equation}
        R=\frac{\mathcal{L}_{\Lambda}^{\rm Kin.}}{\mathcal{L}_{\Lambda}^{\rm Pot.}},
    \end{equation}
    then one sees that in the ultraviolet regime the kinetic contribution dominates over the potential contribution. It should be remembered that the kinetic term here is not the standard quadratic form, but rather the square root of the gradient invariant.

    \item[2.] \textbf{Power-law profiles:} Consider instead power-law profiles
    \begin{equation}
        \hat{\mathcal{T}}(\Lambda^2\Phi)=\hat{\mathcal{T}}_0(\Lambda^2\Phi)^p,
        \qquad
        \hat{\mathcal{V}}(\Lambda^2\Phi)=\hat{\mathcal{V}}_0(\Lambda^2\Phi)^q,
    \end{equation}
    for \(r\neq0\). Then
    \begin{equation}
        \mathcal{L}_{\Lambda}^{\rm Kin.}
        =
        -\hat{\mathcal{T}}_0 \Lambda^{D+1} (\Lambda r)^{2p}\sqrt{X},
        \qquad
        \mathcal{L}_{\Lambda}^{\rm Pot.}
        =
        -\hat{\mathcal{V}}_0 \Lambda^{D} (\Lambda r)^{2q}.
    \end{equation}
In the above configuration, the individual terms diverge in the ultraviolet limit \((\Lambda\to\infty)\) when
\begin{equation}
    D+1>|2p|,
    \qquad
    D>|2q|,
\end{equation}
for the kinetic and potential contributions, respectively. By contrast, both limits remain finite provided \(p,q<0\) and
\begin{equation}
    |2q|+1=|2p|.
\end{equation}
In the infrared regime, both terms diverge under the conditions
\begin{equation}
    p,q<0,
    \qquad
    |2p|>D+1,
    \qquad
    |2q|>D.
\end{equation}

Now consider the ratio
\begin{equation}
    R=\frac{\mathcal{L}_{\Lambda}^{\rm Kin.}}{\mathcal{L}_{\Lambda}^{\rm Pot.}}.
\end{equation}
Then the kinetic term dominates in the ultraviolet limit whenever
\begin{equation}
    |2p|+1>|2q|,
\end{equation}
whereas under the same condition the potential term dominates in the infrared limit. Conversely, if
\begin{equation}
    |2p|+1<|2q|,
\end{equation}
then the potential contribution dominates in the ultraviolet regime, while the kinetic contribution dominates in the infrared regime. Finally, if
\begin{equation}
    |2p|+1=|2q|,
\end{equation}
then both terms contribute with the same scaling at all energy scales \(\Lambda\).
\end{itemize}

In a similar manner, one may consider different combinations of profiles and obtain different ultraviolet and infrared dominance patterns for the kinetic and potential sectors, depending on the transverse size of the tube.

However, the choice of profiles and the overall scaling cannot be completely arbitrary. Physically, one should restrict attention to those choices for which the ultraviolet limit reproduces the point-particle action:
\begin{equation}
    -m\int d\tau
    =
    \lim_{\Lambda\to\infty}\int d^Dx \sqrt{-g}\,\mathcal{L}_\Lambda .
\end{equation}
Let us examine an explicit example in which the overall scaling is initially chosen according to Eq.~\eqref{ScaledLagrangian}, and the tension and potential are taken to have Gaussian profiles, motivated by Lemma 2 and Eq.~\eqref{Lemma2}:
\begin{equation}\label{Gaussianprofile1}
    \hat{\mathcal{T}}(\Lambda^2\Phi)=\hat{\mathcal{T}}_0 e^{-\Lambda^2\Phi},
    \qquad
    \hat{\mathcal{V}}(\Lambda^2\Phi)=\hat{\mathcal{V}}_0 e^{-\Lambda^2\Phi}.
\end{equation}

For simplicity, consider adapted coordinates \(x^\mu=[\xi^a,r]\), with \(a\in[0,D-2]\) and \(\xi^0=\tau\). We further assume that the spatial cross-sections of the leaves are spherically symmetric, so that the induced metric on each leaf may be written as
\begin{equation}
    dS^2\big|_{\Sigma_{\Phi}}
    =
    -h_{\tau\tau}(x)\,d\tau^2+r^2 d\Omega_{D-2}^2,
\end{equation}
where \(d\Omega_{D-2}^2\) is the line element on \(S^{D-2}\), parametrised by the angular coordinates \(\xi^i\), \(i\in[1,D-2]\).

\begin{figure}[ht]
    \centering
    \includegraphics[width=0.8\linewidth]{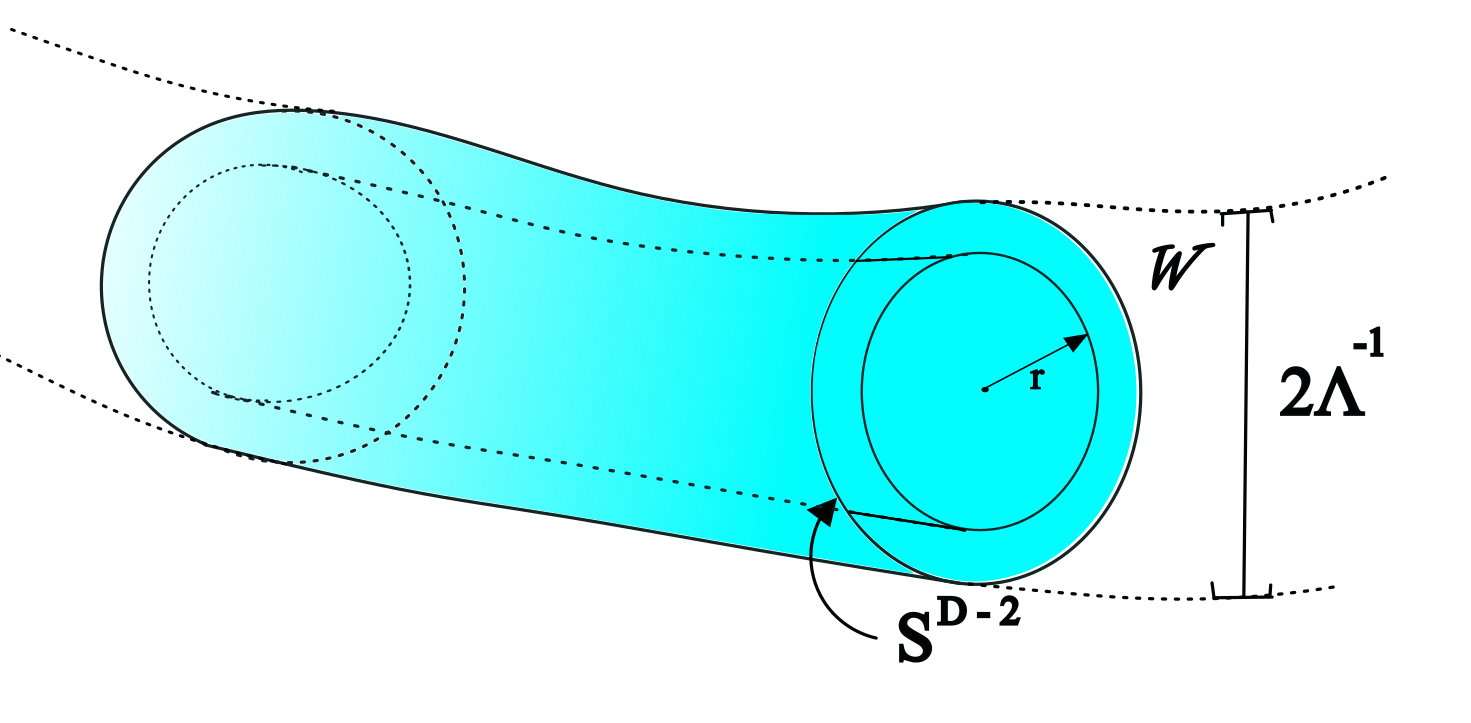}
    \caption{Tube \(\mathcal{W}\), of transverse radius \(\Lambda^{-1}\), with leaves possessing \(S^{D-2}\) spatial symmetry.}
    \label{fig:9}
\end{figure}

The volume element then splits into temporal, radial, and angular parts:
\begin{equation}
    d^Dx\sqrt{-g}
    =
    \sqrt{h_{\tau\tau}}\, d\tau\, dr\, d\Omega_{D-2}.
\end{equation}
In these adapted coordinates, using \(\nabla_\mu\Phi=2r\,N_\mu\) and \(g_{rr}=1\), the square-root term evaluates to
\begin{equation}
    \sqrt{g_{\mu\nu}\nabla^\mu \Phi\nabla^\nu \Phi}
    =
    2r\sqrt{g_{\mu\nu}N^\mu N^\nu}
    =
    2r.
\end{equation}
Substituting this into the rescaled action,
\begin{equation}
    \mathcal{S}_{\Lambda}
    =
    -\int d^Dx \sqrt{-g}\bigg[
    \Lambda ^{D+1}\hat{\mathcal{T}}(\Lambda^2\Phi)\sqrt{\nabla_\mu \Phi \nabla^\mu \Phi}
    +
    \Lambda^D\hat{\mathcal{V}}(\Lambda^2\Phi)
    \bigg],
\end{equation}
we obtain
\begin{equation}
    \mathcal{S}_{\Lambda}
    =
    -\int \sqrt{h_{\tau\tau}}\,d\tau
    \int d\Omega_{D-2}
    \int_{0}^{\Lambda^{-1}} r^{D-2}dr
    \bigg[
    2r\,\Lambda ^{D+1}\hat{\mathcal{T}}_0 e^{-(\Lambda r)^2}
    +
    \Lambda^D\hat{\mathcal{V}}_0 e^{-(\Lambda r)^2}
    \bigg].
\end{equation}

For the moment, let us set \(h_{\tau\tau}=1\). Later in this section, we will generalise the result under the assumption that \(h_{\tau\tau}\) remains analytic in the region \(\Omega\). Using
\begin{equation}
    \int d\Omega_{D-2}
    =
    \frac{2\pi^{\frac{D-1}{2}}}{\Gamma(\frac{D-1}{2})},
\end{equation}
the above integration reduces the action to
\begin{equation}
    \mathcal{S}_{\Lambda}
    =
    -\int d\tau\, 2\pi^{\frac{D-1}{2}}
    \Bigg[
    \Lambda \hat{\mathcal{T}}_0
    \Bigg(
    \frac{D}{D-1}
    -
    \frac{\Gamma(\frac{D}{2},1)}{\Gamma(\frac{D-1}{2})}
    \Bigg)
    +
    \frac{\Lambda \hat{\mathcal{V}}_0}{2}
    \Bigg(
    1-
    \frac{\Gamma(\frac{D-1}{2},1)}{\Gamma(\frac{D-1}{2})}
    \Bigg)
    \Bigg].
\end{equation}
Let us denote the integrand of this reduced action by \(m(\Lambda)\):
\begin{equation}
    m(\Lambda)
    =
    2\pi^{\frac{D-1}{2}}
    \Bigg[
    \Lambda \hat{\mathcal{T}}_0
    \Bigg(
    \frac{D}{D-1}
    -
    \frac{\Gamma(\frac{D}{2},1)}{\Gamma(\frac{D-1}{2})}
    \Bigg)
    +
    \frac{\Lambda \hat{\mathcal{V}}_0}{2}
    \Bigg(
    1-
    \frac{\Gamma(\frac{D-1}{2},1)}{\Gamma(\frac{D-1}{2})}
    \Bigg)
    \Bigg].
\end{equation}
This quantity diverges in the ultraviolet limit \(\Lambda\to\infty\). Hence, with the scaling \eqref{ScaledLagrangian}, the shrinking tube does not reproduce a finite point-particle action.

The origin of this ultraviolet divergence lies in the overall scaling of the Lagrangian density. To remove it, we instead consider the rescaled Lagrangian
\begin{equation}\label{correctlyscaledlagrangian}
    \mathcal{L}_{\Lambda}
    =
    -\Lambda^D \hat{\mathcal{T}}(\Lambda^2\Phi)\sqrt{\nabla_\mu \Phi \nabla^\mu \Phi}
    -\Lambda^{D-1}\hat{\mathcal{V}}(\Lambda^2\Phi),
\end{equation}
so that the rescaled functions are no longer completely dimensionless, but instead satisfy
\begin{equation}
    [\hat{\mathcal{T}}]=[\hat{\mathcal{V}}]=1.
\end{equation}
With this revised scaling, and under the same Gaussian ansatz and metric assumptions, the reduced mass becomes
\begin{equation}\label{correctscaling}
    m(\Lambda)
    =
    2\pi^{\frac{D-1}{2}}
    \Bigg[
    \hat{\mathcal{T}}_0
    \Bigg(
    \frac{D}{D-1}
    -
    \frac{\Gamma(\frac{D}{2},1)}{\Gamma(\frac{D-1}{2})}
    \Bigg)
    +
    \frac{\hat{\mathcal{V}}_0}{2}
    \Bigg(
    1-
    \frac{\Gamma(\frac{D-1}{2},1)}{\Gamma(\frac{D-1}{2})}
    \Bigg)
    \Bigg].
\end{equation}
See Appendix \eqref{A.6Incompletegammafunction} for the definition of the incomplete gamma function \(\Gamma(s,x)\).

For the Gaussian profiles \eqref{Gaussianprofile1}, the revised scaling \eqref{correctlyscaledlagrangian} implies
\begin{equation}
    [\hat{\mathcal{T}}_0]=[\hat{\mathcal{V}}_0]=1.
\end{equation}
In this case, the reduced mass is free from ultraviolet divergence, and it also remains free from infrared divergence. Indeed,
\begin{equation}
    \lim_{\Lambda\to\infty}m(\Lambda)
    =
    m
    =
    \lim_{\Lambda\to0}m(\Lambda).
\end{equation}
It is therefore reasonable to say that the mass information, which in the point-particle picture is sharply concentrated on a worldline with Dirac-delta support, is here encoded instead in the tension and potential profiles with no delta-function support. In this sense, the mass becomes an emergent feature of the ultraviolet limit of the tube.

The above calculation shows that it is possible to recover the point-particle action in the ultraviolet limit, provided the Lagrangian is rescaled appropriately. This realisation is not unique. For example, if one instead chooses
\begin{equation}
    \hat{\mathcal{T}}(\Lambda^2\Phi)
    =
    \hat{\mathcal{T}}_0\sqrt{1-\Lambda r}\,e^{-(\Lambda r)^2},
    \qquad
    \hat{\mathcal{V}}(\Lambda^2\Phi)
    =
    \hat{\mathcal{V}}_0\sqrt{1-\Lambda r}\,e^{-(\Lambda r)^2},
\end{equation}
under the same geometric assumptions, then one obtains
\begin{equation}\label{example2ofmassprofiles}
    m(\Lambda)
    =
    \frac{2\pi^{\frac{D-1}{2}}}{\Gamma(\frac{D-1}{2})}
    \Bigg[
    \hat{\mathcal{T}}_0
    \Bigg(
    \sum_{k=0}^{\infty}
    \frac{(-1)^k}{k!}
    \frac{\Gamma(D+2k)\Gamma(3/2)}{\Gamma(D+2k+3/2)}
    \Bigg)
    +
    \hat{\mathcal{V}}_0
    \Bigg(
    \sum_{k=0}^{\infty}
    \frac{(-1)^k}{k!}
    \frac{\Gamma(D+2k-1)\Gamma(3/2)}{\Gamma(D+2k+1/2)}
    \Bigg)
    \Bigg].
\end{equation}
Since both series are finite, the right-hand side remains finite, and so does the induced mass on the left-hand side. Moreover, the right-hand side carries no residual \(\Lambda\)-dependence, so it is automatically safe in both the ultraviolet and infrared limits. A further important observation is that even if the potential vanishes, the mass information need not be lost; it may still be carried entirely by the tension profile.

It is therefore natural to define, in general, a class of admissible profile functions that yield a finite positive point-particle mass in the ultraviolet limit. This requirement may be written as
\begin{equation}\label{positivitycriterion}
   0
   <
   \lim_{\Lambda\to\infty}\int m(\Lambda)d\tau
   =
   -\lim_{\Lambda\to\infty}\int_{\Omega}d^{D}x\,\mathcal{L}_{\Lambda}
   <
   \infty,
   \qquad
   \forall D.
\end{equation}
This condition is strong, but to extract something genuinely useful from it, one needs an additional assumption about the ambient spacetime. We therefore assume that the spacetime is globally hyperbolic and admits a Cauchy surface. This excludes closed timelike curves and, more importantly for our purposes, allows a \((D-1)+1\) decomposition both topologically and geometrically.

Global hyperbolicity implies that the spacetime manifold is globally diffeomorphic to
\begin{equation}
    \mathcal{M}\simeq \Sigma\times\mathbb{R}_\tau,
\end{equation}
with
\begin{equation}
    \Sigma_{\Phi}\cap\Sigma=S^{D-2},
\end{equation}
where \(\Sigma_{\Phi}\) is a timelike leaf of the tubular foliation \eqref{tubularfoliation}, and \(\Sigma\) is a spacelike Cauchy surface. The cross-section of the tube \eqref{tube} is then diffeomorphic, inside the region \(\Omega\), to
\begin{equation}
    \mathcal{W}\cap\Sigma=S^{D-2}\times \mathbb{R}_{\Phi}.
\end{equation}
Hence the tube itself is diffeomorphic to the swept submanifold
\begin{equation}\label{splittedtube}
    \mathcal{W}
    =
    (\mathcal{W}\cap \Sigma)\times \mathbb{R}_\tau
    =
    S^{D-2}\times \mathbb{R}_{\Phi}\times \mathbb{R}_\tau.
\end{equation}
Accordingly, at any point \(p\in\mathcal{W}\), the tangent space splits as
\begin{equation}\label{Split}
    T_p(\mathcal{W})
    =
    T_p(S^{D-2})
    \oplus
    T_p(\mathbb{R}_{\Phi})
    \oplus
    T_p(\mathbb{R}_\tau).
\end{equation}
Let this split tangent space be spanned by the basis
\begin{equation}
    \{\mathbf{E}_i,\mathbf{N},\mathbf{U}\},
    \qquad
    i\in[1,D-2],
\end{equation}
where \(\mathbf{E}_i\) spans \(T_p(S^{D-2})\), \(\mathbf{N}\) spans \(T_p(\mathbb{R}_{\Phi})\), and \(\mathbf{U}\) spans \(T_p(\mathbb{R}_\tau)\). The dual tangent space then splits as
\begin{equation}
    T_p^*(\mathcal{W})
    =
    T_p^*(S^{D-2})
    \oplus
    T_p^*(\mathbb{R}_{\Phi})
    \oplus
    T_p^*(\mathbb{R}_\tau),
\end{equation}
and is spanned by
\begin{equation}
    \{\mathbf{e}^i,\underline{\mathbf{N}},\underline{\mathbf{U}}\},
\end{equation}
with the duality relations
\begin{equation}
    \langle \mathbf{E}_i,\mathbf{e}^j \rangle=\delta^j_i,
    \qquad
    \langle \mathbf{N},\underline{\mathbf{N}}\rangle=1,
    \qquad
    \langle \mathbf{U},\underline{\mathbf{U}}\rangle=-1.
\end{equation}

Let the metric on \(S^{D-2}\) be
\begin{equation}
    \boldsymbol{\gamma}
    =
    \Phi\,\gamma_{ij}\mathbf{e}^i\otimes \mathbf{e}^j,
\end{equation}
where we have used Lemma 2 with \(\varphi_A\varphi^A=r^2\), so that \(\Phi=r^2\). The canonical \((D-2)\)-area form is then
\begin{equation}
    \mathbf{A}_{D-2}
    =
    \Phi^{\frac{D-2}{2}}\sqrt{\gamma}\,
    \mathbf{e}^1\wedge\cdots\wedge\mathbf{e}^{D-2},
\end{equation}
where \(\gamma=\det(\gamma_{ij})\). The spacetime volume form in the region \(\Omega\) may therefore be written as
\begin{equation}\label{volumeformsplit}
    {\rm Vol}_g(\mathcal{W})
    =
    -\underline{\mathbf{U}}\wedge \underline{\mathbf{N}}\wedge \mathbf{A}_{D-2}.
\end{equation}
The unit normal covector is
\begin{equation}
    \underline{\mathbf{N}}
    =
    \frac{\mathbf{d}\Phi}{\sqrt{g_{\mu\nu}\nabla^\mu \Phi \nabla^\nu \Phi}},
\end{equation}
while the velocity covector is
\begin{equation}
    \underline{\mathbf{U}}
    =
    -\sqrt{h_{\tau\tau}}\,\mathbf{d}\tau,
\end{equation}
with \(h_{\tau\tau}\) the \(\tau\tau\)-component of the induced metric on the leaves of the foliation \eqref{tubularfoliation}.

For a consistently oriented tube, the measure splits according to \eqref{volumeformsplit}, and the spacetime integral becomes
\begin{equation}
    \int_{\mathcal{W}} d^Dx \sqrt{-g}
    =
    -\int_{\mathcal{W}} \underline{\mathbf{U}}\wedge \underline{\mathbf{N}}\wedge \mathbf{A}_{D-2}
    =
    \int_\mathcal{W}
    \Phi^{\frac{D-2}{2}}
    \sqrt{\frac{h_{\tau\tau}}{g_{\mu\nu}\nabla^\mu \Phi \nabla^\nu \Phi}}
    \,d\tau\,d\Phi\,d\Omega_{D-2},
\end{equation}
where \(d\Omega_{D-2}\) is the intrinsic volume form on \(S^{D-2}\).

Using this split measure, the scaled action may be rewritten as
\begin{equation}
    \mathcal{S}_{\Lambda}
    =
    -\int_{\mathcal{W}} d^Dx \sqrt{-g}\,\mathcal{L}_{\Lambda}
    =
    \int_{\mathcal{W}} \mathcal{L}_{\Lambda}
    \big[
    \underline{\mathbf{U}}\wedge \underline{\mathbf{N}}\wedge \mathbf{A}_{D-2}
    \big]
    =
    -\int_\mathcal{W}
    \mathcal{L}_{\Lambda}
    \Phi^{\frac{D-2}{2}}
    \sqrt{\frac{h_{\tau\tau}}{g_{\mu\nu}\nabla^\mu \Phi \nabla^\nu \Phi}}
    \,d\tau\,d\Phi\,d\Omega_{D-2}.
\end{equation}
Substituting the rescaled Lagrangian density \eqref{correctlyscaledlagrangian}, we obtain
\begin{equation}
\mathcal{S}_{\Lambda}
=
\int_{\mathbb{R}_{\tau}}\sqrt{h_{\tau\tau}}\,d\tau
\int_{S^{D-2}}d\Omega_{D-2}
\int_{\mathbb{R}_{\Phi}}d\Phi\,
\Phi^{\frac{D-2}{2}}
\bigg[
\Lambda^{D}\hat{\mathcal{T}}(\Lambda^2\Phi)
+
\frac{\Lambda^{D-1}}{\sqrt{{g_{\mu\nu}\nabla^\mu \Phi \nabla^\nu \Phi}}}\hat{\mathcal{V}}(\Lambda^2\Phi)
\bigg].
\end{equation}
The angular integration can now be carried out, giving
\begin{equation}
    \mathcal{S}_{\Lambda}
    =
    \Omega_{D-2}
    \int_{\mathbb{R}_{\tau}}\sqrt{h_{\tau\tau}}\,d\tau
    \int_{\mathbb{R}_{\Phi}}d\Phi\,
    \Phi^{\frac{D-2}{2}}
    \bigg[
    \Lambda^{D}\hat{\mathcal{T}}(\Lambda^2\Phi)
    +
    \frac{\Lambda^{D-1}}{\sqrt{{g_{\mu\nu}\nabla^\mu \Phi \nabla^\nu \Phi}}}\hat{\mathcal{V}}(\Lambda^2\Phi)
    \bigg].
\end{equation}

To proceed further, it is convenient to return to the adapted coordinates in \(\mathcal{W}\) and again use the Morse--Bott form \(\Phi=r^2\). Then
\begin{equation}
  \sqrt{{g_{\mu\nu}\nabla^\mu \Phi \nabla^\nu \Phi}}=2\sqrt{\Phi}.
\end{equation}
After the substitution
\begin{equation}
    \lambda=\Lambda^2\Phi,
    \qquad
    \lambda\in[0,1],
\end{equation}
the action takes the simpler form
\begin{equation}\label{betteraction}
    \mathcal{S}_{\Lambda}
    =
    \Omega_{D-2}
    \int_{\mathbb{R}_{\tau}}\sqrt{h_{\tau\tau}(\lambda,\tau)}\,d\tau
    \bigg[
    \int_0^1d\lambda \,\lambda^{\frac{D-2}{2}}\hat{\mathcal{T}}(\lambda)
    +
    \frac{1}{2}\int_0^1d\lambda \,\lambda^{\frac{D-3}{2}}\hat{\mathcal{V}}(\lambda)
    \bigg].
\end{equation}
Here
\begin{equation}
    \Omega_{D-2}
    =
    \frac{2\pi^{\frac{D-1}{2}}}{\Gamma(\frac{D-1}{2})}.
\end{equation}

Assume now that \(h_{\tau\tau}\) is real analytic in \(\lambda\). Writing \(h_{\tau\tau}(\lambda,\tau)=h(\lambda,\tau)\) for brevity, we expand near the core \(\lambda=0\) as
\begin{equation}
    h(\lambda,\tau)
    =
    h_0(\tau)
    +
    \sum_{k=1}^{\infty}
    \frac{\lambda^k}{k!}
    \frac{\partial^kh}{\partial\lambda^k}\Bigg|_{\lambda=0}.
\end{equation}
Define
\begin{equation}
    \varepsilon(\lambda,\tau)
    =
    \frac{1}{h_0(\tau)}
    \sum_{k=1}^{\infty}
    \frac{\lambda^k}{k!}
    \frac{\partial^kh}{\partial\lambda^k}\Bigg|_{\lambda=0},
\end{equation}
so that
\begin{equation}
    \sqrt{h(\lambda,\tau)}
    =
    \sqrt{h_0(\tau)}\,\big[1+\varepsilon(\lambda,\tau)\big]^{1/2}.
\end{equation}
Since the series converges for \(0\leq\lambda<1\), we may use the binomial expansion to obtain
\begin{equation}
     \sqrt{h(\lambda,\tau)}
     =
     \sqrt{h_0(\tau)}
     +
     \varepsilon(\lambda,\tau)\sqrt{h_0(\tau)}
     +
     \sqrt{h_0(\tau)}
     \sum_{n=1}^{\infty}
     \frac{(-1)^n\Gamma(2n)}{2^{2n}\Gamma(n+2)\Gamma(n)}
     \varepsilon^{n+1}(\lambda,\tau).
\end{equation}
Substituting this expansion into the action \eqref{betteraction}, we find
\begin{align}
    \mathcal{S}_{\Lambda}
    &=
    m(\Lambda)\int_{\mathbb{R}_s} ds \notag\\
    &\quad
    +
    \Omega_{D-2}
    \sum_{k=1}^{\infty}\frac{1}{k!}
    \int_{\mathbb{R}_\tau}
    \Bigg(
    h_0^{-1/2}(\tau)\frac{\partial^kh}{\partial\lambda^k}\Bigg|_{\lambda=0}
    \Bigg)d\tau
    \int_0^1d\lambda
    \bigg[
    \lambda^{\frac{D-2+2k}{2}}\hat{\mathcal{T}}(\lambda)
    +
    \lambda^{\frac{D-3+2k}{2}}\hat{\mathcal{V}}(\lambda)
    \bigg] \notag\\
    &\quad
    +
    \Omega_{D-2}
    \sum_{n=1}^{\infty}
    \frac{(-1)^n\Gamma(2n)}{2^{2n}\Gamma(n+2)\Gamma(n)}
    \int_{\mathbb{R}_{\tau}}\int_{0}^{1}
    h_0^{1/2}(\tau)\varepsilon^{n+1}(\lambda,\tau)
    \bigg[
    \lambda^{\frac{D-2}{2}}\hat{\mathcal{T}}(\lambda)
    +
    \lambda^{\frac{D-3}{2}}\hat{\mathcal{V}}(\lambda)
    \bigg]
    d\tau\, d\lambda .
\end{align}

For readability, let us rewrite this as

\begin{center}
\fbox{
\parbox{1.0\textwidth}{
\begin{equation}\label{splittedaction}
    \mathcal{S}_{\Lambda}
    =
    m(\Lambda)\int_{\mathbb{R}_s}ds
    +
    \int_{\mathbb{R}_{\tau}}\mu(\tau)\,d\tau .
\end{equation}
}
}
\end{center}

where
\begin{equation}\label{mu}
    \mu(\tau)
    =
    \int_{0}^{1} {\rm A}(\lambda)
    \sum_{k=1}^{\infty}
    \Bigg[
    \frac{c_k(\tau)}{k!}\lambda^k
    +
    h_0^{1/2}(\tau)b_k\varepsilon^{k+1}(\lambda,\tau)
    \Bigg]
    d\lambda,
\end{equation}
with
\begin{align}
    & {\rm A}(\lambda)
    =
    \Omega_{D-2}
    \bigg[
    \lambda^{\frac{D-2}{2}}\hat{\mathcal{T}}(\lambda)
    +
    \lambda^{\frac{D-3}{2}}\hat{\mathcal{V}}(\lambda)
    \bigg],\\
    &c_k(\tau)
    =
    h_0^{-1/2}(\tau)\frac{\partial^kh}{\partial\lambda^k}\Bigg|_{\lambda=0}, \\
    &b_k
    =
    \frac{(-1)^k\Gamma(2k)}{2^{2k}\Gamma(k+2)\Gamma(k)},\\
    &m(\Lambda)
    =
    \int_{0}^{1}{\rm A}(\lambda)\,d\lambda,\\
    &ds
    =
    \sqrt{h_{\tau\tau}(0,\tau)}\,d\tau.
\end{align}

The split action \eqref{splittedaction} now makes it possible to define the admissible class of tension and potential profiles by imposing positivity and finiteness of the ultraviolet mass, namely
\begin{equation}\label{weakcriterion}
    \infty
    >
    \lim_{\Lambda\to\infty}m(\Lambda)
    =
    \Omega_{D-2}\int_0^1 d\lambda\,\lambda^{\frac{D-2}{2}}\hat{\mathcal{T}}(\lambda)
    +
    \Omega_{D-2}\int_0^1 d\lambda\,\lambda^{\frac{D-3}{2}}\hat{\mathcal{V}}(\lambda)
    >0.
\end{equation}
This gives a weak positive class of admissible profiles. A stronger positivity condition may be imposed by demanding
\begin{align}\label{strongpositivityconditions}
    &\infty
    >
    \int_0^1 d\lambda\,\lambda^{\frac{D-2}{2}}\hat{\mathcal{T}}(\lambda)
    >0,\\
    &\infty
    >
    \int_0^1 d\lambda\,\lambda^{\frac{D-3}{2}}\hat{\mathcal{V}}(\lambda)
    \geq 0.
\end{align}

The space of such profiles is large. One may choose different tension and potential profiles satisfying these strong positivity conditions in order to model different geometric distributions of mass inside the world tube \eqref{tube}. This is not unusual in physics: for an extended body of finite mass there is generally no unique density profile, and physical considerations constrain only an admissible class. In the present construction, sharp localisation is avoided and with it the Geroch--Traschen obstruction discussed earlier, while the mass information is encoded instead in the geometry of the tube through the tension and potential profiles. Only in the ultraviolet limit does the mass concentrate onto the central worldline, which is also the Morse--Bott critical set where the tubular foliation becomes singular.

One must, however, keep in mind that the choice of profile affects not only the mass term but also the second contribution in the split action \eqref{splittedaction}, through \(\mu(\tau)\), since \(\mu(\tau)\) depends explicitly on \({\rm A}(\lambda)\). The term involving \(\mu(\tau)\) survives the ultraviolet limit because it carries no explicit \(\Lambda\)-dependence. In particular, if the \(\tau\tau\)-component of the induced metric is constant in \(\lambda\), then
\begin{equation}
    \mu(\tau)=0,
\end{equation}
since in that case all \(\lambda\)-derivatives of \(h_{\tau\tau}\) vanish, and therefore
\begin{equation}
    c_k=0=\varepsilon^{k+1}.
\end{equation}

It is also useful to recall that \(\sqrt{h_{\tau\tau}}\) is the ADM lapse restricted to \(\Sigma\cap\mathcal{W}\), where \(\Sigma\) is the spacelike Cauchy surface and \(\mathcal{W}\) is the tube. Thus, taking
\begin{equation}
    \sqrt{h_{\tau\tau}(\lambda,\tau)}=C(\tau)
\end{equation}
means that the disks
\begin{equation}
    \mathcal{D}=\Sigma\cap\mathcal{W}
\end{equation}
are separated unevenly along the time direction.

\begin{figure}[htbp]
    \centering
    \includegraphics[width=0.3\linewidth]{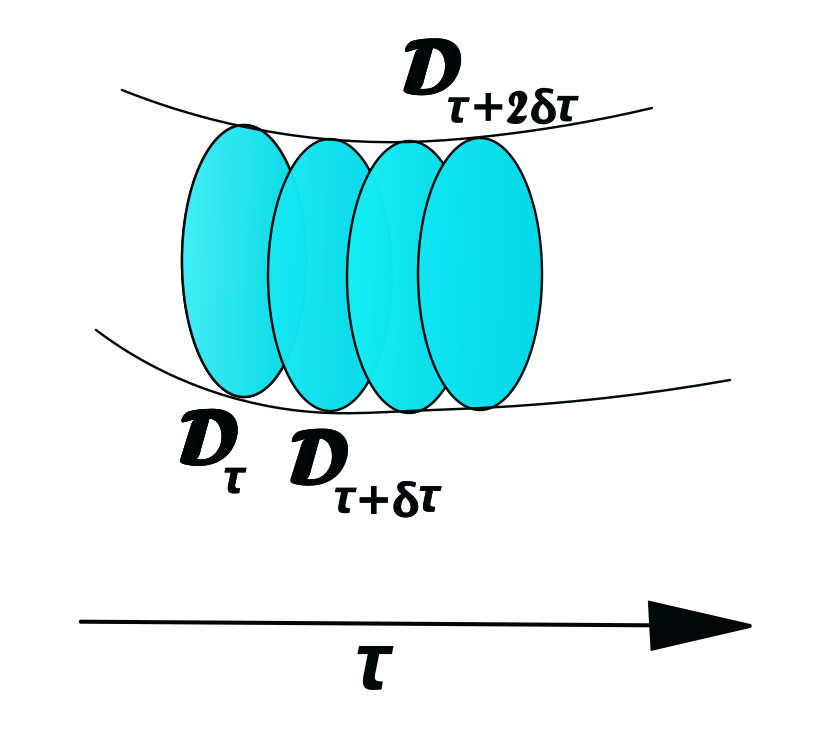}
    \caption{Separated disks for restricted ADM lapse, i.e. \(\sqrt{h_{\tau\tau}}=C(\tau)\).}
    \label{fig:10}
\end{figure}

More precisely,
\begin{equation}
  \int_{\tau'=\tau}^{\tau'=\tau+\delta\tau} C(\tau')\,d\tau'
  =
  d(\mathcal{D}_{\tau},\mathcal{D}_{\tau+\delta\tau})
  \neq
  d(\mathcal{D}_{\tau+\delta\tau},\mathcal{D}_{\tau+2\delta\tau})
  =
  \int_{\tau'=\tau+\delta\tau}^{\tau'=\tau+2\delta\tau} C(\tau')\,d\tau' .
\end{equation}
Thus the \(\mu(\tau)\)-contribution survives only if the restricted ADM lapse depends on the transverse parameter, that is,
\begin{equation}
    \partial_{\lambda}h_{\tau\tau}\neq0.
\end{equation}

At this stage, a natural question arises. When the transverse geometry is integrated out and the reduced action \eqref{splittedaction} is obtained, what exactly is the extra term \(\mu(\tau)\)? Should the ultraviolet limit not reduce directly to the ordinary point-particle action with finite mass? To answer this, rewrite \eqref{splittedaction} in the equivalent form
\begin{equation}
     \mathcal{S}_{\Lambda}
     =
     m(\Lambda)\int_{\mathbb{R}_{\tau}} d\tau\sqrt{g_{\mu\nu}\dot X^\mu\dot X^\nu}
     +
     \int_{\mathbb{R}_{\tau}}\mu(X(\tau))\,d\tau .
 \end{equation}
Varying with respect to the embedding map \(X^\mu(\tau)\), one obtains the effective equation of motion for the core \(\gamma\):
\begin{equation}\label{Forcedequationofmotion}
     m(\Lambda)\bigg[
     \frac{d^2X^\nu}{d\tau^2}
     +
     \Gamma^\nu_{\rho\sigma}
     \frac{dX^\rho}{d\tau}\frac{dX^\sigma}{d\tau}
     \bigg]
     =
     Q^\nu,
     \qquad
     Q^\nu=\nabla^\nu \mu .
 \end{equation}

At first sight, the right-hand side may appear to be a force that can be removed by reparametrisation, reducing the equation to the standard geodesic equation. In general, however, this is not true. Such a redundancy occurs only when the force is parallel to the velocity \(\dot X^\nu\), namely when
\begin{equation}
     \nabla^\nu\mu=f(\tau)\dot X^\nu .
 \end{equation}
For \(\mu\) as defined in Eq.~\eqref{mu}, the direction of \(\nabla^\nu\mu\) is not fixed a priori. It may be decomposed into tangential and normal components with respect to the velocity vector. The tangential part is always proportional to \(\dot X^\nu\) and can be absorbed by reparametrisation, but the normal part cannot. Thus the genuine equation of motion for the core is
 \begin{equation}\label{finalformoftheeom}
      m(\Lambda)\bigg[
      \frac{d^2X^\nu}{d\tau^2}
      +
      \Gamma^\nu_{\rho\sigma}
      \frac{dX^\rho}{d\tau}\frac{dX^\sigma}{d\tau}
      \bigg]
      =
      [\delta^\nu_\rho+ U^\nu U_\rho]Q^\rho
      =
      \hat Q^\nu,
      \qquad
      U^\nu=\dot X^\nu .
 \end{equation}
Hence the core is not, in general, a geodesic curve, but an accelerated trajectory. This acceleration is not produced by an external force field; rather, it is an artifact of the transversely integrated tubular geometry.

Moreover, the force
\begin{equation}
    \hat Q^\nu=[\delta^\nu_\rho+ U^\nu U_\rho]Q^\rho
\end{equation}
does not vanish in general even in the ultraviolet limit, when the tube collapses to a one-dimensional curve:
\begin{equation}
    \lim_{\Lambda\to\infty}
    \Bigg[
    m(\Lambda)\bigg(
    \frac{d^2X^\nu}{d\tau^2}
    +
    \Gamma^\nu_{\rho\sigma}
    \frac{dX^\rho}{d\tau}\frac{dX^\sigma}{d\tau}
    \bigg)
    -
    \hat Q^\nu
    \Bigg]
    =
    m\,U^\rho \nabla_\rho U^\nu-\hat Q^\nu .
\end{equation}
This non-vanishing \(\hat Q^\nu\) is one of the striking features of the theory defined by \eqref{correctlyscaledlagrangian}. It shows that if one regularises a point-particle worldline geometrically by a codimension-zero tube, then the core of that tube, which is also a Morse--Bott critical set where the timelike foliation becomes singular, need not follow a geodesic of the ambient spacetime. Even after the tube is smoothly shrunk to a line, a non-trivial force contribution remains in the effective equation of motion.

To understand the origin of this force more clearly, it is useful to simplify \(\mu(\tau)\). Define the cumulative mass
\begin{equation}
    m(\lambda)=\int_{0}^{\lambda}{\rm A}(y)\,dy.
\end{equation}
Integrating Eq.~\eqref{mu} by parts then gives
\begin{equation}\label{zoomedinmu}
    \mu(\tau)
    =
    m(\Lambda){\rm B}(1,\tau)
    -
    \int_{0}^{1}m(\lambda)\,\partial_{\lambda}{\rm B}(\lambda,\tau)\, d\lambda,
\end{equation}
where
\begin{equation}
    {\rm B}(\lambda,\tau)
    =
    \sum_{k=1}^{\infty}
    \Bigg[
    \frac{c_k(\tau)}{k!}\lambda^k
    +
    h_0^{1/2}(\tau)b_k\varepsilon^{k+1}(\lambda,\tau)
    \Bigg],
\end{equation}
and note that
\begin{equation}
    {\rm B}(0,\tau)=0.
\end{equation}
Equation \eqref{zoomedinmu} displays the dependence of \(\mu\) on the total mass \(m(\Lambda)\), the cumulative mass \(m(\lambda)\), and the transverse derivatives of \(h_{\tau\tau}\) near the core \(\gamma\). The effective force is therefore sourced by both the total mass of the tube and the transverse geometry encoded in the derivatives of the \(\tau\tau\)-component of the induced metric.

This naturally suggests interpreting \(\mu(\tau)\) as a self-force potential and \(\hat Q^\nu\) as a self-force. Thus, within the singularly foliated tubular geometry \eqref{tube}, and for tension and potential profiles belonging to either the weak or strong positive classes \eqref{weakcriterion} and \eqref{strongpositivityconditions}, a canonical self-force correction to the worldline emerges in the ultraviolet limit, provided \(h_{\tau\tau}\in C^k\) with \(k\geq1\).

We do not pursue the self-force analysis further here, since it deserves a separate treatment. Instead, we now return to the full action and turn to the next major question: the form of the stress-energy tensor inside the tube and the corresponding energy conditions. 

\subsection{Stress-Energy Tensor}\label{4.1.2stressenergytensor}

We now couple the foliation dynamics minimally to gravity through the Einstein--Hilbert action,
\begin{equation}
    \mathcal{S}
    =
    \int_{\mathcal{W}} d^Dx\sqrt{-g}\Bigg[
    \frac{M_P^2}{2}{\rm R}
    -\frac{1}{2}\mathcal{T}(\Phi)\bigg(e\, g_{\mu\nu} \nabla^\mu\Phi \nabla^\nu \Phi+e^{-1}\bigg)
    -\mathcal{V}(\Phi)
    \Bigg].
\end{equation}
Metric variation then gives the Einstein equations in the interior region of the tube \(\mathcal{W}\),
\begin{equation}\label{Einstein}
    {\rm R}_{\mu\nu}-\frac{1}{2}g_{\mu\nu}{\rm R}
    =
    \frac{2}{M_P^2}\,{\rm T}_{\mu\nu},
    \qquad
    M_P^2=\frac{1}{4\pi G},
\end{equation}
where we are using the convention
\begin{equation}
    {\rm T}_{\mu\nu}
    =
    -\frac{2}{\sqrt{-g}}\frac{\delta \mathcal{S}_{\rm matter}}{\delta g^{\mu\nu}}.
\end{equation}

For the action \eqref{Ouraction}, before eliminating the auxiliary field \(e\), the stress-energy tensor is
\begin{align}\label{ourstressenergy}
{\rm T}^{(e)}_{\mu\nu}
=
\mathcal{T}(\Phi)\, e\, \nabla_\mu \Phi \nabla_\nu \Phi
- g_{\mu\nu}
\left[
\frac{1}{2}\mathcal{T}(\Phi)\left(
e\, g^{\alpha\beta}\nabla_\alpha \Phi \nabla_\beta \Phi
+ e^{-1}
\right)
+ \mathcal{V}(\Phi)
\right].
\end{align}
It is important to emphasise that this tensor is defined only in the tubular region \(\mathcal{W}\subset\mathcal{M}\). Outside \(\mathcal{W}\), the geometry is taken to be vacuum, so that
\begin{equation}
    {\rm T}^{(e)}_{\mu\nu}=0,
    \qquad
    x\in\mathcal{M}\setminus\mathcal{W}.
\end{equation}

Inside \(\mathcal{W}\), but away from the core, the foliation is regular and all fields and gradients are well defined. At the core, however, the foliation becomes singular because \(\nabla\Phi=0\). Since Eq.~\eqref{ourstressenergy} is off-shell with respect to the auxiliary field \(e\), the stress-energy tensor at the core reduces to
\begin{equation}
    {\rm T}^{(e)}_{\mu\nu}\big|_{\nabla\Phi=0}
    =
    -g_{\mu\nu}\bigg[
    \frac{1}{2}\mathcal{T}(\Phi)\,e^{-1}
    +\mathcal{V}(\Phi)
    \bigg].
\end{equation}
Thus, the off-shell stress-energy remains regular at the core provided \(\mathcal{T}(\Phi)\), \(\mathcal{V}(\Phi)\), and \(e\) remain finite there.

On-shell, the auxiliary field is fixed by the constraint
\begin{equation}\label{onshell}
    e
    =
    \frac{1}{\sqrt{g_{\mu\nu}\nabla^\mu\Phi \nabla^\nu\Phi}}.
\end{equation}
Although \(e\) itself diverges as \(\nabla\Phi\to0\), the combination
\begin{equation}
    e\,\nabla_\mu\Phi\nabla_\nu\Phi
\end{equation}
scales as
\begin{equation}
    \sqrt{g^{\alpha\beta}\nabla_\alpha\Phi\nabla_\beta\Phi},
\end{equation}
and therefore vanishes smoothly at the core. Similarly, \(e^{-1}\to0\) smoothly. Hence the on-shell stress-energy tensor remains finite and takes the limiting form
\begin{equation}
    {\rm T}_{\mu\nu}\big|_{\nabla\Phi=0}
    =
    -g_{\mu\nu}\,\mathcal{V}(\Phi_0),
\end{equation}
where \(\Phi_0\) denotes the value of \(\Phi\) on the core.

Therefore, if
\begin{equation}
    \mathcal{V}(\Phi_0)=0,
\end{equation}
the core behaves as a vacuum line. If instead
\begin{equation}
    \mathcal{V}(\Phi_0)\neq0,
\end{equation}
the core carries an effective vacuum-energy density, so that the local stress-energy there is of cosmological-constant type.

This is one of the main differences from the ordinary point-particle description in general relativity. In the worldline picture, the source is distributional and supported by a Dirac delta on the trajectory. In the present construction, no singular source is introduced from the start; rather, the tube carries a smooth stress-energy tensor all the way to the core.

It is then natural to ask how the mass content of the tube should be characterised. Let \(U^\mu\) be the velocity of a timelike observer moving on one of the leaves \(\Sigma_{\Phi=s}\), with
\begin{equation}
    U^\mu U_\mu=-1.
\end{equation}
The local energy density measured by this observer is
\begin{equation}
    \rho^{(e)}(x)
    =
    {\rm T}^{(e)}_{\mu\nu}U^\mu U^\nu.
\end{equation}
Since the leaves are level sets of \(\Phi\), one has
\begin{equation}
    U^\mu\nabla_\mu\Phi=0.
\end{equation}
Therefore,
\begin{equation}
    \rho^{(e)}(x)
    =
    \frac{1}{2}\mathcal{T}(\Phi)\left(
    e\, g^{\alpha\beta}\nabla_\alpha \Phi \nabla_\beta \Phi
    + e^{-1}
    \right)
    + \mathcal{V}(\Phi).
\end{equation}

A natural quasilocal mass associated with the tube up to the leaf \(\Sigma_{\Phi=s}\) is then
\begin{equation}
    {\rm M}^{(e)}(s)
    =
    \int_{\Phi=\Phi_0}^{\Phi=s} d^{D-1}x\,
    \rho^{(e)}(x).
\end{equation}
Substituting the on-shell relation \eqref{onshell}, this becomes
\begin{equation}\label{totalmass}
    {\rm M}(s)
    =
    \int_{\Phi=\Phi_0}^{\Phi=s} d^{D-1}x
    \bigg[
    \mathcal{T}(\Phi)\sqrt{g^{\mu\nu}\nabla_{\mu}\Phi\nabla_\nu \Phi}
    +\mathcal{V}(\Phi)
    \bigg].
\end{equation}
Likewise, the corresponding spacetime-integrated mass functional for the full tube is
\begin{equation}\label{inertialmass}
    {\rm m}
    =
    \int_{\mathcal{W}} d^D x \sqrt{-g}\,\rho(x)
    =
    \int_{\Phi=\Phi_0}^{\Phi=\Phi_{\rm max}} d^{D}x\sqrt{-g}
    \bigg[
    \mathcal{T}(\Phi)\sqrt{g^{\mu\nu}\nabla_{\mu}\Phi\nabla_\nu \Phi}
    +\mathcal{V}(\Phi)
    \bigg].
\end{equation}
Here the integration extends from the core to the outermost leaf of the tube.

It is worth noting that the spacetime integral of the on-shell energy density closely resembles the action functional itself. This is a consequence of the constrained structure of the theory: once the auxiliary field is eliminated, the distinction between the energy density and the Lagrangian density becomes less pronounced. Nevertheless, the two should not be interpreted as identical in general.

The difference between Eq.~\eqref{totalmass}, evaluated at \(s=\Phi_{\rm max}\), and Eq.~\eqref{inertialmass} may be interpreted as a gravitational contribution in the same broad sense in which one discusses binding or field energy in general relativity. In contrast with the ordinary worldline formulation, where the mass appears as a fixed parameter multiplying a distributional source, the mass of the tube arises from integrating a smooth density profile over the interior region. For this mass to represent ordinary, non-exotic matter, the on-shell density should remain non-negative, which in particular is ensured when
\begin{equation}
    \mathcal{T}(\Phi)\geq0,
    \qquad
    \mathcal{V}(\Phi)\geq0.
\end{equation}
With a smooth stress-energy tensor now available, we may analyse the energy conditions explicitly.

\subsection{Energy Conditions Inside the Tube Environment}\label{4.1.3energycondition}

We now examine the null, weak, strong, and dominant energy conditions for the off-shell and on-shell stress-energy tensor \eqref{ourstressenergy}. Throughout, the identity
\begin{equation}
    U^\mu \nabla_\mu \Phi=0
\end{equation}
holds by construction.

\begin{itemize}
    \item[1.] \textbf{Null energy condition (NEC):} Let \(k^\mu\) be any null vector in the region \(\Omega\), so that
    \begin{equation}
        k_\mu k^\mu=0.
    \end{equation}
    The null energy condition requires
    \begin{equation}
        {\rm T}_{\mu\nu}k^\mu k^\nu\geq0.
    \end{equation}
    For Eq.~\eqref{ourstressenergy},
    \begin{equation}
        {\rm T}^{(e)}_{\mu\nu}k^\mu k^\nu
        =
        \mathcal{T}(\Phi)e\,[k^\mu \nabla_\mu \Phi]^2.
    \end{equation}
    Since \(e>0\) and \([k^\mu \nabla_\mu\Phi]^2\geq0\), the NEC reduces simply to
    \begin{equation}
        \mathcal{T}(\Phi)\geq0.
    \end{equation}
    The same conclusion holds on-shell.

    \item[2.] \textbf{Weak energy condition (WEC):} Let \(U^\mu\) be the four-velocity of a timelike observer inside \(\mathcal{W}\). The weak energy condition requires
    \begin{equation}
        {\rm T}_{\mu\nu}U^\mu U^\nu\geq0.
    \end{equation}
    For Eq.~\eqref{ourstressenergy},
    \begin{equation}
        {\rm T}^{(e)}_{\mu\nu}U^\mu U^\nu
        =
        \frac{1}{2}\mathcal{T}(\Phi)\left(
        e\, g^{\alpha\beta}\nabla_\alpha \Phi \nabla_\beta \Phi
        + e^{-1}
        \right)
        + \mathcal{V}(\Phi).
    \end{equation}
    Hence, throughout the regular region, the WEC is guaranteed if
    \begin{equation}
        \mathcal{T}(\Phi)\geq0,
        \qquad
        \mathcal{V}(\Phi)\geq0.
    \end{equation}
    If \(\mathcal{V}(\Phi)<0\), then the WEC instead requires
    \begin{equation}
        \mathcal{T}(\Phi)\sqrt{g^{\alpha\beta}\nabla_\alpha\Phi\nabla_\beta\Phi}
        \geq
        |\mathcal{V}(\Phi)|.
    \end{equation}
    At the core, on-shell, this reduces to
    \begin{equation}
        \mathcal{V}(\Phi_0)\geq0,
    \end{equation}
    because the kinetic contribution vanishes there.

    \item[3.] \textbf{Strong energy condition (SEC):} For a timelike observer of velocity \(U^\mu\), the strong energy condition is
    \begin{equation}
        \left(
        {\rm T}_{\mu\nu}
        -\frac{1}{D-2}g_{\mu\nu}{\rm T}
        \right)U^\mu U^\nu
        \geq0.
    \end{equation}
    For the present stress-energy tensor one finds
    \begin{equation}
        \left(
        {\rm T}^{(e)}_{\mu\nu}
        -\frac{1}{D-2}g_{\mu\nu}{\rm T}^{(e)}
        \right)U^\mu U^\nu
        =
        -\frac{1}{D-2}e^{-1}\mathcal{T}(\Phi)
        -\frac{2}{D-2}\mathcal{V}(\Phi).
    \end{equation}
    Therefore, if
    \begin{equation}
        \mathcal{T}(\Phi)\geq0,
        \qquad
        \mathcal{V}(\Phi)>0,
    \end{equation}
    the SEC is violated everywhere. If instead
    \begin{equation}
        \mathcal{V}(\Phi)<0,
        \qquad
        \mathcal{T}(\Phi)\geq0,
    \end{equation}
    then the SEC requires
    \begin{equation}
        |\mathcal{V}(\Phi)|
        \geq
        \frac{1}{2e}\mathcal{T}(\Phi).
    \end{equation}
    At the core, on-shell, this becomes
    \begin{equation}
        \mathcal{V}(\Phi_0)\leq0.
    \end{equation}

    \item[4.] \textbf{Dominant energy condition (DEC):} Let \(U^\mu\) be the timelike observer velocity. The dominant energy condition requires the momentum density
    \begin{equation}
        J^\mu=-{\rm T}^\mu{}_\nu U^\nu
    \end{equation}
    to be timelike or null. For our stress-energy tensor,
    \begin{equation}
        J^\mu
        =
        U^\mu\bigg[
        \frac{1}{2}\mathcal{T}(\Phi)\left(
        e\, g^{\alpha\beta}\nabla_\alpha \Phi \nabla_\beta \Phi
        + e^{-1}
        \right)
        + \mathcal{V}(\Phi)
        \bigg].
    \end{equation}
    Thus \(J^\mu\) is parallel to \(U^\mu\), and the DEC reduces to the same positivity requirement as the WEC:
    \begin{equation}
        \frac{1}{2}\mathcal{T}(\Phi)\left(
        e\, g^{\alpha\beta}\nabla_\alpha \Phi \nabla_\beta \Phi
        + e^{-1}
        \right)
        + \mathcal{V}(\Phi)\geq0.
    \end{equation}
\end{itemize}

From these expressions we may summarise the generic behaviour as follows. If
\begin{equation}
    \mathcal{V}(\Phi)>0,
    \qquad
    \mathcal{T}(\Phi)\geq0,
\end{equation}
then the NEC, WEC, and DEC hold, while the SEC is violated. If instead
\begin{equation}
    \mathcal{V}(\Phi)<0,
    \qquad
    \mathcal{T}(\Phi)\geq0,
\end{equation}
then all four conditions can hold only when
\begin{equation}\label{exoticase}
    \mathcal{T}(\Phi)\sqrt{g^{\alpha\beta}\nabla_\alpha \Phi \nabla_\beta \Phi}
    \geq
    |\mathcal{V}(\Phi)|
    \geq
    \frac{1}{2}\mathcal{T}(\Phi)\sqrt{g^{\alpha\beta}\nabla_\alpha \Phi \nabla_\beta \Phi}.
\end{equation}

The violation of the SEC need not be regarded as pathological. Similar violations are common in models with vacuum energy or late-time dark-energy behaviour. In the present construction, the core is especially interesting because it is precisely the one-dimensional timelike set where the tubular foliation becomes singular. At the core, if \(\mathcal{V}(\Phi_0)>0\), then the NEC and WEC hold while the SEC is violated. If \(\mathcal{V}(\Phi_0)<0\), then on-shell the WEC and SEC can both be satisfied at the core only if
\begin{equation}
    \mathcal{V}(\Phi_0)=0.
\end{equation}
Thus the potential term is the essential source of SEC violation at the core.

This feeds directly into the Raychaudhuri equation for timelike geodesic congruences in the tube,
\begin{equation}
    \frac{d\theta}{d\tau}
    =
    -\frac{1}{3}\theta^2-\sigma^2+\omega^2-{\rm R}_{\mu\nu}U^\mu U^\nu.
\end{equation}
When the SEC holds, the usual focusing argument applies; when it fails, defocusing effects become possible. In the present model, demanding SEC preservation throughout the full tube, including the core, strongly constrains the allowed potential and in particular rules out a strictly positive core value.

For convenience, the two qualitatively distinct cases may be summarised as follows.

\noindent\textbf{Case 1:} \(\mathcal{V}(\Phi)>0\)

\begin{table}[h!]
\centering
\small
\renewcommand{\arraystretch}{1.25}
\begin{tabular}{|c|c|c|}
\hline
\textbf{Condition} & \textbf{Region \(\mathcal{W}\) (generic)} & \textbf{Core (\(\nabla\Phi=0\), on-shell)} \\
\hline
\textbf{NEC} 
& Holds if \(\mathcal{T}(\Phi)\geq 0\)
& Holds if \(\mathcal{T}(\Phi_0)\geq 0\) \\
\hline
\textbf{WEC} 
& Holds if \(\mathcal{T}(\Phi)\geq 0\)
& Holds if \(\mathcal{V}(\Phi_0)\geq0\) \\
\hline
\textbf{SEC} 
& Violated
& Violated if \(\mathcal{V}(\Phi_0)>0\) \\
\hline
\textbf{DEC} 
& Holds whenever WEC holds
& Same as core WEC \\
\hline
\end{tabular}
\caption{Energy conditions for \(\mathcal{V}(\Phi)>0\).}
\end{table}

\noindent\textbf{Case 2:} \(\mathcal{V}(\Phi)<0\)

\begin{table}[h!]
\centering
\small
\renewcommand{\arraystretch}{1.25}
\begin{tabular}{|c|c|c|}
\hline
\textbf{Condition} & \textbf{Region \(\mathcal{W}\) (generic)} & \textbf{Core (\(\nabla\Phi=0\), on-shell)} \\
\hline
\textbf{NEC} 
& Holds if \(\mathcal{T}(\Phi)\geq 0\)
& Holds if \(\mathcal{T}(\Phi_0)\geq 0\) \\
\hline
\textbf{WEC} 
& \(\mathcal{T}(\Phi)\sqrt{(\nabla\Phi)^2}\geq |\mathcal{V}(\Phi)|\)
& Requires \(\mathcal{V}(\Phi_0)=0\) \\
\hline
\textbf{SEC} 
& \( |\mathcal{V}(\Phi)| \geq \frac{1}{2e}\mathcal{T}(\Phi)\)
& Requires \(\mathcal{V}(\Phi_0)\leq0\), and together with WEC gives \(\mathcal{V}(\Phi_0)=0\) \\
\hline
\textbf{DEC} 
& Holds whenever WEC holds
& Same as core WEC \\
\hline
\end{tabular}
\caption{Energy conditions for \(\mathcal{V}(\Phi)<0\).}
\end{table}

The potential function therefore plays a decisive role in controlling focusing and defocusing of timelike congruences inside the tube. A positive core potential acts like local vacuum energy and tends to support SEC violation. By contrast, a negative potential can preserve the SEC only within a restricted window. For example, consider
\begin{equation}
    \mathcal{V}(\Phi)
    =
    -\mathcal{V}_0\sqrt{\Lambda^2\Phi}\exp(-\Lambda^2\Phi),
    \qquad
    \mathcal{V}_0>0,
    \qquad
    0\leq\Phi\leq a\Lambda^{-2}.
\end{equation}
Then WEC and SEC can both hold provided
\begin{equation}
    \frac{1}{e}\mathcal{T}(\Phi)
    \geq
    \mathcal{V}_0\sqrt{\Lambda^2\Phi}\exp(-\Lambda^2\Phi)
    \geq
    \frac{1}{2e}\mathcal{T}(\Phi),
\end{equation}
or equivalently,
\begin{equation}
   e\mathcal{V}_0\sqrt{\Lambda^2\Phi}\exp(-\Lambda^2\Phi)
   \leq
   \mathcal{T}(\Phi)
   \leq
   2e\mathcal{V}_0\sqrt{\Lambda^2\Phi}\exp(-\Lambda^2\Phi).
\end{equation}
Violation of the left inequality leads to WEC violation, while violation of the right inequality leads to SEC violation.

For illustration, one may choose the numerical values
\begin{equation}
    \mathcal{V}_0=1,
    \qquad
    \Lambda=1,
    \qquad
    e=1,
    \qquad
    a=2
\end{equation}
in the off-shell description.

\begin{figure}[htbp]
    \centering
    \includegraphics[width=0.8\linewidth]{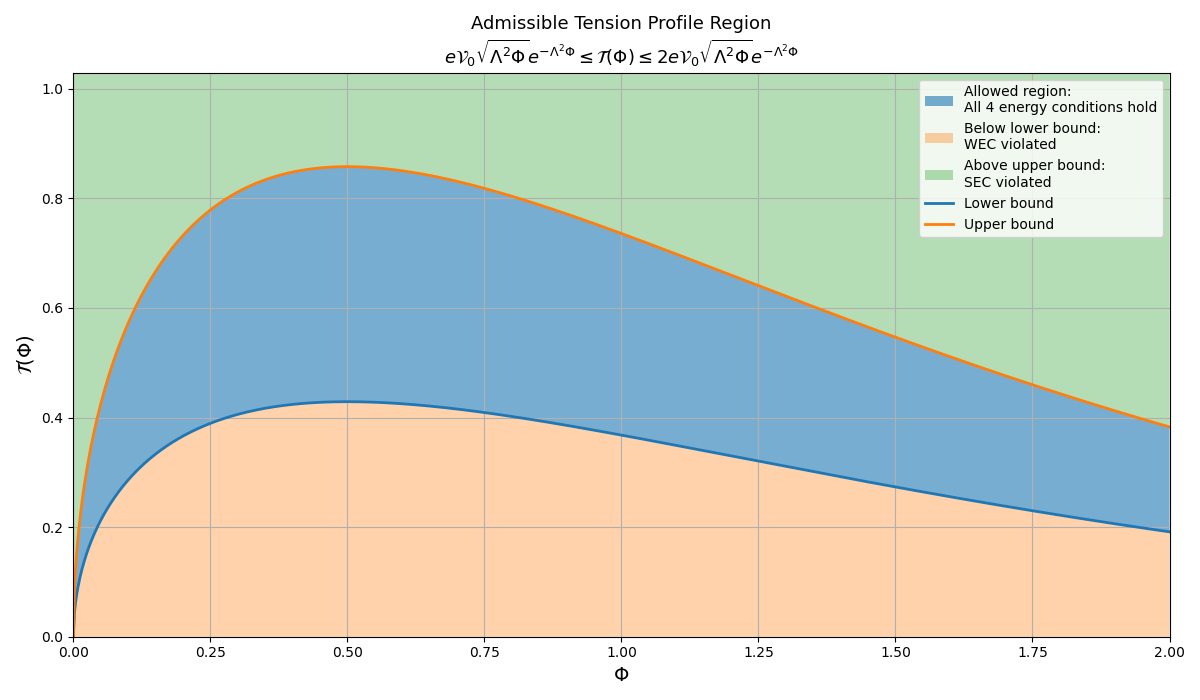}
    \caption{Allowed window for simultaneous WEC and SEC preservation.}
    \label{fig:8}
\end{figure}

The blue region then represents the allowed window for \(\mathcal{T}(\Phi)\) in which all the energy conditions are satisfied for the chosen potential profile.

\newpage
\section{(Constraints) Dynamics and Stability}\label{dynamicsandstability}

The functional form of the on-shell action \eqref{Almostanaction} places it in close analogy with the class of Cuscuton-type actions\cite{Afshordi_2007}. The resemblance is structural rather than interpretational: in the standard cuscuton literature, the scalar is introduced as a non-canonical field whose phase space is degenerate and which, in suitable limits, carries no independent propagating degree of freedom. In the present work, by contrast, the scalar \(\Phi\) is introduced geometrically as the field generating the singular tubular foliation, and the square-root structure emerges from that geometric role.

For comparison, a representative cuscuton action may be written as
\begin{equation}\label{cuscutonaction}
     \mathcal{S}_{\varphi}
     =
     \int d^4x\sqrt{-g}
     \bigg[
     \frac{1}{2}\frac{\partial J}{\partial\varphi}
     \sqrt{g^{\mu\nu} \nabla_\mu \varphi\nabla_\nu\varphi}
     -
     \mathcal{V}(\varphi)
     \bigg].
\end{equation}
Formally, one may identify
\begin{equation}
     \mathcal{T}(\varphi)
     =
     -\frac{1}{2}\frac{\partial J}{\partial\varphi},
     \qquad
     D=4.
\end{equation}
Even so, the physical interpretation of \(\Phi\) in the present construction is different from that of the usual cuscuton field, and this difference will matter in the Hamiltonian and stability analysis.

In this section, we study the Hamiltonian structure of the Lagrangian \eqref{ourlagrangian}, asking whether the foliation field carries independent dynamics in any relevant limit. We also examine the linear stability of the tube, in order to determine whether the configuration is stable both dynamically and gravitationally.


\subsection{Hamiltonian Dynamics and Dirac Constraints}\label{Hamiltoniandynamics}

In this section, we do not couple the action \eqref{Ouraction} to gravity. Instead, we work in a fixed four-dimensional Minkowski background,
\begin{equation}
    g_{\mu\nu}=\mathrm{diag}(-1,+1,+1,+1),
\end{equation}
and consider the off-shell Lagrangian density with auxiliary scalar field \(e\), without imposing any sign restriction on the tension or potential functions:
\begin{equation}\label{ourLagran.}
     \mathcal{L}^{(e)}
     =
     -\frac{1}{2}\mathcal{T}(\Phi)\bigg(e\, g_{\mu\nu} \nabla^\mu\Phi \nabla^\nu \Phi+e^{-1}\bigg)-\mathcal{V}(\Phi).
\end{equation}
In Minkowski coordinates \(x^\mu=[t,x^i]\), the kinetic invariant becomes
\begin{equation}
    X
    =
    g_{\mu\nu}\nabla^\mu\Phi\nabla^\nu\Phi
    =
    -\dot\Phi^2+(\partial_i\Phi)^2.
\end{equation}

We now construct the phase space. Let \(\Pi^{(e)}_{\Phi}\) be the canonical momentum conjugate to \(\Phi\), and let \(\Pi_e\) be the canonical momentum conjugate to the auxiliary field \(e\). The phase space is therefore
\begin{equation}
    \mathcal{P}=\{\Phi,\Pi^{(e)}_{\Phi},e,\Pi_e\},
\end{equation}
with canonical momenta
\begin{equation}
    \Pi^{(e)}_{\Phi}
    =
    \frac{\partial\mathcal{L}^{(e)}}{\partial\dot\Phi}
    =
    \mathcal{T}(\Phi)e\dot\Phi,
    \qquad
    \Pi_e
    =
    \frac{\partial\mathcal{L}^{(e)}}{\partial\dot e}
    =
    0.
\end{equation}
Thus \(\Pi_e\) vanishes identically, confirming that \(e\) is an auxiliary variable. In Dirac notation, the primary constraint is
\begin{equation}
     C_1:\Pi_e\approx0.
\end{equation}

The off-shell Hamiltonian density is
\begin{align}\label{offshellhamiltoniandensity}
     \mathcal{H}^{(e)}
     &=
     \dot\Phi\,\Pi^{(e)}_{\Phi}-\mathcal{L}^{(e)} \notag\\
     &=
     \frac{1}{2e\,\mathcal{T}(\Phi)}\Big[\Pi^{(e)}_{\Phi}\Big]^2
     +\frac{1}{2}\mathcal{T}(\Phi)\bigg[e(\partial_i \Phi)^2+e^{-1}\bigg]
     +\mathcal{V}(\Phi).
\end{align}
This expression is regular throughout the tube, including at the core, provided the potential remains finite there. In particular, at the core, where \(\dot\Phi=0\) and \(\partial_i\Phi=0\),
\begin{equation}
     \mathcal{H}^{(e)}_{\rm core}
     =
     \frac{1}{2}\mathcal{T}(0)e^{-1}+\mathcal{V}(0).
\end{equation}

To recover the on-shell behaviour, one imposes the auxiliary-field constraint
\begin{equation}
     e=\frac{1}{\sqrt{-\dot \Phi^2 +(\partial_i\Phi)^2}}.
\end{equation}
Substituting this into Eq.~\eqref{offshellhamiltoniandensity}, one obtains
\begin{equation}
     \mathcal{H}
     =
     \frac{\mathcal{T}(\Phi)}{2}
     \frac{\dot \Phi^2+(\partial_i\Phi)^2}{\sqrt{-\dot \Phi^2 +(\partial_i\Phi)^2}}
     +
     \frac{\mathcal{T}(\Phi)}{2}\sqrt{-\dot \Phi^2 +(\partial_i\Phi)^2}
     +
     \mathcal{V}(\Phi).
\end{equation}
Near the core, using Lemma 2, one has
\begin{equation}
    \nabla_\mu \Phi =2\varphi_A\nabla_\mu\varphi^A,
\end{equation}
so that
\begin{equation}
    \dot\Phi=2\varphi_A\dot\varphi^A,
    \qquad
    \partial_i\Phi=2\varphi_A\partial_i\varphi^A.
\end{equation}
Writing \(\varphi^A=|\varphi|n^A\) near the core, with \(n^An_A>0\), one finds that both non-potential terms vanish as \(|\varphi|\to0\), since they carry an overall factor of \(|\varphi|\). Therefore,
\begin{equation}
      \mathcal{H}_{\rm core}=\mathcal{V}(0).
\end{equation}
Thus the on-shell and off-shell descriptions agree that the Hamiltonian density remains finite at the core.

The total off-shell Hamiltonian is
\begin{equation}
    {\rm H}^{(e)}
    =
    \int_{\mathcal{W}} d^{D-1}x\, \mathcal{H}^{(e)}
    +
    \int_{\mathcal{W}} d^{D-1}y\,\lambda(y)\Pi_e(y),
\end{equation}
where \(\lambda\) is a Lagrange multiplier enforcing the primary constraint. We now analyse the Dirac constraint algebra in order to determine whether \(\Phi\) behaves purely as a constraining field, as in the cuscuton case, or whether it carries a genuine canonical degree of freedom.

Preservation of the primary constraint gives
\begin{equation}
    \dot\Pi_e
    =
    \{\Pi_e,{\rm H}^{(e)}\}
    \approx0,
\end{equation}
which yields the secondary constraint
\begin{equation}
    C_2:
    \frac{1}{2e^2 \mathcal{T}(\Phi)}\Big[\Pi_\Phi^{(e)}\Big]^2
    -\frac{1}{2} \mathcal{T}(\Phi)\big[\partial_i \Phi\big]^2
    +\frac{1}{2e^2} \mathcal{T}(\Phi)
    \approx0.
\end{equation}
The relevant Poisson brackets are then
\begin{align}
    &\{C_1(x),C_1(y)\}=0,\\
    &\{C_1(x),C_2(y)\}
    =
    \{\Pi_e(x),C_2(y)\}
    =
    -\frac{\delta C_2(y)}{\delta e(x)} \notag\\
    &\qquad\qquad
    =
    \Bigg[
    \frac{1}{e^3 \mathcal{T}(\Phi)}\Big[\Pi_\Phi^{(e)}\Big]^2
    +
    \frac{1}{e^3}\mathcal{T}(\Phi)
    \Bigg]\delta(x-y),\\
    &\{C_2(x),C_2(y)\}
    =
    \int d^{D-1}z
    \bigg[
    \frac{\delta C_2(x)}{\delta \Phi(z)}
    \frac{\delta C_2(y)}{\delta \Pi_\Phi(z)}
    -
    \frac{\delta C_2(x)}{\delta \Pi_\Phi(z)}
    \frac{\delta C_2(y)}{\delta \Phi(z)}
    \bigg],
\end{align}
with
\begin{align}
    \frac{\delta C_2(x)}{\delta \Phi(z)}
    &=
    \Bigg[
    -\frac{\mathcal{T}'}{2e^2\mathcal{T}^2}\Big[\Pi_\Phi^{(e)}\Big]^2
    +
    \frac{1}{2e^2}\mathcal{T}'
    \Bigg]\delta^{D-1}(x-z)
    +
    \partial_i\big[\delta^{D-1}(x-z)\mathcal{T}\partial_i\Phi\big],\\
    \frac{\delta C_2(x)}{\delta \Pi_\Phi^{(e)}(z)}
    &=
    \frac{1}{e^2\mathcal{T}}\Pi_\Phi^{(e)}\delta^{D-1}(x-z).
\end{align}

Away from the core, the Dirac matrix \(C_{ij}=\{C_i,C_j\}\) is generically non-degenerate, since in particular
\begin{equation}
    \{C_1,C_2\}\neq0.
\end{equation}
Thus \(C_1\) and \(C_2\) are second-class constraints in the bulk of the tube. The local degree-of-freedom count is therefore
\begin{equation}
    N_{\rm bulk}
    =
    \frac{1}{2}\big[2N-2N_{\rm FC}-N_{\rm SC}\big]
    =
    \frac{1}{2}\big[2\times2-0-2\big]
    =
    1.
\end{equation}

At the core, however, the on-shell limit sends \(e\to\infty\), so that
\begin{equation}
    \{C_1(x),C_2(y)\}\to0.
\end{equation}
Likewise, the \(C_{22}\) bracket also vanishes in the same limit. Hence the rank of the Dirac matrix drops at the core, signalling a degeneration of the canonical structure there. In this reduced analysis, the local degree-of-freedom count collapses to
\begin{equation}
    N_{\rm core}=0.
\end{equation}
Therefore the fields \(\Phi\) and \(e\) become non-dynamical at the core in the canonical sense.

This behaviour persists in the ultraviolet limit of the tube, where \(\Lambda\to\infty\) and the tube shrinks to a timelike curve. It should not be viewed as a pathology. Rather, it reflects the fact that the longitudinal mode freezes out, while the only nontrivial bulk deformation is the transverse, radial mode associated with the breathing of the leaves. This is consistent with the geometric discussion of congruences of hypersurfaces in Chapter \eqref{chapter2}, particularly in subsection \eqref{congruenceofsubmanifolds}.

In this sense, the present construction extends the cuscuton-like picture to a setting with a critical submanifold at which \(\nabla\Phi=0\): the field becomes completely frozen on the critical set, while retaining a single transverse degree of freedom in the bulk. This matches the usual cuscuton intuition that the square-root sector is non-propagating in the ordinary dynamical sense.

\subsection{Perturbative Stability of the Tube}\label{4.2.2perturbativestability}

Before turning to perturbations, we first compute the equation of motion for the scalar field \(\Phi\). The off-shell variation of the action \eqref{Ouraction} with respect to \(\Phi\) gives
\begin{equation}\label{rawequations}
    \nabla_\mu\!\big[\mathcal{T}(\Phi)e\nabla^\mu\Phi\big]
    -
    \frac{1}{2}\mathcal{T}'(\Phi)\big(e\,\nabla_\mu\Phi\nabla^\mu\Phi+e^{-1}\big)
    -
    \mathcal{V}'(\Phi)
    =
    0.
\end{equation}
On-shell, using Eq.~\eqref{onshell}, one has
\begin{equation}
    e\nabla^\mu\Phi=N^\mu,
    \qquad
    \nabla_\mu N^\mu=\hat K=K,
\end{equation}
where \(\hat K\) is the mean extrinsic curvature of the timelike leaves for \(\Phi\neq0\). Furthermore, for any smooth function \(F(\Phi)\),
\begin{equation}\label{normalderivative}
    \nabla_\mu F(\Phi)=\frac{dF}{d\Phi}\nabla_\mu\Phi
    \qquad \Rightarrow \qquad
    N^\mu\nabla_\mu F
    =
    \frac{1}{e}\frac{dF}{d\Phi},
\end{equation}
since
\begin{equation}
    N^\mu\nabla_\mu\Phi=e^{-1}.
\end{equation}
Substituting these relations into Eq.~\eqref{rawequations}, the terms involving \(\mathcal{T}'\) cancel, leaving the on-shell equation of motion
\begin{equation}\label{SoapEquationofmotion}
     \hat K = \frac{1}{\mathcal{T}}\frac{d\mathcal{V}}{d\Phi}.
\end{equation}
Hence each leaf of the tubular foliation is a constant-mean-curvature hypersurface,
\begin{equation}
    \hat K\big|_{\Phi=c}=\mathrm{constant}.
\end{equation}
The mean extrinsic curvature is therefore sourced by the tension and the gradient of the potential.

If one instead rewrites the field equation in second-order form before imposing the geometric interpretation, one finds a nonlinear wave-type equation. The principal symbol is proportional to the tangential projector onto the leaves, and is therefore degenerate in the normal direction. This is the same qualitative mechanism by which cuscuton-type models lose an ordinary propagating scalar mode .

\vspace{2mm}

\noindent\textbf{\underline{Perturbative analysis}}

There are two distinct perturbative questions to address. First, one may perturb the scalar field directly in the action and ask about ghost, gradient, and tachyonic instabilities. Second, one may perturb the embedding of the timelike leaves themselves and study the resulting Jacobi equation for normal deformations.

\vspace{2mm}

\noindent\textbf{A. Field perturbations}

Let the Lagrangian density be a general function \(\mathcal{L}(X,\Phi)\), with
\begin{equation}
    X=\nabla_\mu\Phi\nabla^\mu\Phi,
\end{equation}
in a fixed background metric. Consider a perturbation of the form
\begin{equation}
    \Phi(x)=\Phi_0(x)+\pi(x),
\end{equation}
so that
\begin{equation}
    X=X_0+\delta X+\delta^2X,
\end{equation}
with
\begin{equation}
    \delta X=2\nabla_\mu\Phi_0\nabla^\mu\pi,
    \qquad
    \delta^2X=\nabla_\mu\pi\nabla^\mu\pi,
    \qquad
    X_0=\nabla_\mu\Phi_0\nabla^\mu\Phi_0.
\end{equation}
Expanding \(\mathcal{L}(X,\Phi)\) about \((X_0,\Phi_0)\) gives
\begin{equation}
    \mathcal{L}=\mathcal{L}_0+\mathcal{L}_1+\mathcal{L}_2+\cdots,
\end{equation}
where
\begin{align}
    &\mathcal{L}_0=\mathcal{L}(X_0,\Phi_0),\\
    &\mathcal{L}_1=\mathcal{L}_X\,\delta X+\pi\,\mathcal{L}_\Phi,\\
    &\mathcal{L}_2=\mathcal{L}_X\,\delta^2X+\frac{1}{2}\mathcal{L}_{XX}(\delta X)^2+\pi\,\mathcal{L}_{\Phi X}\delta X+\frac{1}{2}\pi^2\mathcal{L}_{\Phi\Phi}.
\end{align}
Correspondingly, the action expands as
\begin{equation}
    \mathcal{S}=\mathcal{S}_0+\mathcal{S}_1+\mathcal{S}_2.
\end{equation}
The linear term \(\mathcal{S}_1\) reproduces the background equation of motion. The quadratic term \(\mathcal{S}_2\) controls the perturbative stability.

For the square-root theory, the coefficient of the quadratic kinetic term is
\begin{equation}
    \mathcal{L}_X+2X_0\mathcal{L}_{XX}.
\end{equation}
For
\begin{equation}
    \mathcal{L}(X,\Phi)=-\mathcal{T}(\Phi)\sqrt{X}-\mathcal{V}(\Phi),
\end{equation}
one finds
\begin{equation}
    \mathcal{L}_X=\frac{\mathcal{T}(\Phi_0)}{2\sqrt{X_0}},
    \qquad
    \mathcal{L}_{XX}=-\frac{\mathcal{T}(\Phi_0)}{4X_0^{3/2}},
\end{equation}
and therefore
\begin{equation}
    \mathcal{L}_X+2X_0\mathcal{L}_{XX}=0.
\end{equation}
Thus the perturbation has no ordinary quadratic kinetic term. Equivalently, the formal sound speed
\begin{equation}
    c_s^2=\frac{\mathcal{L}_X}{\mathcal{L}_X+2X\mathcal{L}_{XX}}
\end{equation}
diverges, which is again the standard cuscuton pattern [web:164][web:163].

The remaining quadratic sector is then governed by the spatial-gradient and mass terms. After integrating by parts, the reduced perturbation equation takes the Helmholtz form
\begin{equation}
   \nabla_i\nabla^i \pi-\hat m_{\rm eff}^2\pi=0,
\end{equation}
with
\begin{equation}
    \hat m_{\rm eff}^2
    =
    \frac{\mathcal{L}_{\Phi\Phi}+\frac{d}{dt}\big[\dot\Phi_0\mathcal{L}_{X\Phi}\big]}{2\mathcal{L}_X}.
\end{equation}
For the present model this becomes
\begin{equation}
    \hat m_{\rm eff}^2
    =
    \frac{3\mathcal{T}''(\Phi_0)X_0+2\mathcal{V}''(\Phi_0)\sqrt{X_0}}{\mathcal{T}(\Phi_0)},
\end{equation}
assuming \(X_0>0\). Stability against exponential growth then requires
\begin{equation}
    \hat m_{\rm eff}^2\geq0.
\end{equation}
This places additional restrictions on the admissible tension and potential profiles, beyond the weak and strong positivity conditions of Eqs.~\eqref{weakcriterion} and \eqref{strongpositivityconditions}.

To summarise, this first perturbative analysis shows that the field is non-dynamical in the ordinary temporal sense. The Dirac analysis yielded one bulk canonical degree of freedom, but the quadratic perturbation analysis shows that this mode behaves as an instantaneous constraint rather than a conventional wave mode. In that sense, the theory is again cuscuton-like.

\vspace{4mm}

\noindent\textbf{B. Normal deformations of the leaves}

We now perturb the embedding maps \(X^\mu(\xi^a)\) defining the intrinsic geometry of each leaf:
\begin{equation}
    \mathbf{x}=\mathbf{X}+\delta\mathbf{X}.
\end{equation}
The deformation may be decomposed into tangential and normal parts. Tangential deformations are pure diffeomorphisms on the leaf, so only the normal deformation is physically relevant. Denoting the normal deformation amplitude by \(\zeta\), the Jacobi equation for a constant-mean-curvature timelike hypersurface is
\begin{equation}\label{jacobiiequation}
    \hat \nabla_a \hat\nabla^a \zeta
    +
    \bigg[
    \hat K_{ab}\hat K^{ab}
    +
    {\rm R}_{\mu\nu}N^\mu N^\nu
    \bigg]\zeta
    =0,
\end{equation}
or equivalently
\begin{equation}
    [\square+M^2]\zeta=0,
\end{equation}
with
\begin{equation}
    M^2=\hat K_{ab}\hat K^{ab}+{\rm R}_{\mu\nu}N^\mu N^\nu.
\end{equation}

To determine the sign of \(M^2\), one combines the bulk Einstein equation, the scalar field equation \eqref{SoapEquationofmotion}, and the scalar Gauss relation \eqref{Scalargaussfornormal}. After using the trace-reversed Einstein equations in the bulk and on the hypersurface, one arrives at the reduced relation
\begin{equation}\label{reducedscalarrelation}
    \frac{D-1}{D-3}{\rm R}_{\mu\nu} N^\mu N^\nu +2\hat{K}_{ab}\hat{K}^{ab}
    =
    \hat K^2
    +\frac{8{\rm T}}{M_p^2(D-2)(D-3)}
    -N^\sigma \nabla_\sigma K
    + \hat D_b a^b -a_b a^b.
\end{equation}
Since
\begin{equation}
    N^\sigma \nabla_\sigma K=\frac{1}{e}\frac{dK}{d\Phi},
\end{equation}
and on-shell the normal lapse coincides with the auxiliary field, the acceleration term reduces to
\begin{equation}
    \hat D_b a^b-a_b a^b=-\frac{1}{e}D_\mu D^\mu e.
\end{equation}
Rearranging the Gauss relation then gives
\begin{equation}\label{Reducedeffectivemass}
    {\rm R}_{\mu\nu} N^\mu N^\nu +\hat{K}_{ab}\hat{K}^{ab}
    =
    \frac{\hat K^2}{2}
    +\frac{4{\rm T}}{M_p^2(D-2)(D-3)}
    -\frac{1}{2e}\frac{d K}{d\Phi}
    -\frac{1}{e}D_\mu D^\mu e
    -\frac{2}{D-3}{\rm R}_{\mu\nu} N^\mu N^\nu.
\end{equation}
For the stress-energy tensor \eqref{ourstressenergy}, this becomes
\begin{equation}
     M^2
     =
     \frac{\hat K^2}{2}
     -4\bigg[
     \frac{2(D-1)e^{-1}\mathcal{T}+3\mathcal{V}}{M_p^2(D-2)(D-3)}
     \bigg]
     -\frac{1}{2e}\frac{d K}{d\Phi}
     -\frac{1}{e}D_\mu D^\mu e.
\end{equation}

We now analyse the spectrum of the Jacobi equation
\begin{equation}\label{newjacobi}
    [\square+M^2]\zeta=0
\end{equation}
in a locally Minkowskian background with spherical symmetry in the transverse directions, so that the interior geometry is locally \(\mathbb{R}\times S^{D-2}\). The Laplace--Beltrami operator splits as
\begin{equation}\label{newlaplacebeltrami}
    \square=-\square_t+\square_{S^{D-2}}.
\end{equation}
Expanding the deformation in spherical harmonics,
\begin{equation}\label{sphericalharmo}
    \zeta(t,\Theta)=\sum_l\sum_m q_{lm}(t)Y_{lm}(\Theta),
\end{equation}
one finds
\begin{equation}
    \ddot q_{lm}+(\lambda_l-M^2)q_{lm}=0,
\end{equation}
where the spherical eigenvalues are
\begin{equation}
    \lambda_l=\frac{l(l+D-3)}{r^2},
    \qquad
    l=0,1,2,\dots.
\end{equation}
The mode \(q_{lm}\) is oscillatory and stable when
\begin{equation}
    \lambda_l-M^2>0.
\end{equation}
Thus the sufficient condition for stability of all modes, including the zero mode, is
\begin{equation}
    M^2<0.
\end{equation}

This criterion imposes constraints on the internal structure of \(M^2\). A sufficient set of conditions is
\begin{equation}
    \mathcal{T}(\Phi)>0,
    \qquad
    \mathcal{V}(\Phi)>0,
\end{equation}
or, for negative potential,
\begin{equation}
    \mathcal{T}(\Phi)>\frac{3e|\mathcal{V}(\Phi)|}{2(D-1)}.
\end{equation}
If the mean extrinsic curvature decreases radially away from the core, one further requires
\begin{equation}
    \hat{K}^2+\bigg|\frac{1}{2e}\frac{dK}{d\Phi}\bigg|
    <
    4\bigg[
    \frac{2(D-1)e^{-1}\mathcal{T}+3\mathcal{V}}{M_p^2(D-2)(D-3)}
    \bigg],
\end{equation}
up to the sign choices for \(\mathcal{T}\) and \(\mathcal{V}\). For minimal leaves, \(\hat K=0\), which corresponds to \(\mathcal{V}=\mathrm{constant}\), the stability condition simplifies considerably: ignoring the normal covariant jerk term \(D_\mu D^\mu e\), one needs only
\begin{equation}
    \mathcal{T}(\Phi)>0.
\end{equation}

Finally, these perturbative stability conditions must be compatible with the null and weak energy conditions discussed in subsection \ref{4.1.3energycondition}, so that the ultraviolet limit of the tube corresponds to ordinary rather than exotic matter. This leaves \(\mathcal{T}>0\) as the basic necessary condition, irrespective of the sign of the potential. For \(\mathcal{V}>0\), the SEC is violated throughout the tube, while for \(\mathcal{V}<0\) there remains a narrow window in which all four energy conditions hold:
\begin{equation}
    \frac{\mathcal{T}}{e}\geq |\mathcal{V}|\geq\frac{\mathcal{T}}{2e}.
\end{equation}
Violation of the left inequality destroys the WEC, whereas violation of the right inequality destroys the SEC. We will return to the implications of this behaviour in the next chapter.

 \chapter*{Discussion and Conclusion\\
\begin{center}
\textit{\small\textbf{``A worldline is too thin to encode gravitating matter''}}
\end{center}}

The quote above captures the central problem motivating this thesis: the incompatibility of higher-codimension concentrated sources with Einstein's equation. As emphasized by Geroch and Traschen \cite{GEROCHandtraschen}, classical general relativity does not admit a satisfactory treatment of generic higher-codimension distributional matter sources, such as the worldline of a massive particle, as smooth and well-behaved gravitational sources. Motivated by this obstruction, the present thesis takes a codimension-zero tube, rather than a worldline, as the more fundamental notion of a source. The aim was not to smear matter by hand, but to replace the singular notion of localization by a geometric thickening that remains smooth in the bulk and becomes singular only in a controlled Morse--Bott sense at the core.

To implement this idea rigorously, a substantial amount of mathematical groundwork was required, and Chapter \ref{chapter2} was devoted to these preliminaries. In this chapter, we summarize the conceptual development of the theory, the principal physical results, and the open directions that naturally emerge from them. In particular, we address three broad questions: why this framework was introduced, how it was constructed, and what it ultimately implies.

\vspace{2mm}

\noindent\textbf{A. The construction}

\vspace{2mm}

Using the geometric preliminaries established in Chapter \ref{chapter2}, Chapter \ref{chapter3} first developed a tubular neighbourhood around a timelike curve (see Section \eqref{constructionandgeometry}) and then foliated the interior of this neighbourhood by timelike codimension-one leaves (see Definition \eqref{tubularfoliation}). The immediate goal of Chapter \ref{chapter3} was therefore geometric: to construct a tube intrinsically on the spacetime manifold. During this construction, however, it became clear that an everywhere regular foliation-generating scalar field cannot by itself describe the core of the tube, which is a one-dimensional timelike curve. To resolve this, we proposed that the scalar field defining the foliation should be of Morse--Bott type (see Lemma 2, Eq.~\eqref{Lemma2}) and explicitly allowed its gradient to vanish on the core. As a result, the core emerged as a one-dimensional timelike Morse--Bott critical set (see Definition \eqref{morsebottcriticalset}), where the foliation becomes singular in a controlled manner, thereby completing the geometric construction consistently with the definition of the tube in Eq.~\eqref{tube}.

A second objective of Chapter \ref{chapter3} was to understand both the intrinsic and extrinsic geometry of the leaves inside the tube. In this regard, the scalar Gauss relation for timelike leaves, Eq.~\eqref{Scalargaussfornormal}, turned out to be especially important. Although it arose naturally as part of the geometry, it later became one of the key ingredients in the perturbative stability analysis carried out in Chapter \ref{chapter4}.

A purely geometric construction, however, was not sufficient to establish the tube as a gravitational source. To replace the worldline by a tube at the dynamical level, it was necessary to introduce an action principle based on the foliation-generating scalar field and then couple it to the Einstein--Hilbert action. This was done systematically in Chapter \ref{chapter4}. The Lagrangian density we introduced, Eq.~\eqref{ourlagrangian}, was not chosen ad hoc; rather, it was motivated by the Nambu--Goto action for a stack of space-filling codimension-one branes and then recast in a Polyakov-type form using an auxiliary scalar field. Once coupled to gravity, the theory produced Einstein's equations with the tube as a smooth gravitational source, quantified by a stress-energy tensor without delta-function support, see Eq.~\eqref{ourstressenergy}. Because the tube is canonically codimension zero, the construction is compatible with the Geroch--Traschen obstruction from the outset.

With the action in hand, several further questions became unavoidable. One had to understand the Hamiltonian structure and count the local degrees of freedom (Section \eqref{Hamiltoniandynamics}), analyse perturbative stability in order to identify conditions under which small fluctuations remain under control, and study the ultraviolet and infrared limits of the theory. These last limits were especially important because the ultraviolet limit had to reproduce an effective point-particle action, while the theory also needed to remain well behaved at large transverse scales.

\vspace{2mm}

\noindent\textbf{B. Main results}

\vspace{2mm}

One of the earliest and most important observations was that identifying the foliation-generating scalar field \(\Phi\) as a Morse--Bott function fixes its mass dimension. As discussed in Chapter \ref{chapter3}, this leads to
\begin{equation}
    [\Phi]=-2.
\end{equation}
Because the kinetic term in the on-shell action \eqref{Almostanaction} contains a square root, this nonstandard scaling forces the tension profile to carry a high mass dimension,
\begin{equation}
    [\mathcal{T}]=D+1.
\end{equation}
This immediately shows that the theory is intrinsically scale-sensitive, with the relevant scale set not arbitrarily but by the inverse transverse size of the tube, as discussed around Eq.~\eqref{scale}.

After introducing the appropriate rescaling, see Eq.~\eqref{correctscaling}, we showed that the tube action can be reduced smoothly to a worldline action by integrating out the transverse directions and taking the ultraviolet limit in which the transverse size shrinks to zero. This reduction works for a broad admissible class of tension and potential profiles. The result is conceptually important: the mass that would ordinarily be introduced as a parameter concentrated on the worldline is instead encoded geometrically in the tension and potential profiles of the tube. In this sense, the worldline mass is not fundamental but emergent from the ultraviolet limit of the finite-thickness description, as discussed in subsection \eqref{4.1.1UVIR}.

A particularly striking result appeared when the correctly rescaled theory \eqref{correctlyscaledlagrangian} was analysed in the ultraviolet limit with a general \(\tau\tau\)-component of the induced metric. In that case, the reduced action did not collapse to the standard point-particle action alone: an additional non-redundant term survived, see Eq.~\eqref{splittedaction}. As a consequence, the effective equation of motion for the core was not purely geodesic, but contained an additional force term, Eq.~\eqref{finalformoftheeom}. This force depends on the total mass, the integrated tension and potential profiles, and the transverse geometry of the tube through the condition
\begin{equation}
    \partial_\lambda h_{\tau\tau}(\tau,\lambda)\neq0,
    \qquad
    h_{\tau\tau}\in C^k,\quad k\geq1.
\end{equation}

This force is not sourced by any external matter field, which strongly suggests interpreting it as a self-force. The reason is structural. The ambient metric \(g_{\mu\nu}\) inside the tube is determined by Einstein's equations with the tube stress-energy tensor \eqref{ourstressenergy} as its source. That bulk metric induces a metric on each leaf,
\begin{equation}
    h_{ab}(\lambda,\xi)
    =
    g_{\mu\nu}(X)\,\partial_aX^\mu(\lambda,\xi)\partial_bX^\nu(\lambda,\xi),
\end{equation}
and, generically,
\begin{equation}
    \partial_\lambda h_{ab}\neq0.
\end{equation}
Through this dependence, the induced geometry sources the reduced potential \(\mu(\tau)\) in Eq.~\eqref{mu}, while the gradient \(\nabla_\nu\mu\) in turn governs the effective motion of the core through Eq.~\eqref{finalformoftheeom}. In this way the geometry sourced by the tube feeds back into the motion of the tube's own core, completing a self-consistent loop. This is the principal reason for interpreting \(\mu\) as a self-force potential and \(\hat Q_\nu=\nabla_\nu\mu\) as a self-force. A detailed comparison with the established self-force literature \cite{Harte_2012,Harte_2015,Pound_2010,Pound_2015}, however, lies beyond the scope of this thesis and remains an important direction for future work.

A complementary line of investigation concerned the stress-energy content of the tube itself. In Section \eqref{4.1.2stressenergytensor}, we derived the stress-energy tensor associated with the action \eqref{Ouraction} and analysed its behaviour throughout the interior of the tube. One important result was that the stress-energy tensor remains finite at the core. Moreover, at the core it takes a form closely resembling vacuum-energy stress tensors familiar from de Sitter and anti-de Sitter geometries, depending on the sign of the potential. For \(\mathcal{V}>0\), the core behaves like a one-dimensional de Sitter-type vacuum line; for \(\mathcal{V}<0\), it behaves like an anti-de Sitter-type core; and for \(\mathcal{V}=0\), the core is effectively vacuum.

The stress-energy tensor also determines the energy conditions and therefore controls the focusing properties of timelike and null congruences inside the tube. Since the stress-energy tensor is generically non-vanishing throughout the interior, and may vanish at the core only if \(\mathcal{V}(\Phi_0)=0\), it was necessary to examine whether the geometry inside the tube inevitably develops focusing and caustics. We found that for a broad class of profiles, especially
\begin{equation}
    \mathcal{T}>0,\qquad \mathcal{V}\geq0,
\end{equation}
the strong energy condition is violated throughout the tube, including at the core, while the null and weak energy conditions remain satisfied. On the other hand, for negative potential there exists a narrower admissible window in which all energy conditions may hold, provided in particular that the potential vanishes at the core.

This result is important in light of the singularity theorems \cite{Hawking:1973uf,Hawking:1970zqf}. In the standard Hawking--Penrose framework, the strong energy condition is one of the ingredients that drive the focusing of timelike congruences and ultimately force causal geodesic incompleteness under suitable global assumptions. The violation of the strong energy condition therefore obstructs the usual focusing argument and may help prevent singular structure formation in the tube interior. In this sense, SEC violation is not merely a formal curiosity; it is directly tied to the possibility that the tube avoids gravitational collapse and remains regular inside.

This interpretation is reinforced by the perturbative analysis of the intrinsic geometry of the leaves. In subsection \eqref{4.2.2perturbativestability}, we studied small normal deformations \(\zeta\) of the leaves and found that they obey a Klein--Gordon-type Jacobi equation with an effective mass term, see Eq.~\eqref{jacobiiequation}. Spectral analysis showed that stable oscillatory behaviour is favoured when
\begin{equation}
    M^2\leq0.
\end{equation}
Because the sign of \(M^2\) is not fixed a priori, it was necessary to combine Einstein's equation \eqref{Einstein}, the scalar Gauss relation \eqref{Scalargaussfornormal}, and the scalar field equation \eqref{SoapEquationofmotion} in order to express it in a more useful form. This analysis indicated that the class
\begin{equation}
    \mathcal{T}>0,\qquad \mathcal{V}\geq0
\end{equation}
supports stable oscillatory intrinsic deformations. The perturbative treatment of leaf deformations, which helped identify this safe region of profile space, was inspired in part by the membrane-deformation literature, especially the work of Capovilla \cite{Capovilla}.

In Section \eqref{dynamicsandstability}, we also analysed ordinary field perturbations of the scalar theory itself. There we found that, at quadratic order, the coefficient of the kinetic term vanishes, which formally implies an infinite squared sound speed. This indicates that the foliation-generating field is constraint-like rather than dynamically propagating in the usual sense, even though the Hamiltonian analysis in the bulk gives
\begin{equation}
    N_{\rm bulk}=1.
\end{equation}
The correct interpretation is that the allowed deformation mode is not a standard propagating degree of freedom with finite signal speed, but an instantaneous constraint mode adjusting the configuration. The situation becomes even more restrictive at the core, where the Dirac matrix degenerates and the relevant Poisson brackets vanish. As a result, the constraint becomes first-class there and the effective degree-of-freedom count drops to
\begin{equation}
    N_{\rm core}=0.
\end{equation}
Thus the canonical structure itself distinguishes between the bulk of the tube and the Morse--Bott critical core.

There remain many aspects of tube dynamics that have not been addressed in this thesis. For example, explicit solutions of Einstein's equation in the tube interior have not yet been constructed. Likewise, because the exterior of the tube is vacuum, an important open question concerns the Israel junction conditions on the outermost leaf, and how they determine the globally continuous spacetime metric. These issues lie beyond the scope of the present work. The aim here was more modest but foundational: to construct timelike tubes, Eq.~\eqref{tube}, in such a way that sharply localized gravitating matter on a one-dimensional worldline is avoided geometrically rather than physically smeared. In that goal, the thesis succeeds. Moreover, it yields the additional and unexpected result that self-force-like corrections emerge canonically from the geometry of the tube and survive in the ultraviolet worldline limit.

\vspace{2mm}

\noindent\textbf{C. Open directions}

\vspace{2mm}

One immediate application concerns regular black-hole models. Since the strong energy condition is violated inside the tube for a large class of admissible tension and potential profiles, one may model a regular black-hole interior by treating the outermost leaf of the tube as the surface swept out by the horizon, while the interior is identified with a singularly foliated region possessing a de Sitter core, that is,
\begin{equation}
    \mathcal{V}(\Phi_0)>0.
\end{equation}
This is conceptually close to the logic of regular black-hole geometries, beginning with the original proposal of Bardeen \cite{1968qtr..conf...87B,2767662}. In the Bardeen metric,
\begin{equation*}
    ds^2=-f(r)\,dt^2+f^{-1}(r)\,dr^2+r^2d\Omega_2^2,
\end{equation*}
with
\begin{equation}
    f(r)=1-\frac{2Mr^2}{(r^2+l_0^2)^{3/2}},
\end{equation}
the small-\(r\) region behaves effectively like a de Sitter core,
\begin{equation*}
    f(r)\sim 1-\frac{\Lambda_c r^2}{3},
\end{equation*}
thereby avoiding the central singularity. Since subsection \eqref{4.1.3energycondition} showed that the tube can naturally support a de Sitter-type core when \(\mathcal{V}>0\), the present framework may provide a canonical geometric regularization mechanism for regular black-hole interiors. One may also investigate the thermodynamics of such tube-based regular black holes. A fuller exploration of this possibility is left for future work; see also \cite{konoplya2023bardeenspacetimequantumcorrected,Hayward_2006,1992GReGr..24..235D,ansoldi2008sphericalblackholesregular}.

A second and equally intriguing direction is cosmology. In this thesis we have considered a single isolated tube, but the framework may be extended to a many-body or continuum setting: one could study a tube dust, a homogeneous tube fluid, or even a tube condensate. The central question would then be how such an ensemble contributes to cosmological expansion. Does it behave effectively like matter, radiation, dark energy, dynamical dark energy, or some more exotic fluid? Conceptually, this would amount to replacing the standard picture of a universe filled with particle worldlines by a universe filled with causally admissible tubes of finite transverse size. Exploring this possibility would be a natural next step in understanding the large-scale implications of the tube paradigm.



 \appendix
 \appendixpage
 \addappheadtotoc


\chapter{Appendix} 
\section{Distribution}\label{A.1DISTRIBUTION}

Let $\mathcal{M}$ be a $C^{\infty}$ D-dimensional space-time manifold. Let ${\mathbf{t}}$ be the test field of weight $-1$, with a compact support on $\mathcal{M}$, such that the integral over $\mathcal{M}$ of the scalar test field can be carried without any additional volume element on $\mathcal{M}$, and the contraction of the test fields with any smooth tensor field and its derivative via any smooth derivative operator all yields test fields\cite{GEROCHandtraschen}. A distribution on the manifold $\mathcal{M}$ is defined as a linear mapping from the vector space of test fields of a given index structure to the real numbers satisfying the continuity condition (one can refer to ref.\cite{GEROCHandtraschen} for more details). An important class of distribution is defined in the following way. Let, $\mu_{a\cdots c}^{\ \ \  b\cdots d }$ be a smooth tensor field on $\mathcal{M}$. Now, this tensor field give rise to a distribution $\hat \mu_{a\cdots c}^{\ \ \ b\cdots d }$, with the action
\begin{equation}
    \hat \mu_{a\cdots c}^{\ \ \ b\cdots d }{*} t^{a\cdots c}_{\ \ \ b\cdots d } =\int \mu_{a\cdots c}^{\ \ \  b\cdots d } t^{a\cdots c}_{\ \ \ b\cdots d } 
\end{equation}
Consider a timelike worldline $\gamma\subset\mathcal{M}$, parametrized by proper time $\tau$, with embedding $x^\mu=z^\mu(\tau)$, and four velocity
$u^\mu=\frac{dz^\mu}{d\tau}$.

The stress-energy tensor associated with a point particle of mass m is formally written as
\begin{equation}
    T^{\mu\nu}(x)
    =
    m\int_\gamma
    u^\mu u^\nu
    \delta^{(D)}(x-z(\tau))\,d\tau .
\end{equation}
To show that \(T^{\mu\nu}\) defines a distribution in the sense of Geroch and Traschen, let $t_{\mu\nu}$ be a smooth compactly supported test field of weight \(-1\). Then define the action
\begin{equation}
\begin{aligned}
    \hat T^{\mu\nu} * t_{\mu\nu}
    &=
    \int_{\mathcal M}
    T^{\mu\nu}(x)t_{\mu\nu}(x)
    \\
    &=
    m\int_{\mathcal M}
    \int_\gamma
    u^\mu u^\nu
    \delta^{(D)}(x-z(\tau))
    t_{\mu\nu}(x)
    \,d\tau .
\end{aligned}
\end{equation}
Interchanging the order of integration and using the defining property of the Dirac delta yields
\begin{equation}
    \hat T^{\mu\nu} * t_{\mu\nu}
    =
    m\int_\gamma
    u^\mu u^\nu
    t_{\mu\nu}(z(\tau))
    \,d\tau .
\end{equation}
This quantity is finite because: $t_{\mu\nu}$ is smooth, has compact support, and the intersection of the support of $t_{\mu\nu}$ with the worldline $\gamma$ is compact. Furthermore, the map
\begin{equation*}
    t_{\mu\nu}\mapsto
m\int_\gamma
u^\mu u^\nu
t_{\mu\nu}(z(\tau))
\,d\tau
\end{equation*}
is linear and continuous in the test-field topology. Hence, $T^{\mu\nu}$ defines a tensor distribution on $\mathcal M$. Therefore, the point-particle stress-energy tensor is distributional in the sense of Geroch and Traschen.

\section{Lie Dragging:}\label{A.2LIEDRAGGING}
Lie dragging for various mathematical objects is defines as,
\begin{itemize}

    \item [1.]\textbf{Scalar field:}
    \begin{equation}
        \mathcal{L}_X f
        =
        X^\mu \partial_\mu f
    \end{equation}
\item [2] \textbf{Contravariant vector:}
    \begin{equation}
        (\mathcal{L}_X V)^\mu
        =
        X^\nu \partial_\nu V^\mu
        -
        V^\nu \partial_\nu X^\mu
    \end{equation}
\item[3] \textbf{Covariant vector (one-form):}
    \begin{equation}
        (\mathcal{L}_X \omega)_\mu
        =
        X^\nu \partial_\nu \omega_\mu
        +
        \omega_\nu \partial_\mu X^\nu
    \end{equation}
 \item[4] \textbf{Metric tensor:}
    \begin{equation}
        (\mathcal{L}_X g)_{\mu\nu}
        =
        \nabla_\mu X_\nu
        +
        \nabla_\nu X_\mu
    \end{equation}
\item [5]\textbf{Volume element:}
    \begin{equation}
        \mathcal{L}_X \sqrt{-g}
        =
        (\nabla_\mu X^\mu)\sqrt{-g}
    \end{equation}

\end{itemize}

\section{Integrability conditions\label{A.3INTEGRABILITY}}

\noindent Consider the one form $\underline{\mathbf U}=U_\mu \mathbf{d}x^\mu$ and its exterior derivative $d \underline{\mathbf{U}}$,
\begin{equation}
    d\underline{\mathbf{U}} = \frac{1}{2} [\nabla_\mu U_\nu -\nabla_\nu U_\mu] \mathbf{d}x^\mu \wedge \mathbf{d}x^\nu
\end{equation}
forming a 3-form using the wedge product, 
\begin{equation}
    \underline{\mathbf{U}}\wedge d\underline{\mathbf{U}} =   \frac{1}{2}U_\mu \Big[\nabla_\rho U_\nu -\nabla_\nu U_\rho\Big]\mathbf{d}x^\mu \wedge \mathbf{d}x^\rho \wedge \mathbf{d}x^\nu
\end{equation}
under the full anti-symmetrization for all the three indices we have,
\begin{equation}
    \underline{\mathbf{U}}\wedge d\underline{\mathbf{U}} = \frac{1}{3!}[U_\mu \nabla_\rho U_\nu
+ U_\rho \nabla_\nu U_\mu
+ U_\nu \nabla_\mu U_\rho
- U_\mu \nabla_\nu U_\rho
- U_\rho \nabla_\mu U_\nu
- U_\nu \nabla_\rho U_\mu]\mathbf{d}x^\mu \wedge \mathbf{d}x^\rho \wedge \mathbf{d}x^\nu
\end{equation}
The, vanishing of this 3 form is precisely the integrability condition $U_{[\mu}\nabla_\nu U_{\rho ]}=0$, which if satisfied, implise the non-vorticity of the flow and vanishing of the $\hat \omega_{\mu\nu}$ type of tensor.
\section{Equivalence of the Co-dim(1) Brane Actions}\label{A.4Equaivalentactions}
We show that the Nambu--Goto action for a codimension-one brane,
\begin{equation}
S = -\mathcal{T} \int d^{D-1}\xi\sqrt{-h}
\end{equation}
can be rewritten as a spacetime integral involving a scalar field $\Phi$ whose zero level set defines the hypersurface $\Sigma$. Consider a scalar field $\Phi$ such that the hypersurface is given by $\Phi(x)=0$ with $\nabla_\mu \Phi \neq 0$ on the surface. The Nambu--Goto action is
\begin{equation}
\mathcal{S} = -\mathcal{T} \int_{\Sigma} d^{D-1}\xi \sqrt{-h}
\end{equation}
To rewrite this as a spacetime integral introduce coordinates $(\xi^a, \lambda)$ with $\lambda=\Phi(x)$. The spacetime measure can be written as
\begin{equation}
d^D x \sqrt{-g} = d^{D-1}\xi \ d\lambda \ \sqrt{-g}
\end{equation}
Using the standard identity
\begin{equation}
\sqrt{-g} = \frac{\sqrt{-h}}{\sqrt{|\nabla_\mu \Phi \nabla^\mu \Phi|}}
\end{equation}
one finds
\begin{align}
\int d^D x \sqrt{-g}  \delta(\Phi) \sqrt{|\nabla_\mu \Phi \nabla^\mu \Phi|}
&= \int d^{D-1}\xi d\lambda \ \frac{\sqrt{-h}}{\sqrt{|\nabla \Phi|^2}} \delta(\lambda) \sqrt{|\nabla \Phi|^2} \\
&= \int d^{D-1}\xi d\lambda \ \sqrt{-h} \delta(\lambda) \\
&= \int d^{D-1}\xi \sqrt{-h}
\end{align}
\begin{equation}
\int_{\Sigma} d^{D-1}\xi \sqrt{-h}
= \int d^D x \sqrt{-g} \ \delta(\Phi) \sqrt{|\nabla_\mu \Phi \nabla^\mu \Phi|}
\end{equation}
and the action becomes
\begin{equation}
\mathcal{S} = -\mathcal{T} \int d^D x \sqrt{-g} \ \delta(\Phi) \sqrt{|\nabla_\mu \Phi \nabla^\mu \Phi|}
\end{equation}

\section{Morse-Bott Function}\label{A.5Morsebottfunctions}
Here is the list of definitions, Let, $f:\mathcal{M\to \mathbb R}$ be a smooth function, and $\mathcal{M}$ be a $D$ dimensional space-time manifold, then
\begin{itemize}
    \item[1.]\textbf{Critical Point:} A point $p\in \mathcal{M}$ is critical point of $f$, if the induced map $d_p f: T_p(\mathcal{M})\to T_{f(p)}(\mathbb{R})$ vanishes. Which means for any local chart $(\mathcal{U},\mathbf x)$ for $p\in \mathcal{U}$, the gradient of $f$ vanishes, or $\nabla f|_p=0$. With that $f(p)$ is called the critical value.
    \item[2.]\textbf{Hessian of $f$:} Let $p$ be the critical point of $f$, then the hessian $\mathbf H_p$ is defined as a bilinear form on $T_p(\mathcal{M})$ such that,
    $u,v\in T_p(\mathcal{M})$ , $\mathbf H_p(u,v)=U_p(V(f))$, with $U,V$ being the vector field extensions of vectors $u,v$. For any local chart $(\mathcal{U},\mathbf x)$ for $p\in \mathcal{U}$, the hessian is represented by the matrix,
    \begin{equation}
         H_{\mu\nu} =\nabla_\mu\nabla_\nu f.
    \end{equation}
    \item[3.]\textbf{Nullity and Index:} The index of $f$ at point $p$ is defined as the index of $\mathbf{H_p}$, namely the maximal dimension of the a subspace of $T_p(\mathcal{M})$ on which $\mathbf H_p$ is negative definite. The nullity of $f$ at $p$ is the nullity of $\mathbf H_p$, namely the dimension of subspace of $T_p(\mathcal{M})$ consisting vector $v\in T_p(\mathcal{M})$. such that $\mathbf H_p(v,w)=0$ for all $w\in T_p(\mathcal{M})$ or $u^\mu v^\nu H_{\mu\nu}=0$.
    \item[4.]\textbf{Non-degenerate critical point:} A critical point $p$ is said to be non-degenerate if the nullity of  $f$ at $p$ is trivial, which means the Hessian is non-singular at that point .  If all the critical point of $f$ are non-degenerate, then $f$is called a morse function.
    \item [5.]\textbf{Non-degenerate Critical Submanifold:} Let, $C_f$ be the set of critical points of $f$, and $C$ be the connected components of $C_f$, then $C\subset\mathcal{M}$ is said to be a non-degenerate critical submanifold  if $C$ is connected, and for all points $p\in C$ the induced symmetric form $ \mathbf{h}_pf$ on the normal space $N_p(C)$, is non-degenerate. $\mathbf{h}_pf$ is non-degenerate if and only if $T_p(C)=\rm Ker(H_pf)$, which means $\mathbf H_pf$ is non-degenerate in the direction normal to $C$ at point $p$. In such case $f$ is a Morse-Bott function.
    \item[6.]\textbf{Morse-Bott Lemma}, Let, $f:\mathcal{M}\to \mathbb R$ be a Morse-Bott function and $C$ a connected component of $C_f$, of dimension $k$ as a manifold. Then, for $p\in C$, there exist a local-coordinate system $\mathcal{U},\varphi:\mathcal{U}\to V\subset \mathbb R^k\times \mathbb R^{D-k}$, containing $p$ such that $\varphi(p)=0$, $\varphi(\mathcal{U}\cap C)=\{(x,y)\in V:y=0\}$ and the identity $f=f(C)-y_1^2-y_2^2\cdots-y_\lambda^2+\cdots y_{D-k}^2$, where $\lambda$ is the index of $\mathbf{h}_pf$.
\end{itemize} 
\section{Incomplete-Gamma function and Series Convergence}\label{A.6Incompletegammafunction}
The incomplete gamma function is defined as \cite{abramowitz1965handbook},
\begin{equation}
    \Gamma(a,x) =\int_x^{\infty} t^{a-1}e^{-t}dt
\end{equation}
which is upper incomplete, where as lower incomplete gamma function is defined as,
\begin{equation}
    \gamma(a,x) =\int_0^{x} t^{a-1}e^{-t}dt
\end{equation}
such that,
\begin{equation}
    \Gamma(a) =\gamma(a,x)+\Gamma(a,x)
\end{equation}
Consider the mass we obtained after integrating the transverse geometry of the tube for Gaussian tension and potential profiles 
\begin{equation}
      m(\Lambda)  =\frac{2\pi^{\frac{D-1}{2}}}{\Gamma(\frac{D-1}{2})}\Bigg[\hat{\mathcal{T}}_0\Bigg(\sum_{k=0}^{\infty} \frac{(-1)^k}{k!}\frac{\Gamma(D+2k)\Gamma(3/2)}{\Gamma(D+2k+3/2)}\Bigg)\ +\ \hat{\mathcal{V}}_0\Bigg(\sum_{k=0}^{k=\infty}\frac{(-1)^{k}}{k!}\frac{\Gamma(D+2k-1)\Gamma(3/2)}{\Gamma(D+2k+1/2)}\Bigg)\Bigg]
\end{equation}
Now, to prove that the mass is finite, we need to prove that the series on the RHS is finite, to prove that. Consider the first series
\begin{equation}
S_1
=
\sum_{k=0}^{\infty}
\frac{(-1)^k}{k!}
\frac{\Gamma(D+2k)\Gamma(3/2)}
{\Gamma(D+2k+3/2)}.
\end{equation}
let,
\begin{equation}
a_k
=
\frac{1}{k!}
\frac{\Gamma(D+2k)\Gamma(3/2)}
{\Gamma(D+2k+3/2)}.
\end{equation}
Using Stirling's approximation
\begin{equation}
\Gamma(z)
\sim
\sqrt{2\pi}\,
z^{z-\frac12}e^{-z},
\qquad z\to\infty,
\end{equation}
we obtain
\begin{equation}
\Gamma(D+2k)
\sim
\sqrt{2\pi}
(D+2k)^{D+2k-\frac12}
e^{-(D+2k)},
\end{equation}
and
\begin{equation}
\Gamma(D+2k+3/2)
\sim
\sqrt{2\pi}
(D+2k+3/2)^{D+2k+1}
e^{-(D+2k+3/2)}.
\end{equation}
Therefore,
\begin{equation}
\frac{\Gamma(D+2k)}
{\Gamma(D+2k+3/2)}
\sim
e^{3/2}
\frac{
(D+2k)^{D+2k-\frac12}
}{
(D+2k+3/2)^{D+2k+1}
}.
\end{equation}
For large $k$,
\begin{equation}
D+2k+3/2 \sim D+2k,
\end{equation}
hence
\begin{equation}
\frac{\Gamma(D+2k)}
{\Gamma(D+2k+3/2)}
\sim
(D+2k)^{-3/2}.
\end{equation}
Thus the general term behaves as
\begin{equation}
a_k
\sim
\frac{1}{k!}\frac{1}{(D+2k)^{3/2}}.
\end{equation}
For sufficiently large $k$,
\begin{equation}
(D+2k)^{3/2}\sim k^{3/2},
\end{equation}
and therefore
\begin{equation}
a_k
\sim
\frac{1}{k!\,k^{3/2}}.
\end{equation}
Since the factorial grows faster than any polynomial,
\begin{equation}
\sum_{k=0}^{\infty}
\frac{1}{k!\,k^{3/2}}
\end{equation}
converges absolutely. Similarly, the second series also behaves in the similar fashion
which makes it absolutely summable. Thus both series converge absolutely, implying that $m(\Lambda)$ remains finite, $\square$.



\printbibliography  

@book{Lee2013,
  author       = {Lee, John M.},
  title        = {Introduction to Smooth Manifolds},
  edition      = {2nd},
  series       = {Graduate Texts in Mathematics},
  volume       = {218},
  year         = {2013},
  publisher    = {Springer},
  address      = {New York, NY},
  isbn         = {978-1-4419-9981-8},
  doi          = {10.1007/978-1-4419-9982-5}
}

@book{mukherjee2015differential,
  title={Differential Topology},
  author={Mukherjee, A.},
  isbn={9783319190457},
  url={https://books.google.co.in/books?id=aD8PCgAAQBAJ},
  year={2015},
  publisher={Springer International Publishing}
}

@article{Whitney1,
 ISSN = {0003486X, 19398980},
 URL = {http://www.jstor.org/stable/1968482},
 author = {Hassler Whitney},
 journal = {Annals of Mathematics},
 number = {3},
 pages = {645--680},
 publisher = {[Annals of Mathematics, Trustees of Princeton University on Behalf of the Annals of Mathematics, Mathematics Department, Princeton University]},
 title = {Differentiable Manifolds},
 urldate = {2026-01-18},
 volume = {37},
 year = {1936}
}

@inproceedings{Vergara:2000D/,
  author = "Vergara, J. David",
  title = "{Canonical formulation of the Conformal p-brane}",
  doi = "10.22323/1.005.0039",
  booktitle = "Proceedings of Third Latin American Symposium on High Energy Physics {\textemdash} PoS(silafae-III)",
  year = 2000,
  volume = "005",
  pages = "039"
}

@incollection{Einstein1916English,
  author       = {Einstein, A.},
  title        = {The foundation of the general theory of relativity},
  booktitle    = {The Road to Relativity},
  editor       = {Gutfreund, H. and Renn, J.},
  publisher    = {Princeton University Press},
  year         = {2015},
  pages        = {179--226},
  note         = {English translation of ``Die Grundlage der allgemeinen Relativit{\"a}tstheorie'' (Annalen der Physik, 1916)}
}

@book{Wald:1984rg,
    author = "Wald, Robert M.",
    title = "{General Relativity}",
    doi = "10.7208/chicago/9780226870373.001.0001",
    publisher = "Chicago Univ. Pr.",
    address = "Chicago, USA",
    year = "1984"
}

@book{Weinberg:1972kfs,
    author = "Weinberg, Steven",
    title = "{Gravitation and Cosmology}: {Principles and Applications of the General Theory of Relativity}",
    isbn = {9780471925675},
    publisher = "John Wiley and Sons",
    address = "New York",
    year = "1972"
}

@book{spivak1975comprehensive,
  title={A Comprehensive Introduction to Differential Geometry},
  author={Spivak, M.},
  number={v. 1},
  isbn={9780914098843},
  lccn={73076372},
  series={A Comprehensive Introduction to Differential Geometry},
  url={https://books.google.co.in/books?id=HFTvAAAAMAAJ},
  year={1975},
  publisher={Publish or Perish, Incorporated}
}

@article{Capovilla_1995,
   title={Large deformations of relativistic membranes: A generalization of the Raychaudhuri equations},
   volume={52},
   ISSN={0556-2821},
   url={http://dx.doi.org/10.1103/PhysRevD.52.1072},
   DOI={10.1103/physrevd.52.1072},
   number={2},
   journal={Physical Review D},
   publisher={American Physical Society (APS)},
   author={Capovilla, Riccardo and Guven, Jemal},
   year={1995},
   month=jul, pages={1072–1081} }

@article{Capovilla,
   title={Geometry of deformations of relativistic membranes},
   volume={51},
   ISSN={0556-2821},
   url={http://dx.doi.org/10.1103/PhysRevD.51.6736},
   DOI={10.1103/physrevd.51.6736},
   number={12},
   journal={Physical Review D},
   publisher={American Physical Society (APS)},
   author={Capovilla, Riccardo and Guven, Jemal},
   year={1995},
   month=jun, pages={6736–6743} }

@book{Poisson:2009pwt,
    author = "Poisson, Eric",
    title = "{A Relativist's Toolkit: The Mathematics of Black-Hole Mechanics}",
    doi = "10.1017/CBO9780511606601",
    publisher = "Cambridge University Press",
    month = "12",
    year = "2009"
}

@misc{BlauGRNotes,
  author       = {Mathias Blau},
  title        = {Lecture Notes on General Relativity},
  howpublished = {\url{http://www.blau.itp.unibe.ch/newlecturesGR.pdf}},
  note         = {University of Bern lecture notes},
  year         = {2015},
}

@article{Barack_2018,
   title={Self-force and radiation reaction in general relativity},
   volume={82},
   ISSN={1361-6633},
   url={http://dx.doi.org/10.1088/1361-6633/aae552},
   DOI={10.1088/1361-6633/aae552},
   number={1},
   journal={Reports on Progress in Physics},
   publisher={IOP Publishing},
   author={Barack, Leor and Pound, Adam},
   year={2018},
   month=nov, pages={016904} }

@book{Misner:1973prb,
    author = "Misner, Charles W. and Thorne, K. S. and Wheeler, J. A.",
    title = "{Gravitation}",
    isbn = {9780716703440},
    publisher = "W. H. Freeman",
    address = "San Francisco",
    year = "1973"
}

@inbook{Pound_2015,
   title={Motion of Small Objects in Curved Spacetimes: An Introduction to Gravitational Self-Force},
   ISBN={9783319183350},
   ISSN={2365-6425},
   url={http://dx.doi.org/10.1007/978-3-319-18335-0_13},
   DOI={10.1007/978-3-319-18335-0_13},
   booktitle={Equations of Motion in Relativistic Gravity},
   publisher={Springer International Publishing},
   author={Pound, Adam},
   year={2015},
   pages={399–486} }

@article{GEROCHandtraschen,
  title = {Strings and other distributional sources in general relativity},
  author = {Geroch, Robert and Traschen, Jennie},
  journal = {Phys. Rev. D},
  volume = {36},
  issue = {4},
  pages = {1017--1031},
  numpages = {0},
  year = {1987},
  month = {8},
  publisher = {American Physical Society},
  doi = {10.1103/PhysRevD.36.1017},
  url = {https://link.aps.org/doi/10.1103/PhysRevD.36.1017}
}

@book{oliva2002geometric,
  title={Geometric Mechanics},
  author={Oliva, W.M.},
  number={no. 1798},
  isbn={9783540442424},
  lccn={02032396},
  series={Geometric Mechanics},
  url={https://books.google.co.in/books?id=CBwkHrEpQ3oC},
  year={2002},
  publisher={Springer}
}

@book{petersen2006riemannian,
  title={Riemannian Geometry},
  author={Petersen, P.},
  isbn={9780387294032},
  lccn={97005786},
  series={Graduate Texts in Mathematics},
  url={https://books.google.gm/books?id=9cekXdo52hEC},
  year={2006},
  publisher={Springer New York}
}

@book{jost2013riemannian,
  title={Riemannian Geometry and Geometric Analysis},
  author={Jost, J.},
  isbn={9783662223857},
  lccn={97045047},
  series={Universitext},
  url={https://books.google.co.in/books?id=VRz2CAAAQBAJ},
  year={2013},
  publisher={Springer Berlin Heidelberg}
}

@book{o1983semi,
  title={Semi-Riemannian Geometry With Applications to Relativity},
  author={O'Neill, B.},
  isbn={9780080570570},
  series={Pure and Applied Mathematics},
  url={https://books.google.co.in/books?id=CGk1eRSjFIIC},
  year={1983},
  publisher={Academic Press}
}

@misc{gourgoulhon200731formalismbasesnumerical,
      title={3+1 Formalism and Bases of Numerical Relativity}, 
      author={Eric Gourgoulhon},
      year={2007},
      eprint={gr-qc/0703035},
      archivePrefix={arXiv},
      primaryClass={gr-qc},
      url={https://arxiv.org/abs/gr-qc/0703035}, 
}

@article{Lawson1974,
  author  = {Lawson, H. Blaine Jr.},
  title   = {Foliations},
  journal = {Bulletin of the American Mathematical Society},
  volume  = {80},
  number  = {3},
  pages   = {369--418},
  year    = {1974},
  doi     = {10.1090/S0002-9904-1974-13432-4},
  mrnumber = {0343289},
  note    = {MSC (1970): Primary 57D30}
}

@inbook{submersion,

title = {Pseudo-Riemannian Submersions},
booktitle = {Pseudo-Riemannian Geometry and Applications},
chapter = {},
pages = {227-240},
doi = {10.1142/9789814329644_0011},
URL = {https://www.worldscientific.com/doi/abs/10.1142/9789814329644_0011},
eprint = {https://www.worldscientific.com/doi/pdf/10.1142/9789814329644_0011},

}

@misc{laurentgengoux2024invitationsingularfoliations,
      title={An invitation to singular foliations}, 
      author={Camille Laurent-Gengoux and Ruben Louis and Leonid Ryvkin},
      year={2024},
      eprint={2407.14932},
      archivePrefix={arXiv},
      primaryClass={math.DG},
      url={https://arxiv.org/abs/2407.14932}, 
}

@article{Mansouri:1996ps,
    author = "Mansouri, R. and Khorrami, M.",
    title = "{Equivalence of Darmois-Israel and distributional methods for thin shells in general relativity}",
    eprint = "gr-qc/9608029",
    archivePrefix = "arXiv",
    doi = "10.1063/1.531740",
    journal = "J. Math. Phys.",
    volume = "37",
    pages = "5672--5683",
    year = "1996"
}

@article{Israel:1966rt,
    author = "Israel, W.",
    title = "{Singular hypersurfaces and thin shells in general relativity}",
    doi = "10.1007/BF02710419",
    journal = "Nuovo Cim. B",
    volume = "44S10",
    pages = "1",
    year = "1966",
    note = "[Erratum: Nuovo Cim.B 48, 463 (1967)]"
}

@article{Afshordi_2007,
   title={Causal field theory with an infinite speed of sound},
   volume={75},
   ISSN={1550-2368},
   url={http://dx.doi.org/10.1103/PhysRevD.75.083513},
   DOI={10.1103/physrevd.75.083513},
   number={8},
   journal={Physical Review D},
   publisher={American Physical Society (APS)},
   author={Afshordi, Niayesh and Chung, Daniel J. H. and Geshnizjani, Ghazal},
   year={2007},
   month=Apr }

@article{Codimmorsebott,
    author = {Scárdua, Bruno and Seade, José},
    title = {Codimension 1 foliations with Bott–Morse singularities II},
    journal = {Journal of Topology},
    volume = {4},
    number = {2},
    pages = {343-382},
    year = {2011},
    month = {01},
    issn = {1753-8416},
    doi = {10.1112/jtopol/jtr004},
    url = {https://doi.org/10.1112/jtopol/jtr004},
    eprint = {https://academic.oup.com/jtopol/article-pdf/4/2/343/2743680/jtr004.pdf},
}

@misc{scardua2006codimensionfoliationsbottmorsesingularities,
      title={Codimension one foliations with Bott-Morse singularities I}, 
      author={Bruno Scardua and Jose Seade},
      year={2006},
      eprint={math/0608585},
      archivePrefix={arXiv},
      primaryClass={math.DG},
      url={https://arxiv.org/abs/math/0608585}, 
}

@article{POLYAKOV1981207,
title = {Quantum geometry of bosonic strings},
journal = {Physics Letters B},
volume = {103},
number = {3},
pages = {207-210},
year = {1981},
issn = {0370-2693},
doi = {https://doi.org/10.1016/0370-2693(81)90743-7},
url = {https://www.sciencedirect.com/science/article/pii/0370269381907437},
author = {A.M. Polyakov},
}

@book{Peskin:1995ev,
    author = "Peskin, Michael E. and Schroeder, Daniel V.",
    title = "{An Introduction to quantum field theory}",
    doi = "10.1201/9780429503559",
    isbn = {9780201503975, 9780429503559, 9780429494178},
    publisher = "Addison-Wesley",
    address = "Reading, USA",
    year = "1995"
}

@article{Brans:1961sx,
    author = "Brans, C. and Dicke, R. H.",
    editor = "Hsu, Jong-Ping and Fine, D.",
    title = "{Mach's principle and a relativistic theory of gravitation}",
    doi = "10.1103/PhysRev.124.925",
    journal = "Phys. Rev.",
    volume = "124",
    pages = "925--935",
    year = "1961"
}

@article{Harte_2012,
   title={Mechanics of extended masses in general relativity},
   volume={29},
   ISSN={1361-6382},
   url={http://dx.doi.org/10.1088/0264-9381/29/5/055012},
   DOI={10.1088/0264-9381/29/5/055012},
   number={5},
   journal={Classical and Quantum Gravity},
   publisher={IOP Publishing},
   author={Harte, Abraham I},
   year={2012},
   month=Feb, pages={055012} 
   }

@inbook{Harte_2015,
   title={Motion in Classical Field Theories and the Foundations of the Self-force Problem},
   ISBN={9783319183350},
   ISSN={2365-6425},
   url={http://dx.doi.org/10.1007/978-3-319-18335-0_12},
   DOI={10.1007/978-3-319-18335-0_12},
   booktitle={Equations of Motion in Relativistic Gravity},
   publisher={Springer International Publishing},
   author={Harte, Abraham I.},
   year={2015},
   pages={327–398} }

@article{Pound_2010,
   title={Self-consistent gravitational self-force},
   volume={81},
   ISSN={1550-2368},
   url={http://dx.doi.org/10.1103/PhysRevD.81.024023},
   DOI={10.1103/physrevd.81.024023},
   number={2},
   journal={Physical Review D},
   publisher={American Physical Society (APS)},
   author={Pound, Adam},
   year={2010},
   month=Jan }

@book{Hawking:1973uf,
    author = "Hawking, Stephen W. and Ellis, George F. R.",
    title = "{The Large Scale Structure of Space-Time}",
    doi = "10.1017/9781009253161",
    isbn = {9781009253161},
    publisher = "Cambridge University Press",
    series = "Cambridge Monographs on Mathematical Physics",
    month = "2",
    year = "2023"
}

@article{Hawking:1970zqf,
    author = "Hawking, S. W. and Penrose, R.",
    title = "{The Singularities of gravitational collapse and cosmology}",
    doi = "10.1098/rspa.1970.0021",
    journal = "Proc. Roy. Soc. Lond. A",
    volume = "314",
    pages = "529--548",
    year = "1970"
}

@INPROCEEDINGS{1968qtr..conf...87B,
       author = {{Bardeen}, James},
        title = "{Non-singular general relativistic gravitational collapse}",
    booktitle = {Proceedings of the 5th International Conference on Gravitation and the Theory of Relativity},
         year = 1968,
        month = sep,
        pages = {87},
       adsurl = {https://ui.adsabs.harvard.edu/abs/1968qtr..conf...87B}
}

@article{2767662,
    author = "Bardeen, James",
    title = "{Nonsingular general relativistic gravitational collapse}"
}

@misc{konoplya2023bardeenspacetimequantumcorrected,
      title={Bardeen spacetime as a quantum corrected Schwarzschild black hole: Quasinormal modes and Hawking radiation}, 
      author={R. A. Konoplya and D. Ovchinnikov and B. Ahmedov},
      year={2023},
      eprint={2307.10801},
      archivePrefix={arXiv},
      primaryClass={gr-qc},
      url={https://arxiv.org/abs/2307.10801}, 
}

@article{Hayward_2006,
   title={Formation and Evaporation of Nonsingular Black Holes},
   volume={96},
   ISSN={1079-7114},
   url={http://dx.doi.org/10.1103/PhysRevLett.96.031103},
   DOI={10.1103/physrevlett.96.031103},
   number={3},
   journal={Physical Review Letters},
   publisher={American Physical Society (APS)},
   author={Hayward, Sean A.},
   year={2006},
   month=Jan }

@ARTICLE{1992GReGr..24..235D,
       author = {{Dymnikova}, Irina},
        title = "{Vacuum nonsingular black hole}",
      journal = {General Relativity and Gravitation},
         year = 1992,
        month = mar,
       volume = {24},
       number = {3},
        pages = {235-242},
          doi = {10.1007/BF00760226},
       adsurl = {https://ui.adsabs.harvard.edu/abs/1992GReGr..24..235D}
}

@misc{ansoldi2008sphericalblackholesregular,
      title={Spherical black holes with regular center: a review of existing models including a recent realization with Gaussian sources}, 
      author={Stefano Ansoldi},
      year={2008},
      eprint={0802.0330},
      archivePrefix={arXiv},
      primaryClass={gr-qc},
      url={https://arxiv.org/abs/0802.0330}, 
}

@book{abramowitz1965handbook,
  title={Handbook of Mathematical Functions: With Formulas, Graphs, and Mathematical Tables},
  author={Abramowitz, M. and Stegun, I.A.},
  isbn={9780486612720},
  lccn={lc65012253},
  series={Applied mathematics series},
  url={https://books.google.co.in/books?id=MtU8uP7XMvoC},
  year={1965},
  publisher={Dover Publications}
}

@article{Wald:2009ue,
    author = "Wald, Robert M.",
    editor = "Blanchet, Luc and Spallicci, Alessandro and Whiting, Bernard",
    title = "{Introduction to Gravitational Self-Force}",
    eprint = "0907.0412",
    archivePrefix = "arXiv",
    primaryClass = "gr-qc",
    journal = "Fundam. Theor. Phys.",
    volume = "162",
    pages = "253--262",
    year = "2011"
}

@article{Dixon,
 ISSN = {00804630},
 URL = {http://www.jstor.org/stable/2416466},
 author = {W. G. Dixon},
 journal = {Proceedings of the Royal Society of London. Series A, Mathematical and Physical Sciences},
 number = {1519},
 pages = {499--527},
 publisher = {The Royal Society},
 title = {Dynamics of Extended Bodies in General Relativity. I. Momentum and Angular Momentum},
 urldate = {2026-05-21},
 volume = {314},
 year = {1970}
}

@ARTICLE{papapetrou,
       author = {{Papapetrou}, A.},
        title = "{Spinning Test-Particles in General Relativity. I}",
      journal = {Proceedings of the Royal Society of London Series A},
         year = 1951,
        month = oct,
       volume = {209},
       number = {1097},
        pages = {248-258},
          doi = {10.1098/rspa.1951.0200},
       adsurl = {https://ui.adsabs.harvard.edu/abs/1951RSPSA.209..248P}
}

@article{Dirac,
    author = {Dirac, Paul Adrien Maurice},
    title = {Classical theory of radiating electrons},
    journal = {Proceedings of the Royal Society of London. A. Mathematical and Physical Sciences},
    volume = {167},
    number = {929},
    pages = {148-169},
    year = {1938},
    month = {08},
    issn = {0080-4630},
    doi = {10.1098/rspa.1938.0124},
    url = {https://doi.org/10.1098/rspa.1938.0124},
    eprint = {https://royalsocietypublishing.org/rspa/article-pdf/167/929/148/35258/rspa.1938.0124.pdf},
}

@article{Duff:1987cs,
    author = "Duff, M. J. and Inami, T. and Pope, C. N. and Sezgin, E. and Stelle, K. S.",
    title = "{Semiclassical Quantization of the Supermembrane}",
    reportNumber = "CERN-TH-4731/87, IC/87/74",
    doi = "10.1016/0550-3213(88)90316-1",
    journal = "Nucl. Phys. B",
    volume = "297",
    pages = "515--538",
    year = "1988"
}

@book{Zwiebach_2009, place={Cambridge}, edition={2}, title={A First Course in String Theory}, publisher={Cambridge University Press}, author={Zwiebach, Barton}, year={2009}}

\end{document}